\documentclass[journal]{IEEEtran}

\usepackage{afterpage}
\usepackage{algorithm}
\usepackage{algorithmic}
\usepackage{amsmath}
\usepackage{amssymb}
\usepackage{cite}
\usepackage{color}
\usepackage{dblfloatfix}
\usepackage{graphicx}
\usepackage{empheq}
\usepackage{epstopdf}
\usepackage{float}
\usepackage{soul}
\usepackage{subcaption}
\usepackage{placeins}
\usepackage[normalem]{ulem}
\usepackage{url}

\DeclareGraphicsExtensions{.eps,.pdf}

\renewcommand\vec{\mathbf}

\newcommand{\overbar}[1]{\mkern 1.5mu\overline{\mkern-1.5mu#1\mkern-1.5mu}\mkern 1.5mu}

\title{Regularized Dual Averaging Image Reconstruction for Full-Wave Ultrasound Computed Tomography}

\author{
	\IEEEauthorblockN{Thomas P. Matthews\IEEEauthorrefmark{1}}
	\IEEEauthorblockN{Kun Wang\IEEEauthorrefmark{1}}
	\IEEEauthorblockN{Cuiping Li\IEEEauthorrefmark{2}}
	\IEEEauthorblockN{Neb Duric\IEEEauthorrefmark{2}\IEEEauthorrefmark{3}}
	\IEEEauthorblockN{Mark A. Anastasio\IEEEauthorrefmark{1}} \\
	\IEEEauthorblockA{\IEEEauthorrefmark{1}Department of Biomedical Engineering, School of Engineering and Applied Science, \\ Washington University in St. Louis, St. Louis, MO 63130} \\
	\IEEEauthorblockA{\IEEEauthorrefmark{2}Delphinus Medical Technologies, Plymouth, MI 48170} \\
	\IEEEauthorblockA{\IEEEauthorrefmark{3}Karmanos Cancer Institute, Wayne State University, Detroit, MI 48201}
}

\begin{document}

	\maketitle
		
	\begin{abstract}
		Ultrasound computed tomography (USCT) holds great promise for breast cancer screening. Waveform inversion-based image reconstruction methods account for higher order diffraction effects and can produce high-resolution USCT images, but are computationally demanding. Recently, a source encoding technique was combined with stochastic gradient descent to greatly reduce image reconstruction times. However, this method bundles the stochastic data fidelity term with the deterministic regularization term. This limitation can be overcome by replacing stochastic gradient descent (SGD) with a structured optimization method, such as the regularized dual averaging (RDA) method, that exploits knowledge of the composition of the cost function. In this work, the dual averaging method is combined with source encoding techniques to improve the effectiveness of regularization while maintaining the reduced reconstruction times afforded by source encoding. It is demonstrated that each iteration can be decomposed into a gradient descent step based on the data fidelity term and a proximal update step corresponding to the regularization term. Furthermore, the regularization term is never explicitly differentiated, allowing non-smooth regularization penalties to be naturally incorporated. The wave equation is solved by use of a time-domain method. The effectiveness of this approach is demonstrated through computer-simulation and experimental studies. The results suggest that the dual averaging method can produce images with less noise and comparable resolution to those obtained by use of stochastic gradient descent.
	\end{abstract}

\begin{IEEEkeywords}
Ultrasound computed tomography, waveform inversion, sound speed imaging, image reconstruction
\end{IEEEkeywords}

	\section{Introduction}
	Ultrasound computed tomography (USCT) shows promise for a number of applications including breast cancer screening \cite{li_vivo_2009,ruiter_realization_2011,schreiman_ultrasound_1984, wiskin_non-linear_2012,greenleaf_quantitative_1977}. USCT is ideally suited to breast imaging as it offers novel tissue contrasts that can help differentiate benign masses from tumors \cite{greenleaf_quantitative_1977}. It has several potential advantages over conventional imaging methods, as it is radiation-free, breast-compression-free, and relatively inexpensive. In addition, ultrasound imaging may offer some advantages over mammography for the detection of breast cancer in women with dense breasts \cite{duric_detection_2007,kolb_comparison_2002}. A variety of studies have been reported demonstrating the application of USCT to breast imaging \cite{carson_breast_1981,zalev_clinical_2012,xia_enhancement_2013,duric_detection_2007,andre1997high,johnson_apparatus_1999,manohar_concomitant_2007,sandhu_frequency_2015,bosch_breast_2014,bosch_high-resolution_2015}, with clinical measurements of breast cancer patients having already been performed \cite{duric_clinical_2014,schreiman_ultrasound_1984}. While USCT has several potential contrast mechanisms, in this study we focus on the estimation of the sound speed distribution. 
	
	%Typically, in USCT, a collection of ultrasonic transducers are placed around an object of interest. A subset of the transducers, known as emitters, emit an ultrasonic pulse, one after the other. Each pulse propagates through the object, and the resulting pressure wave is recorded by all of the transducers. This gives rise to a collection of pressure measurements, corresponding to different tomographic views about the object. The image reconstruction problem is then to estimate the sound speed from this collection of pressure measurements.
	
	Most USCT image reconstruction methods are based on linearized solutions
 to the acoustic wave equation \cite{schreiman_ultrasound_1984,duric_detection_2007,hormati_robust_2010,quan2007sound,huthwaite_high-resolution_2011,norton1982correcting,andre1997high}.
 %\cite{wiskin_non-linear_2012,lavarello_density_2010,zhang_efficient_2012,pratt_sound-speed_2007,hormati_robust_2010,lavarello_density_2009,huthwaite_high-resolution_2011,simonetti_multiple_2006,hesford_fast_2010}.
% Ray-based methods, which are based on geometric acoustics, are a popular example. This approach consists of two major steps. First, the time-of-flight (TOF) between each pair of transducers is estimated from the measured pressure data. Second, an integral geometry model is used to estimate the sound speed from the TOF data \cite{li_improved_2009}. This approach ignores higher-order diffraction effects, which limits the resolution of the reconstructed images \cite{pratt_sound-speed_2007,roy_breast_2013,bates_manageable_1991}.
While such methods can possess computational efficient implementations, the spatial resolution of the resulting images can be severely limited
by neglection of acoustic diffraction effects in the imaging model.
 This can hinder breast cancer screening where the ability to identify  small tumors and
  fine features to distinguish cancerous and benign lesions is of great importance. 
To circumvent the limitations of linearized methods, waveform inversion methods seek to directly invert the acoustic wave
equation without relying on linearizations \cite{wiskin_non-linear_2012,lavarello_density_2010,zhang_efficient_2012,pratt_sound-speed_2007,hormati_robust_2010,lavarello_density_2009,hesford_fast_2010,herraiz_full-wave_2016}.
Because they can accurately account for the acoustic wave physics, waveform inversion methods can produce high resolution images; however, these non-linear methods are computationally burdensome and generally correspond to non-convex optimization problems.
Waveform inversion methods can be classified
 by whether they solve the wave equation by use of a time-domain method or a frequency-domain method. While frequency-domain methods have been successfully applied to USCT image reconstruction \cite{sandhu_frequency_2015}, here we focus on time-domain methods \cite{mulder_time-_2002,colton_inverse_1998}.

 Recently, an approach that combines waveform inversion with source encoding, which alleviates much of the computational burden, was proposed \cite{wang_waveform_2015, zhang_efficient_2012, krebs_fast_2009}. In \cite{wang_waveform_2015}, the sound speed distribution was estimated by solving an optimization problem,
 where the cost function consisted of two terms.
 The first term is a data fidelity term. For this term, the pressure at the transducer locations is calculated based on the current estimate of the sound speed and an acoustic model described by the acoustic wave equation. 
This term quantifies how closely this estimated pressure matches the measured pressure. As described below, when the source encoding technique is employed,
the data fidelity term corresponds to the expectation of a random quantity.
 The second term is a deterministic regularization term, which is used to incorporate \textit{a priori} information about the image. This optimization problem was solved by use of stochastic gradient descent. Under this approach, the stochastic data fidelity term and the deterministic regularization term are treated jointly as part of a single cost function. This approach ignores information about the structure of the cost function and requires use of a differentiable regularization function\cite{xiao_dual_2010}.
	
	Here, we propose use of a structured optimization method, known as the regularized dual averaging method (RDA), that considers the two terms in the cost function separately \cite{xiao_dual_2010, nesterov_primal-dual_2009}. This approach can mitigate the impact of the stochastic data fidelity on the deterministic regularization term and result in more effective regularization that offers superior trade-offs between image resolution and noise variance by exploiting the structure of the cost function. It also provides the opportunity, for the first time, to employ non-smooth
penalties in the waveform inversion cost function, which can be designed to exploit certain sparseness properties of the object \cite{starck2010sparse,baraniuk2007compressive,bian_evaluation_2010}.  
	
	The remainder of the paper is organized as follows. In Section II, a discrete-to-discrete USCT imaging model and the waveform with source encoding method are reviewed. Stochastic gradient descent is discussed briefly. In Section III, the regularized dual averaging method and its application to USCT image reconstruction are described. Computer-simulation studies and experimental results are presented in Sections IV and V, respectively. Finally, the paper concludes with a summary in Section VI.

	\section{Background} \label{sec:background}
		\subsection{Discrete-to-discrete USCT imaging model}	
		While digital imaging systems are naturally described by a continuous-to-discrete (C-D) imaging model\cite{barrett_foundations_2004}, it is typically necessarily to approximate this model as a discrete-to-discrete (D-D) mapping in order to facilitate use of iterative image reconstruction algorithms. For simplicity, the D-D model is presented directly.
			
A canonical 2D USCT imaging system that employs a circular transducer array \cite{anastasio2012emerging} that surrounds the object is considered.	
%		In USCT, a collection of ultrasonic transducers are placed around an object of interest. 
Ultrasound pulses are transmitted through the object and measured by the transducers. Often, only one transducer will emit a pulse at a given time, with the pressure being recorded by all other transducers. A subset of the transducers will each serve as the emitter in turn, leading to a collection of measurements corresponding to different views of the object. The propagation of the ultrasound waves is governed by the acoustic wave equation, which can be solved by a numerical wave equation solver. This solver can be formulated as a D-D mapping as described below. In this study, the wave equation was solved by the k-space pseudo-spectral method \cite{tabei_k-space_2002,mast_k-space_2001,treeby_k-wave:_2010}.
			
		Let $\vec{c} \in \mathbb{R}^N$ denote the finite-dimensional representation, in a pixel basis, of
the sought-after  sound speed distribution. Here, $N$ is the number of pixels in the simulation grid employed by the numerical wave solver. The propagation of the pressure wave through the object when the $m$-th transducer is the emitter can be denoted
		\begin{align}
			\vec{g}_m = \mathbf{M} \mathbf{H}\left(\vec{c}\right) \vec{s}_m , \label{eqn:dd_model}
		\end{align}
		where $\vec{s}_m \in \mathbb{R}^{NL}$ is the emitted pulse, $\vec{g}_m \in \mathbb{R}^{N^{rec} L}$ is the pressure at each transducer, $\mathbf{H}\left(\vec{c}\right) \in \mathbb{R}^{NL \times NL}$ is the operator that denotes the action of the wave equation, $\mathbf{M} \in \mathbb{R}^{N^{rec} L \times NL}$ is a sampling matrix that computes the pressure at the transducer locations from the pressure over the entire simulation grid, $L$ is the number of time points employed by the wave solver, and $N^{rec}$ is the number of transducers acting as receivers. The notation $\mathbf{H}\left(\vec{c}\right)$ is used to emphasize the dependence of $\mathbf{H}$ on the sound speed $\vec{c}$. 
		
		An estimate of the sound speed can be obtained by solving the penalized least-squares optimization problem:
		\begin{align}
			\hat{\vec{c}} = \arg\min_{\vec{c}} \frac{1}{2} \sum_{m = 0}^{M-1} \| \underline{\vec{g}_m} - \mathbf{M} \mathbf{H}\left(\vec{c}\right) \vec{s}_m \|^2_2 + \lambda \mathcal{R}\left(\vec{c}\right) , \label{eqn:seq_cost_fnc}
		\end{align}
		where $M$ is the total number of views, $\underline{\vec{g}_m}$ is the measured pressure at each transducer, $\mathcal{R}\left(\vec{c}\right)$ is a regularization function, and $\lambda$ is a regularization parameter, which controls the relative weight of the regularization term. The first term in Eqn.~(\ref{eqn:seq_cost_fnc}), known as the data fidelity term, is a non-convex function of $\mathbf{c}$, while the regularization function is assumed to be a convex function.
		
		This approach can produce high resolution images, but it is computationally very expensive. Each evaluation of the cost function requires the wave equation to be solved $M$ times. This high computational cost has limited the wide-spread use of time-domain-based waveform inversion methods.
		
		\subsection{Waveform inversion with source encoding}
		Recently, a source encoding technique has been employed to efficiently find the solution of Eqn.~(\ref{eqn:seq_cost_fnc}) \cite{wang_waveform_2015,zhang_efficient_2012}. In the waveform inversion with source encoding (WISE) method \cite{wang_waveform_2015}, Eqn.~(\ref{eqn:seq_cost_fnc}) is reformulated as the stochastic optimization problem
		\begin{align}
			\hat{\vec{c}} = \arg\min_{\vec{c}} \mathbf{E}_\vec{w} \left\lbrace \frac{1}{2}  \|  \underline{\vec{g}_w} - \mathbf{M} \mathbf{H}\left(\vec{c}\right) \vec{s}_w \|^2_2 \right\rbrace + \lambda \mathcal{R}\left(\vec{c}\right) , \label{eqn:stoc_cost_fnc}
		\end{align} 
		where $\vec{w}$ is a random encoding vector, $\mathbf{E}_\vec{w}$ denotes the expectation with respect to $\vec{w}$, and
		%\begin{empheq}[box=\colorbox{yellow}]{align}
		\begin{align}
			\underline{\vec{g}_w} &= \sum_{m = 0}^{M-1} \left[\vec{w}\right]_m \underline{\vec{g}_m}  \\ \underline{\vec{s}_w} &= \sum_{m = 0}^{M-1} \left[\vec{w}\right]_m \underline{\vec{s}_m}
		\end{align}
		%\end{empheq}
		are the encoded measured pressure data and the encoded source term, respectively. Here, $\mathbf{w}$ is chosen according to a Rademacher distribution as suggested by \cite{van_leeuwen_seismic_2011}. Under this formulation, evaluating the cost function for a particular choice of $\mathbf{w}$ requires the wave equation to be solved only once. When the number of views is large, this can substantially reduce the computational time needed to reconstruct an image.	The gradient of the data fidelity term is calculated using an adjoint state method as described in \cite{wang_waveform_2015}. This approach allows the gradient to be estimated by solving the acoustic wave equation only one additional time (on top of what is already needed to evaluate the cost function). Knowledge of the gradient allows use of a variety of optimization algorithms.
		
		In \cite{wang_waveform_2015}, Eqn.~(\ref{eqn:stoc_cost_fnc}) was solved by use of the stochastic gradient descent (SGD) method, as described in Algorithm \ref{alg:sgd}. In that approach, at each iteration, the gradient of the cost function is evaluated for a single realization of the encoding vector. The update step for the $(k+1)$-th iteration for SGD is given by \cite{sra_optimization_2012}
		%\begin{empheq}[box=\colorbox{yellow}]{multline}
		\begin{multline}
		%\begin{split}
			\vec{c}_{k+1} = \arg\min_\vec{c} \Big\lbrace \left\langle \nabla_\vec{c} f\left(\vec{c}_k, \vec{w}_k\right) , \vec{c} \right\rangle + \\ \frac{1}{2\alpha_k} \| \vec{c} - \vec{c}_k \|_2^2 + \lambda \mathcal{R}\left(\vec{c}\right) \Big\rbrace 
		%\end{split}
		\end{multline}
		%\end{empheq}
		or equivalently,
		%\begin{empheq}[box=\colorbox{yellow}]{align}
		\begin{align}
			\vec{c}_{k+1} &= \vec{c}_k - \alpha_k \left( \nabla_\vec{c} f\left(\vec{c}_k, \vec{w}_k\right) + \lambda \nabla_\vec{c} \mathcal{R}\left(\vec{c}_k\right) \right),
		\end{align}
		%\end{empheq}
		where $\langle \cdot, \cdot \rangle$ denotes the standard Euclidean inner product, $\alpha_k$ is the step size, $\nabla_\vec{c}$ is the gradient with respect to $\vec{c}$, and 
		\begin{align}
			f\left(\vec{c}, \vec{w}\right) \equiv \frac{1}{2}  \|  \underline{\vec{g}_w} - \mathbf{M} \mathbf{H}\left(\vec{c}\right) \vec{s}_w \|^2_2 .
		\end{align} 
		
		This approach has several limitations. First, it fails to exploit the structure of the objective function. Namely, SGD treats the cost function as a black-box, ignoring potentially useful information about the nature of the cost function. For example, in the problem above, the cost function consists of two terms: a stochastic, but differentiable data fidelity term and a deterministic regularization term. In SGD, this knowledge is ignored, and the gradients of the stochastic and deterministic terms are lumped together. Second, it assumes that all terms in the cost function are differentiable. This is not true of many sparsity-promoting regularization functions, such as the $\ell_1$-norm and the total-variation (TV) semi-norm. In \cite{wang_waveform_2015}, the TV semi-norm was approximated by a smoothed, differentiable version through the introduction of a small smoothing parameter. While this approach can be effective, modifications to other non-smooth regularization functions could be more challenging. Third, it fails to exploit information from previous iterations. For SGD, at each iteration, only the gradient for a single realization of the encoding vector is considered when determining the search direction. When combined with a line search for choosing the step size, this can lead to overfitting \cite{schraudolph_stochastic_2007}. In this case, the line search method will choose a large step that effectively minimizes the cost function evaluated for a single realization of the encoding vector, but which increases, or less effectively minimizes, the cost function evaluated for a large number of realizations. This problem can be overcome by use of a fixed step size, at the expense of slowing the convergence rate.
		
		\begin{algorithm}
			\caption{Stochastic gradient descent (SGD) \label{alg:sgd}}
			\algsetup{indent=2em}
			\begin{algorithmic}[1]
				\REQUIRE $\vec{c}_0$, $\lambda$
				\ENSURE $\hat{\vec{c}}$
				  \STATE {$k \gets 0$} \COMMENT{$k$ is the algorithm iteration number.}
				  \WHILE {stopping criterion is not satisfied}
				  \STATE{Draw $\vec{w}_k$ according to chosen distribution.}
				  \STATE{Calculate $\vec{G}_k \gets \nabla_\vec{c} f\left(\vec{c}_k , \vec{w}_k \right) + \lambda \nabla_\vec{c} \mathcal{R}\left(\vec{c}_k\right)$}
				  \STATE{Choose step size $\alpha_k$}
				  \STATE{$\vec{c}_{k+1} \gets \vec{c}_{k} - \alpha_k \vec{G}_k$}
  				  \STATE{$k \gets k+1$}
				  \ENDWHILE
				  \STATE{$\hat{\vec{c}} \gets \vec{c}_k$}
			\end{algorithmic}
		\end{algorithm}
		
	\section{Regularized Dual Averaging Method}
		The dual averaging method is a primal-dual optimization method originally developed by Nesterov \cite{nesterov_primal-dual_2009}. Xiao \cite{xiao_dual_2010} later extended this approach to include regularization. It can be employed to solve optimization problems of the same form as given in Eqn.~(\ref{eqn:stoc_cost_fnc}). Here, we review the RDA method and detail its application to waveform inversion. Our presentation is similar to that of Xiao and Nesterov \cite{xiao_dual_2010,nesterov_primal-dual_2009}, but differs in several respects due to differences in the target application. In particular, the data fidelity term of our cost function is non-convex. This affects how the step size, or weights for each gradient term, must be chosen. Further, for clarity, we do not attempt to describe the most general form of the RDA method, but merely one that has proven effective for waveform inversion. For the dual averaging method, as described in Algorithm~\ref{alg:rda}, the update step for the $(k+1)$-th iteration is given by
		\begin{align}
			\vec{c}_{k+1} = \arg\min_\vec{c} \left\lbrace \left\langle \overbar{\vec{G}}_k , \vec{c} \right\rangle + \frac{1}{2 \mu_k} \|\vec{c} - \vec{c}_{0}\|^2_2 + \lambda \mathcal{R}\left(\vec{c}\right) \right\rbrace , \label{eqn:dual_update_step}
		\end{align} 
		where $\overbar{\vec{G}}_k$ is the average gradient of the data fidelity term over all past iterations, and $\mu_k > 0$ is a scalar. This differs from the update step for SGD in two key ways. First, instead of considering the gradient at a single point, the average gradient is employed. Second, the proximal term, $\frac{1}{2} \| \vec{c} - \vec{c}_0\|^2_2$, does not depend on the iteration number. In these ways, the RDA method is able to incorporate non-local information when determining the estimate of the object for the next iteration. 
		
		In the case of simple averaging, the average gradient is given by
		%\begin{empheq}[box=\colorbox{yellow}]{align}
		\begin{align}
			\overbar{\vec{G}}_k = \frac{1}{k+1} \sum_{i = 0}^{k} \nabla_\vec{c} f\left(\vec{c}_i, \vec{w}_i\right) .
		\end{align}
		%\end{empheq}
		A weighted average of gradient estimates can also be considered, as suggested by \cite{nesterov_primal-dual_2009}. In this case,
		\begin{align}
			\overbar{\vec{G}}_k = \frac{1}{\sum_{i = 0}^{k} \alpha_i} \sum_{i = 0}^{k} \alpha_i \nabla_\vec{c} f\left(\vec{c}_i, \vec{w}_i\right) ,
		\end{align}
		where $\left\lbrace \alpha_i \right\rbrace_{i=0}^k$ are weights for each of the gradient estimates. Here, the weights are chosen using a line search. Since the search direction is given by the average gradient, adjusting this weight will determine the contribution of the most recent gradient estimate to the search direction. As a result, the direction of the line search changes as this parameter is adjusted. This is in contrast to most line search methods, where the direction is fixed and only the magnitude of the update is affected. As the search direction incorporates information from multiple iterations corresponding to different realizations of the encoding vector, this approach is less prone to overfitting than SGD with a line search. A detailed description of this approach is provided in Appendix \ref{sec:linesearch}.

	If $\mathcal{R}$ is convex, the dual averaging update step can be written in terms of the proximity operator of $\mathcal{R}$ as 
		\begin{align}
			\vec{c}_{k+1} = \text{prox}_{\lambda \mu_k \mathcal{R}} \left( \vec{c}_{0} - \mu_k \overbar{\vec{G}}_k \right) ,
		\end{align}
		where the proximity operator is defined as \cite{parikh_proximal_2013}
		%\begin{empheq}[box=\colorbox{yellow}]{align}
		\begin{align}
			\text{prox}_{\lambda R} \left(\mathbf{x}\right) \equiv \min_{\mathbf{y}} \left\lbrace R\left(\mathbf{y}\right) + \frac{1}{2 \lambda} \|\mathbf{x} - \mathbf{y}\|^2 \right\rbrace.
		\end{align}
		%\end{empheq}
		From this expression, it becomes clear that the update step for the dual averaging method can be divided into two parts. First, a reference value is updated based on a weighted sum of all past gradient estimates. Second, regularization is incorporated by use of the associated proximity operator. In this way, the stochastic estimates of the gradient of the data fidelity term are treated separately from the deterministic regularization term. Averaging the gradient estimates obtained over several iterations may help minimize the impact of the variance of the gradient estimates. Further, since the regularization term is not explicitly differentiated, non-smooth penalties can be easily incorporated through use of the corresponding proximity operator.
		
		\begin{algorithm}
			\caption{Regularized dual averaging (RDA) method \label{alg:rda}}
			\algsetup{indent=2em}
			\begin{algorithmic}[1]
				\REQUIRE $\vec{c}_0$, $\lambda$
				\ENSURE $\hat{\vec{c}}$
				\STATE{$k \gets 0$} \COMMENT{$k$ is the algorithm iteration number.}
				\STATE{$A_{-1} \gets 0$}
				\WHILE {stopping criterion is not satisfied}
				\STATE{Draw $\vec{w}_k$ according to chosen distribution.}
				\STATE{Calculate $\vec{G}_k \gets \nabla_\vec{c} f\left(\vec{c}_k , \vec{w}_k \right)$}
				\STATE{Choose weight $\alpha_k > 0$} \COMMENT{Unweighted case: $\alpha_k = 1$}
				\STATE{$A_{k} \gets A_{k-1} + \alpha_k$}
				\STATE{$\overbar{\vec{G}}_k \gets \left(1 - \frac{\alpha_k}{A_k}\right) \overbar{\vec{G}}_{k-1} + \frac{\alpha_k}{A_k} \vec{G}_k$} \COMMENT{Compute weighted average of gradient.}
				\STATE{Choose $\mu_k$} \COMMENT{For example, $\mu_k = \gamma A_k$, where $\gamma > 0$ is a constant.}
				\STATE{$\vec{c}_{k+1} \gets \vec{c}_{0} - \mu_k \overbar{\vec{G}}_k$}
				\STATE{$ \vec{c}_{k+1} \gets \text{prox}_{\lambda \mu_k \mathcal{R}}\left(\vec{c}_{k+1}\right) $}
				\STATE{$k \gets k+1$}
				\ENDWHILE
				\STATE{$\hat{\vec{c}} \gets \vec{c}_k$}
			\end{algorithmic}
		\end{algorithm}
		
		Unless otherwise noted, the regularization function is chosen to be the total-variation (TV) semi-norm of the sound speed. The TV semi-norm has been shown to be effective at mitigating noise while preserving sharp edges \cite{chambolle_introduction_2010}. The proximity operator of the TV semi-norm is computed using the fast gradient projection method described in \cite{beck_fast_2009-1,chambolle_algorithm_2004}. Using this approach, the computational cost of applying the proximity operator is much less than that of computing the gradient, so that the computational cost of the RDA method is approximately the same as SGD on a per-iteration basis.
		
		The sequence $\left\lbrace \mu_k \right\rbrace$ determines the amount by which the algorithm steps in the search direction. Here, we choose $\mu_k = \gamma A_k$, where $A_k = \sum_{i = 0}^k \alpha_i$ and $\gamma > 0$ is a constant. In the this case, line 10 in Algorithm \ref{alg:rda} becomes
		%\begin{empheq}[box=\colorbox{yellow}]{align}
		\begin{align}
			\vec{c}_{k+1} \gets \vec{c}_{0} - \gamma \sum_{i=0}^{k} \alpha_i \vec{G}_i .
		\end{align}
		%\end{empheq}
		The constant $\gamma$ should be chosen to be sufficiently small to insure convergence. In the unweighted case, $\gamma$ could be chosen to be the inverse of the Lipschitz constant of the gradient of the data fidelity term. It could be similarly chosen in the weighted case as the inverse of the product of the Lipschitz constant and the maximum allowable weight of the gradient, $\alpha_{max}$.
		
	\section{Computer-Simulation Studies}
		\subsection{Methods}
		Two-dimensional computer-simulation studies were performed to compare USCT image reconstruction methods based on SGD and RDA. Studies were performed for two numerical phantoms: (1) a numerical breast phantom (shown in Fig.~\ref{fig:breast_phant}) and (2) a low-contrast phantom with two homogeneous bars (shown in Fig.~\ref{fig:bar_phant}). The first was employed to establish the potential utility of the proposed approaches for USCT breast imaging, and the second was employed to perform a bias-variance analysis comparing SGD and RDA. For both phantoms, the same measurement geometry, excitation pulse, and numerical simulation methods were employed. 
		
			\subsubsection{Measurement Geometry}
\begin{figure}[ht]
				\centering
				\includegraphics[width=0.3\textwidth]{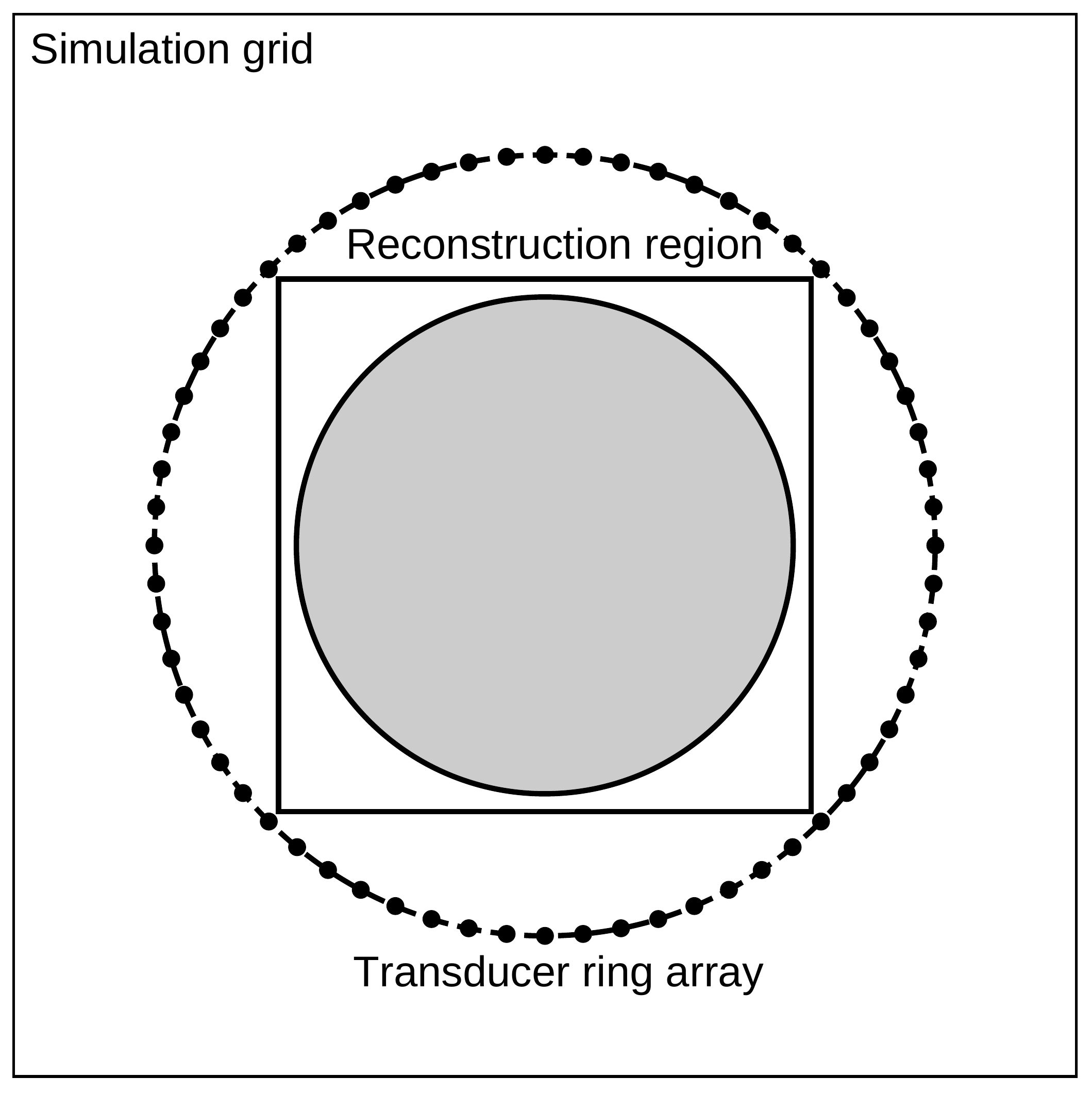}
				\caption{A schematic of the measurement geometry. The measurement system consists of a circular ring array of ultrasonic transducers. These transducers are located in a larger rectangular simulation grid, over which the acoustic wave equation is solved. Within the ring array is a smaller rectangular region representing the reconstructed image. The estimated sound speed distribution is calculated within the gray circular field-of-view within that region.} \label{fig:meas_geo}
			\end{figure}

			The measurement system consisted of a circular transducer array with a radius of 110 mm and 256 evenly distributed elements. This geometry was chosen to match an existing USCT imaging system \cite{duric_breast_2013,duric_clinical_2014,li_clinical_2008}. The wavefield data were simulated for 256 views using the first-order k-space method as described below \cite{tabei_k-space_2002, wang_waveform_2015, treeby_k-wave:_2010}. For each view, one transducer served as the emitter and the pressure was recorded by all 256 transducers. All transducers were modeled as point emitters and receivers. A schematic of this measurement geometry is shown in Fig.~\ref{fig:meas_geo}.
			
			\subsubsection{Excitation pulse}
			The excitation pulse was given by
			\begin{align}
				s\left(t\right) =  \exp\left(-\frac{\left(t - t_c\right)^2}{2 \sigma^2}\right) \sin\left(2 \pi f_c t\right) ,
			\end{align}
			where $f_c = 0.8~\text{MHz}$ is the central frequency, and $t_c = 3.2~\mu s$ and $\sigma = 0.75~\mu s$ are the center and width of a Gaussian window, respectively. This corresponds to roughly three cycles. Since the transducers are treated as point emitters, when nearest neighbor interpolation is employed, the source term for the $m$-th view is given simply by
			\begin{align}
				s_m\left(\vec{r}, t\right) = s\left(t\right) \delta\left(\vec{r} - \vec{r}_m\right) ,
			\end{align} 
			where $\vec{r}_m$ is the location of the pixel nearest to the emitter for the $m$-th view.
			
			\subsubsection{Numerical phantoms}
\begin{figure}[htbp!]
				\centering
				\begin{subfigure}[b]{0.22\textwidth}
					\includegraphics[width=\textwidth]{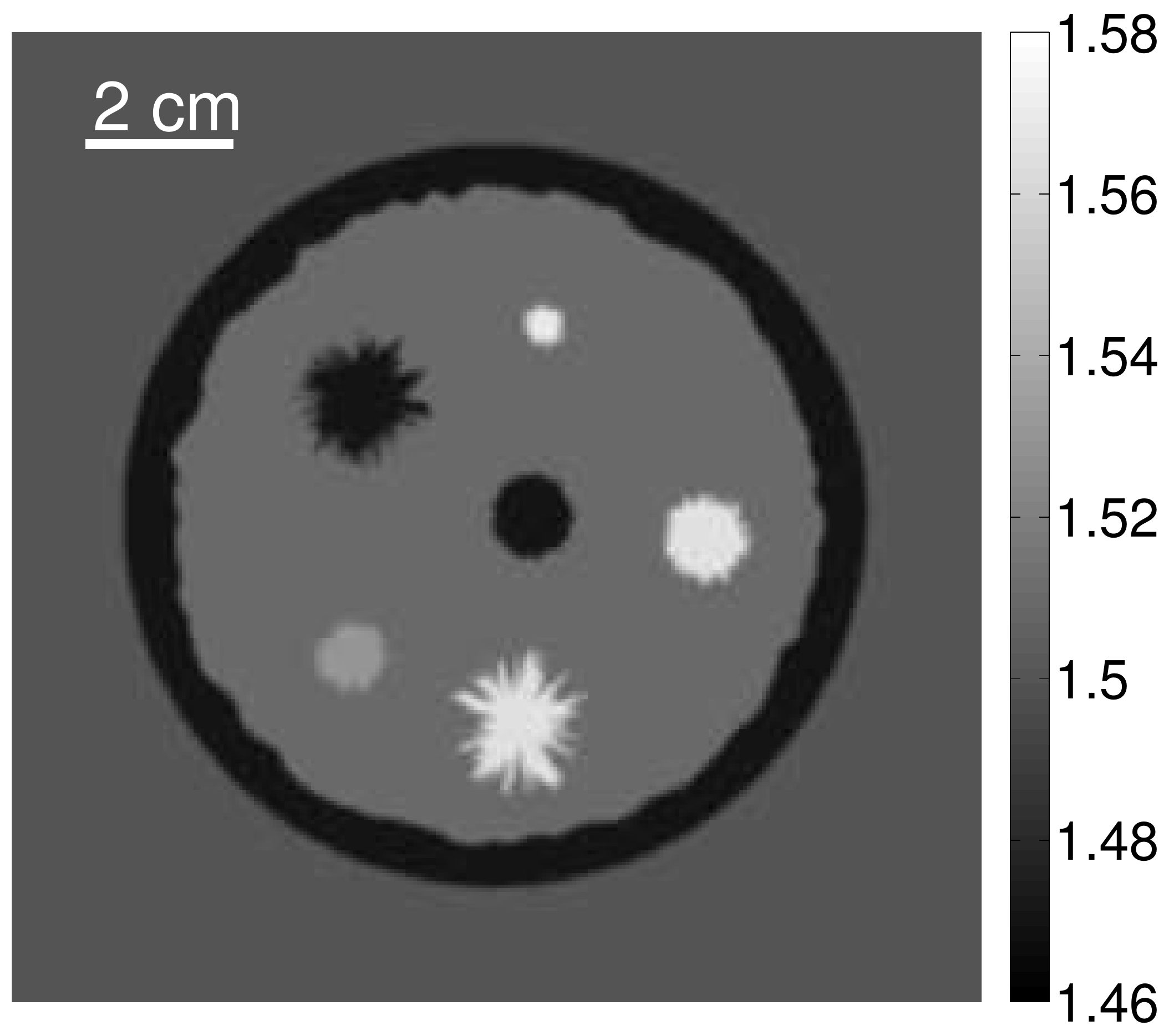}
					\caption{} \label{fig:breast_phant}
				\end{subfigure}
				%\hspace*{1mm}
				\begin{subfigure}[b]{0.225\textwidth}
					\includegraphics[width=\textwidth]{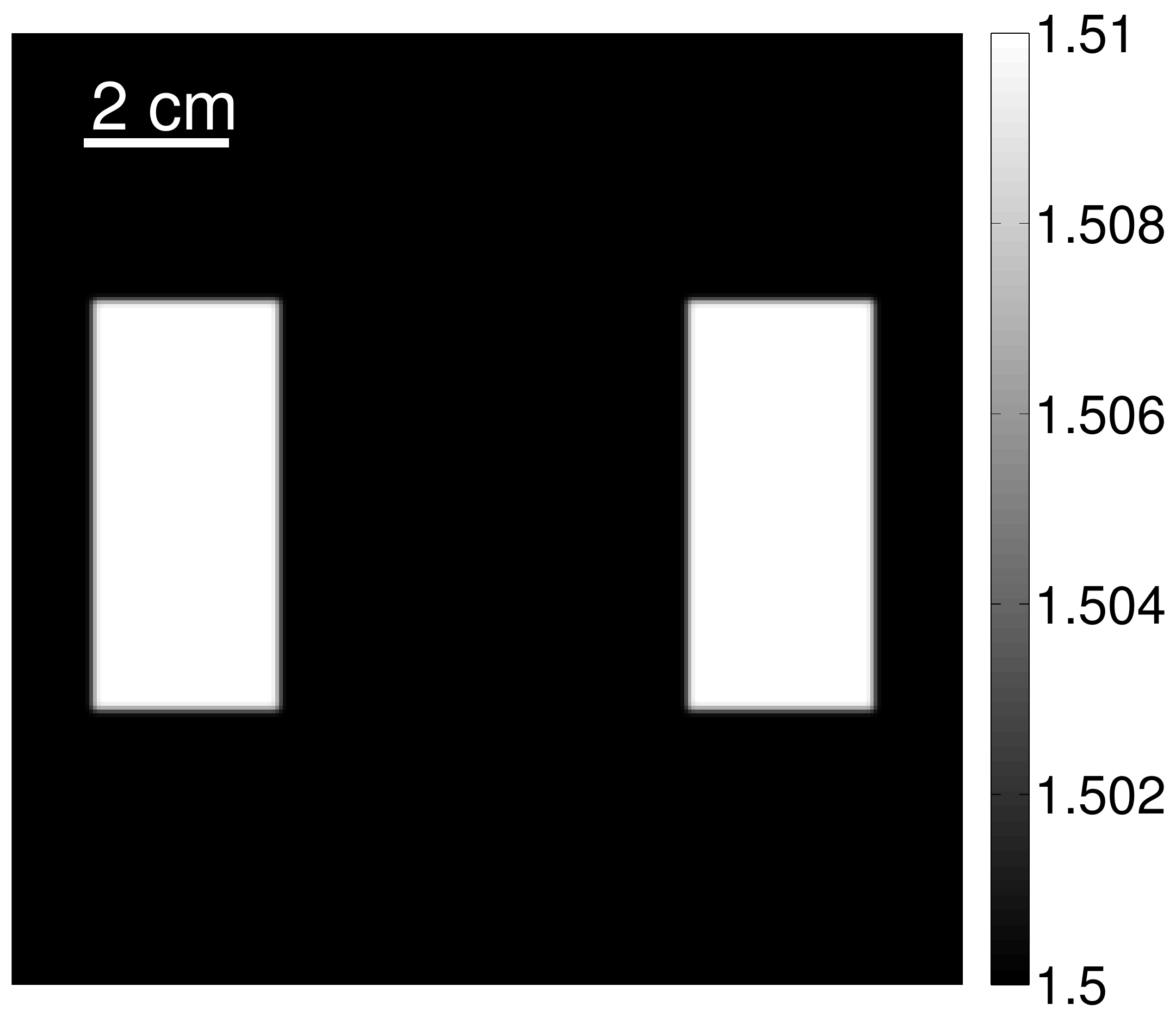}
					\caption{}	\label{fig:bar_phant}
				\end{subfigure}
				\caption{Sound speed distribution of (a) the numerical breast phantom and (b) the low-contrast two bar phantom employed in the bias-variance analysis, given in units of mm/$\mu$s.}
			\end{figure}

			The numerical breast phantom had a radius of 49~mm and was composed of 8 structures representing adipose tissues, parenchymal breast tissues, cysts, benign tumors, and malignant tumors (See Fig.~\ref{fig:breast_phant}). A detailed description of the numerical breast phantom can be found in \cite{wang_waveform_2015}.	A phantom consisting of two low-contrast bars was created for the bias-variance analysis (see Fig.~\ref{fig:bar_phant}). The bars were placed far apart to minimize their influence on one another in the reconstructed images.
			
			\subsubsection{Simulation of pressure data} \label{sec:pressure_data}
			In order to avoid inverse crime \cite{colton_inverse_2013}, two related methods were employed to simulate the measured pressure. When generating the pressure data recorded by each transducer, the wave equation was solved by use of a first-order pseudo-spectral method \cite{tabei_k-space_2002}. In this method, when attenuation and dispersion are neglected, the acoustic wave propagation is modeled by two coupled first-order differential equations:
			\begin{align}
			\frac{\partial \vec{u}\left(\vec{r},t\right)}{\partial t} &= - \nabla p\left(\vec{r},t\right) \\
			\frac{1}{c\left(\vec{r}\right)^2}\frac{\partial p\left(\vec{r},t\right)}{\partial t} &= - \nabla \cdot \vec{u}\left(\vec{r}, t\right) + 4 \pi \int_{0}^{t} dt^\prime s\left(\vec{r}, t^\prime\right) ,
			\end{align}
			where $\vec{u}\left(\vec{r}, t\right)$ is the acoustic particle velocity and $p\left(\vec{r},t\right)$ is the acoustic pressure.
			The calculation domain was of size $512 \times 512~\text{mm}^2$, sampled on a $2048 \times 2048$ uniform Cartesian grid with a spacing of 0.25~mm. Nearest-neighbor interpolation was employed to place the transducers on the grid points. The pressure was simulated for $3600$ time points at a sampling rate of $20$~MHz. Additive Gaussian white noise was added to the pressure data. The noise had zero mean and a standard deviation of 5\% of the maximum pressure amplitude received by the transducer opposite the emitter for a homogeneous medium.
			
			When reconstructing the sound speed images, the operator $\mathbf{H}\left(\mathbf{c}\right)$ was computed by use of the second-order pseudo-spectral k-space method \cite{mast_k-space_2001}. This method solves a single second-order differential equation:
			\begin{align}
				\nabla^2 p\left(\vec{r}, t\right) - \frac{1}{c\left(\vec{r}\right)^2} \frac{\partial^2 p\left(\vec{r}, t\right)}{\partial t^2} = -4 \pi s\left(\vec{r}, t\right) .
			\end{align}
			Here, the calculation domain was of size $512 \times 512~\text{mm}^2$, sampled on a $1024 \times 1024$ uniform Cartesian grid with a spacing of 0.5~mm. The number of time points and sampling rate were reduced to $1800$ and $10$~MHz, respectively. These reconstruction parameters are summarized in Table~\ref{table:recon_pars}. Both wave solvers were implemented using NVIDIA's CUDA platform \cite{_cuda_2015}. These pseudo-spectral k-space methods were chosen for their high numerical accuracy for coarse spatial sampling rates \cite{mast_k-space_2001,tabei_k-space_2002}.
					
			\subsubsection{Bias-variance analysis}
			The statistical properties of the images produced by the two methods were compared by use of a bias-variance analysis. The measured pressure was simulated as described above. Five different noise realizations were generated, each with 5\% noise. Images were reconstructed for each noise realization for six different regularization parameter values by use of both SGD with a constant step size and the unweighted RDA method. Each pixel in the reconstructed images can be treated as a random variable $\hat{c}_i$ (for the $i$-th pixel), whose true value in the original phantom is $c_i$. Due to the long reconstruction times (approx. 1 hr for 250 iterations), it was not feasible to reconstruct images for a large number of noise realizations. Instead, each reconstructed image was divided into several regions, which were treated as independent samples for the purposes of this analysis. Specifically, each bar in the reconstructed image was divided into 10 identical regions. Corresponding pixels in these regions were treated as having arisen from additional noise realizations. This yielded a total of 100 samples per regularization parameter value. In other words, if the set $\hat{C}_i$ contains the values of the $i$-th pixel for the five noise realizations, an augmented set $\tilde{C}_i$ was created such that
			%\begin{empheq}[box=\colorbox{yellow}]{align}
			\begin{align}
				\tilde{C}_i = \bigcup_{j=1}^{N_c} \hat{C}_{\mathcal{I}_i\left(j\right)} ,
			\end{align}
			%\end{empheq}
			where $N_c$ is the total number of regions (20) and $\mathcal{I}_i$ is an iterator that gives the indices of all pixels (across regions) that correspond to the $i$-th pixel. The bias for a pixel was calculated by averaging these 100 samples and computing the difference between the average value and the corresponding value in the true phantom:
			%\begin{empheq}[box=\colorbox{yellow}]{align} 
			\begin{align}
				\text{Bias}_i = \frac{1}{N_s} \sum_{\hat{c} \in \tilde{C}_i} \hat{c} - c_i ,
			\end{align}
			%\end{empheq}
			where $N_s$ is the total number of samples. A summary measure of the bias was calculated by computing the $\ell_2$-norm of the bias values for each pixel. The sample variance of each pixel across all samples was computed as 
			%\begin{empheq}[box=\colorbox{yellow}]{align}
			\begin{align}
				\text{Var}_i = \frac{1}{N_s - 1} \sum_{\hat{c} \in \tilde{C}_i}\left(\hat{c} - \frac{1}{N_s} \sum_{\hat{c} \in \tilde{C}_i} \hat{c}\right)^2 .
			\end{align}
			%\end{empheq}
			The average variance for the pixels was computed as a summary measure. It should be noted that corresponding pixels in different regions may not have the same expected values and variances. In spite of this, the above bias and variance measures still provide insight into the ability of the two reconstruction algorithms to mitigate noise.
		
			\subsection{Images reconstructed by use of SGD}
\begin{figure}[htbp!]
				\centering
				\includegraphics[width=0.4\textwidth]{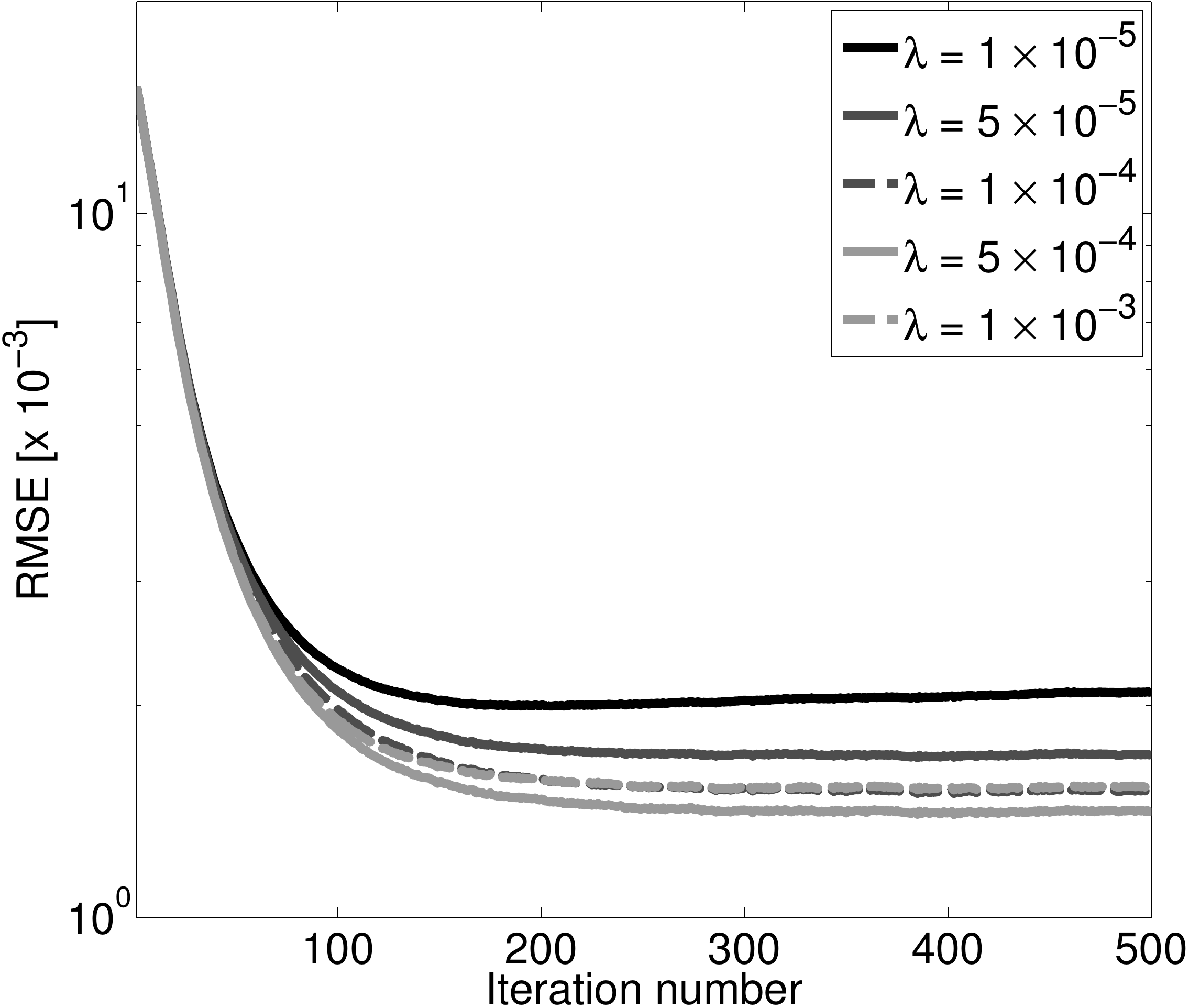}
				\caption{Plot of RMSE versus the number of iterations for images reconstructed by use of SGD with a constant step size of 0.1 for several regularization parameter values.} \label{fig:sgd_reg_param_conv}
			\end{figure}

			In order to provide a clear and fair point of comparison of the RDA and SGD methods, USCT image reconstruction from noisy data by use of SGD was first considered and optimized. 	
%			\begin{figure}[ht!]
%				\centering
%				\begin{subfigure}[b]{0.225\textwidth}
%					\includegraphics[width=\textwidth]{recon499_wisesgd_ls_tv5e-4}
%					%\vspace*{2mm}
%					\caption{} \label{fig:sgd_ls_const_comp__ls}
%				\end{subfigure}
%				\begin{subfigure}[b]{0.225\textwidth}
%					\includegraphics[width=\textwidth]{recon499_wisesgd_const01_tv5e-4}
%					%\vspace*{2mm}
%					\caption{} \label{fig:sgd_ls_const_comp__const}
%				\end{subfigure}
%				\begin{subfigure}[b]{0.38\textwidth}
%					\includegraphics[width=\textwidth]{profile_sgd_ls_const_comparison}
%					\caption{Profiles at y = -6.5 mm through the reconstructed images.} \label{fig:sgd_ls_const_comp__prof}
%				\end{subfigure}
%				\caption{Images reconstructed by use of SGD using (a) a line search and (b) a constant step size of 0.1, shown after 500 iterations for a regularization parameter value of $5 \times 10^{-4}$. The insets in the upper right corners of each image correspond to a zoomed-in image of the inclusion located at about 7 o'clock. The larger images are shown in a grayscale window of $[1.47, 1.58]~ \text{mm}/\mu \text{s}$, while the insets are  shown in a grayscale window of $[1.50, 1.53]~ \text{mm}/\mu \text{s}$.} \label{fig:sgd_ls_const_comp}
%			\end{figure}		
			As seen in Fig.~\ref{fig:sgd_reg_param_conv}, the above choice of $5 \times 10^{-4}$ for the regularization parameter value results in the most accurate reconstructed image for SGD, as quantified by the root-mean-square-error (RMSE). As such, this value will be taken as the optimal value for SGD-based USCT image reconstruction and will be employed in all future comparisons with the results obtained by use of the RDA method.
			
\begin{figure}[htbp!]
				\centering
				\begin{subfigure}[b]{0.245\textwidth}
					\includegraphics[width=\textwidth]{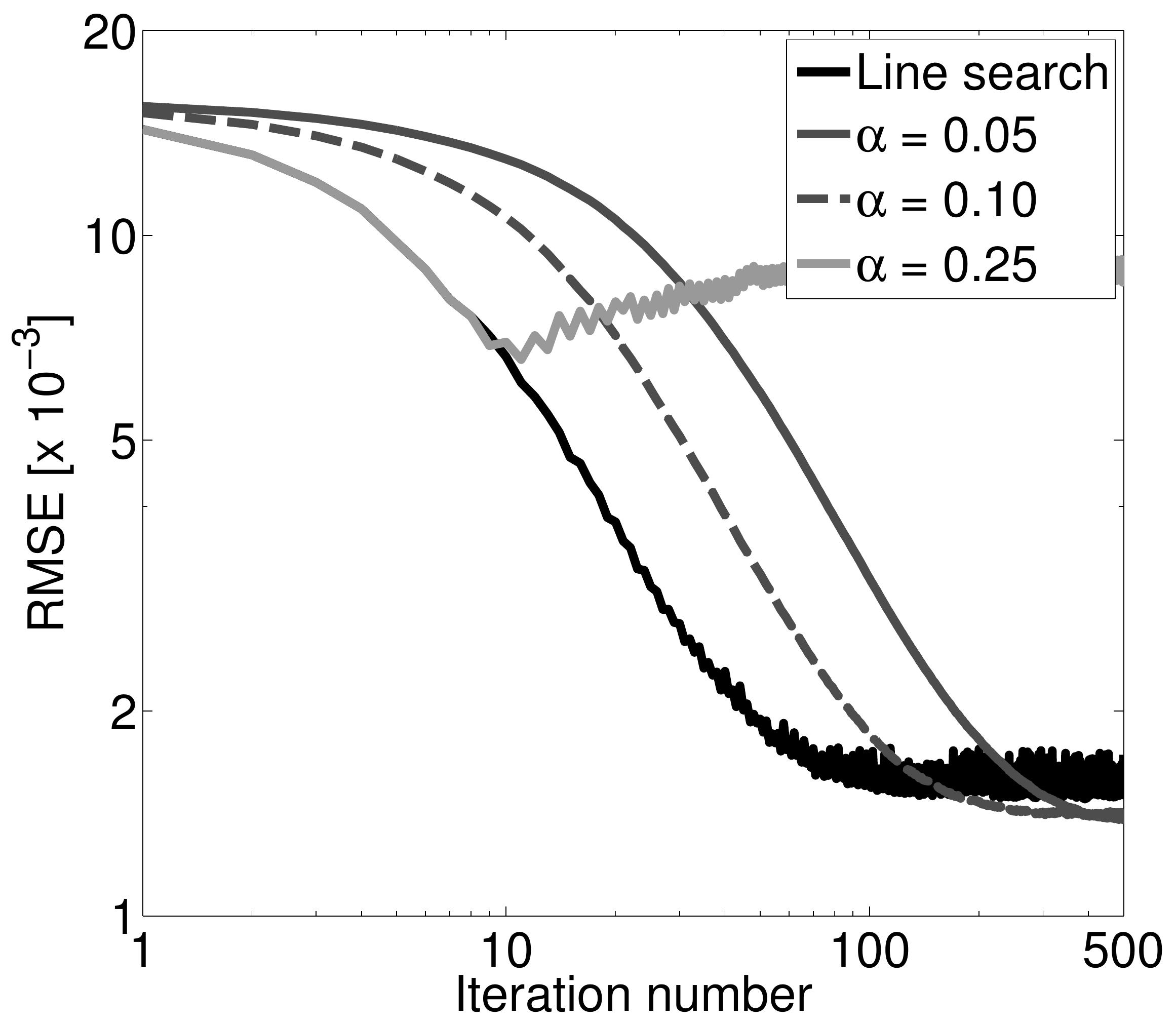}
					\caption{}
				\end{subfigure}%
				\begin{subfigure}[b]{0.245\textwidth}
					\includegraphics[width=\textwidth]{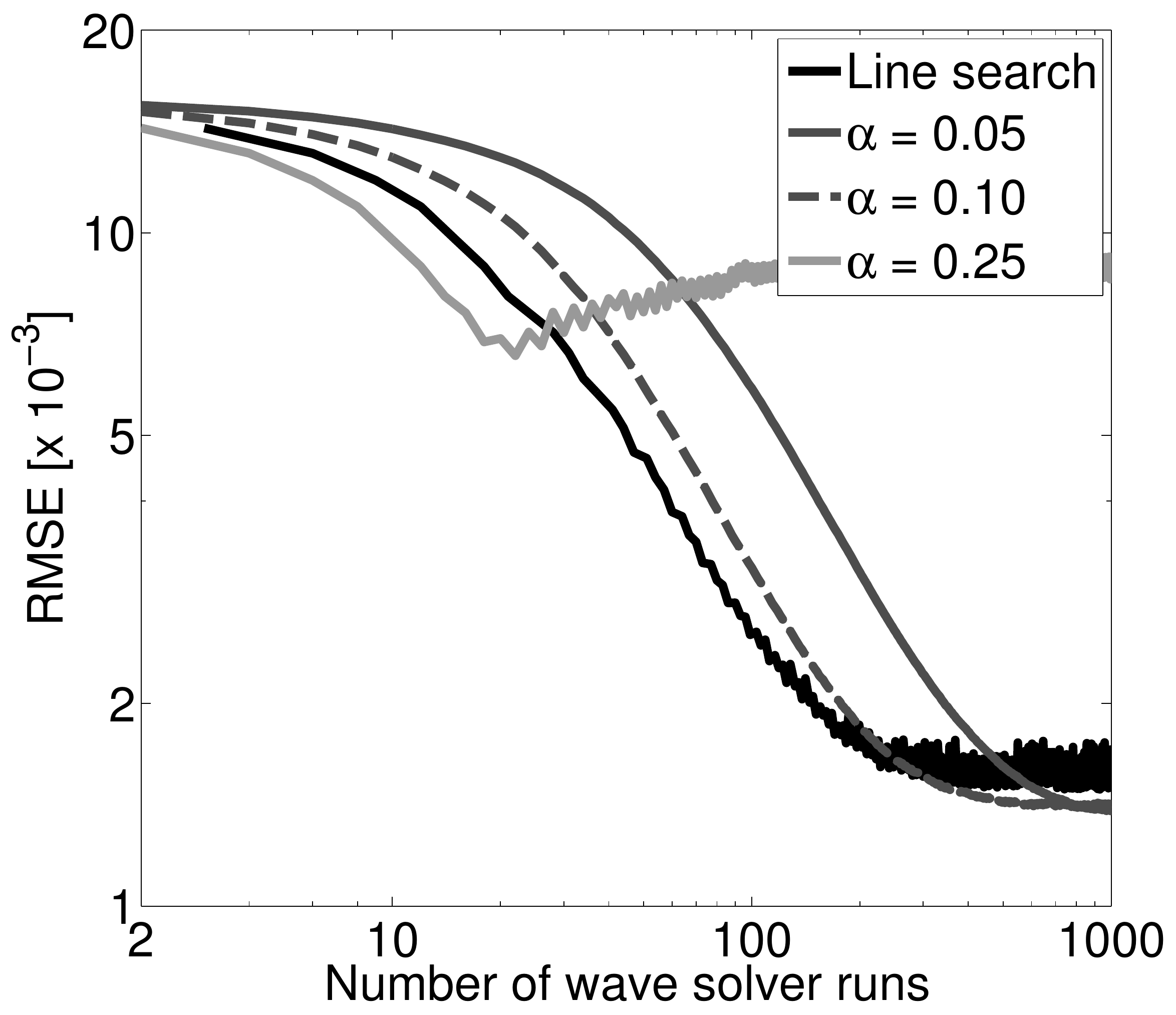}
					\caption{}
				\end{subfigure}
				\caption{Plot of RMSE versus (a) the number of iterations and (b) the number of wave equation solver runs for images reconstructed by use of SGD with a line search and with several constant step size values for a regularization parameter value of $5 \times 10^{-4}$.}
				\label{fig:sgd_stepsize_conv}
			\end{figure}

			A similar methodology was employed to choose the optimal step size for the SGD method. Several constant step sizes were compared with use of a line search method. As seen in Fig.~\ref{fig:sgd_stepsize_conv}, when a constant step size is too large, the optimization algorithm will diverge. However, when the step size is small, the convergence of the optimization algorithm will be slow. Use of a line search method can provide fast convergence, but as mentioned above, can result in reduced image quality. Since use of a line search introduces an additional computational cost, the convergence of these approaches are given both in terms of iteration number and the number of times the wave equation must be solved, referred to here as wave solver runs. Every step size considered as part of the line search will add one additional wave solver run. However, even when this additional computational effort is accounted for, use of a line search can still produce a more accurate reconstructed image for a given level of computational effort than use of a constant step size (at least, up to some threshold level of total computational effort). In addition, it removes the need to wisely choose the step size, a task which is often accomplished through trial-and-error. From Fig.~\ref{fig:sgd_stepsize_conv}, it can be seen that of the constant step size results, a step size of 0.1 produces the fastest convergence rate while still resulting in an accurate reconstructed image.
			
			In Section \ref{sec:background}, it was suggested that use of a line search method may have a negative impact on the obtained solution for SGD. This is demonstrated in Fig.~\ref{fig:sgd_stepsize_conv}. Here, it is seen that the line search method results in oscillations in the RMSE of the reconstructed image, while use of a constant step size produces a smoother convergence curve with fewer jumps. Also, note that the final RMSE is lower for the constant step size method (RMSE~=~$1.42 \times 10^{-3}$) than for the line search method (RMSE~=~$1.73 \times 10^{-3}$).
			
			\subsection{Images reconstructed by use of RDA}
\begin{figure*}[htbp!]
				\begin{subfigure}[b]{0.25\textwidth}
					\includegraphics[width=\textwidth]{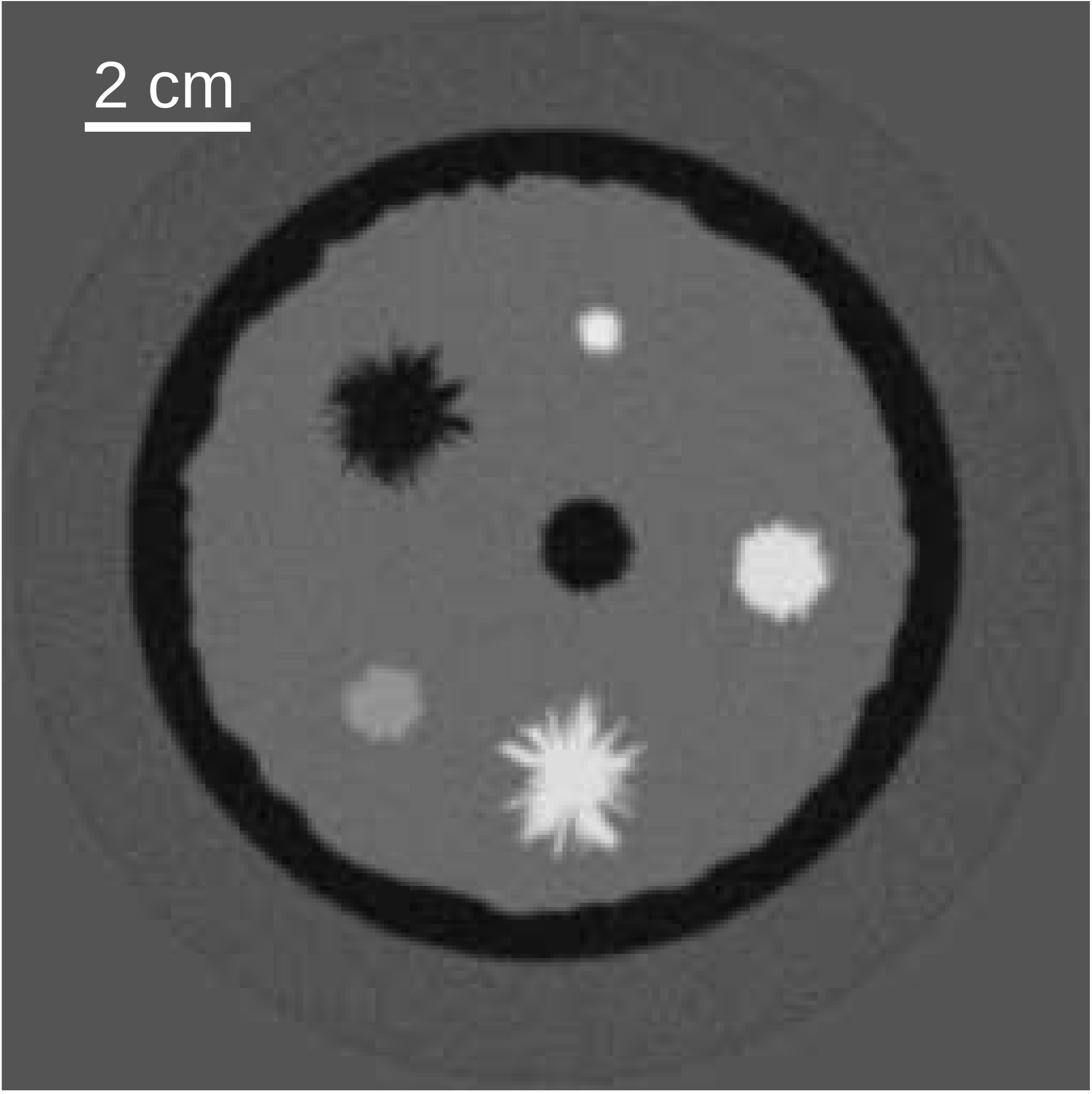}
					\caption{}
				\end{subfigure}%
				\begin{subfigure}[b]{0.25\textwidth}
					\includegraphics[width=\textwidth]{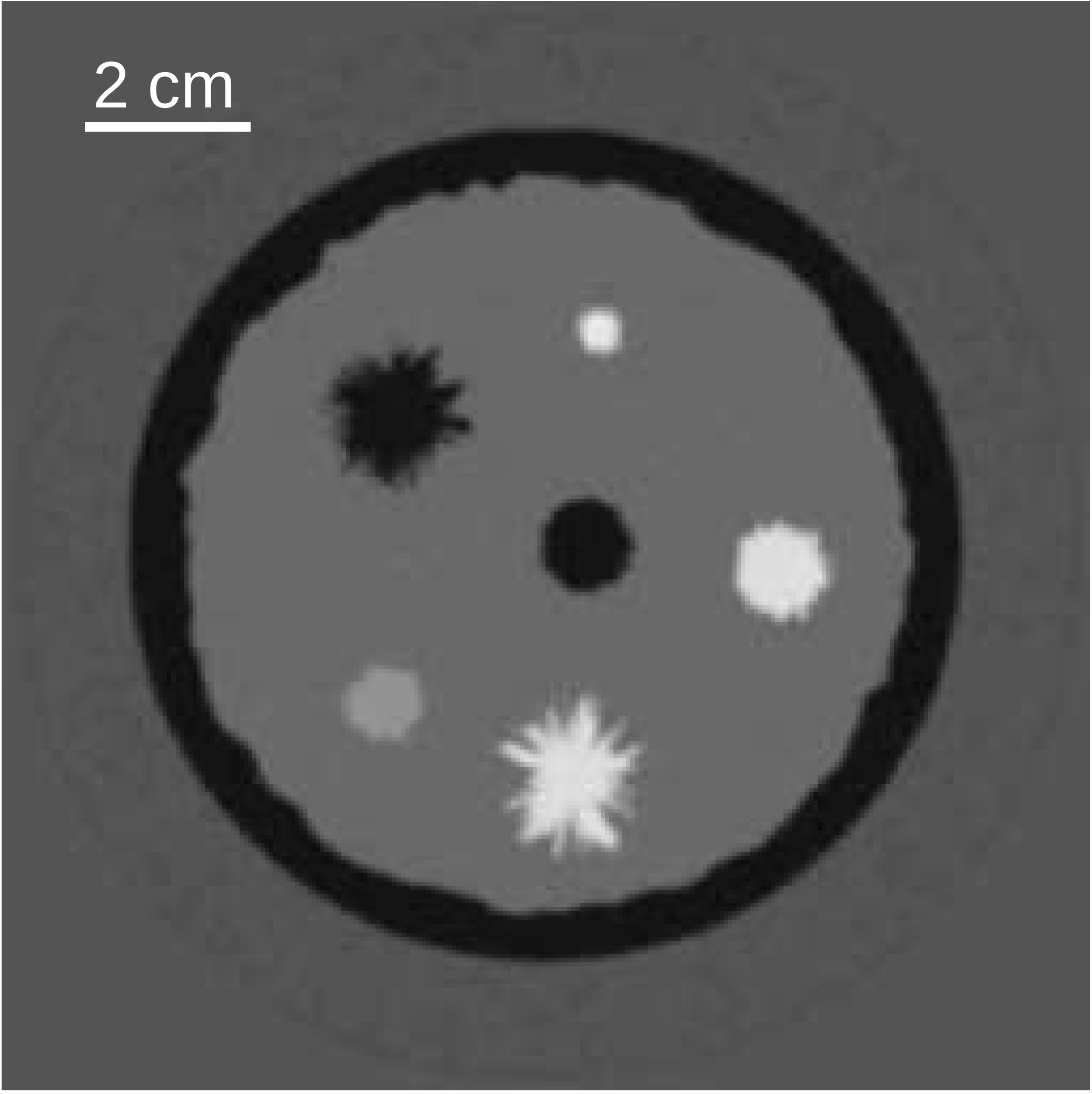}
					\caption{}
				\end{subfigure}%
				\begin{subfigure}[b]{0.25\textwidth}
					\includegraphics[width=\textwidth]{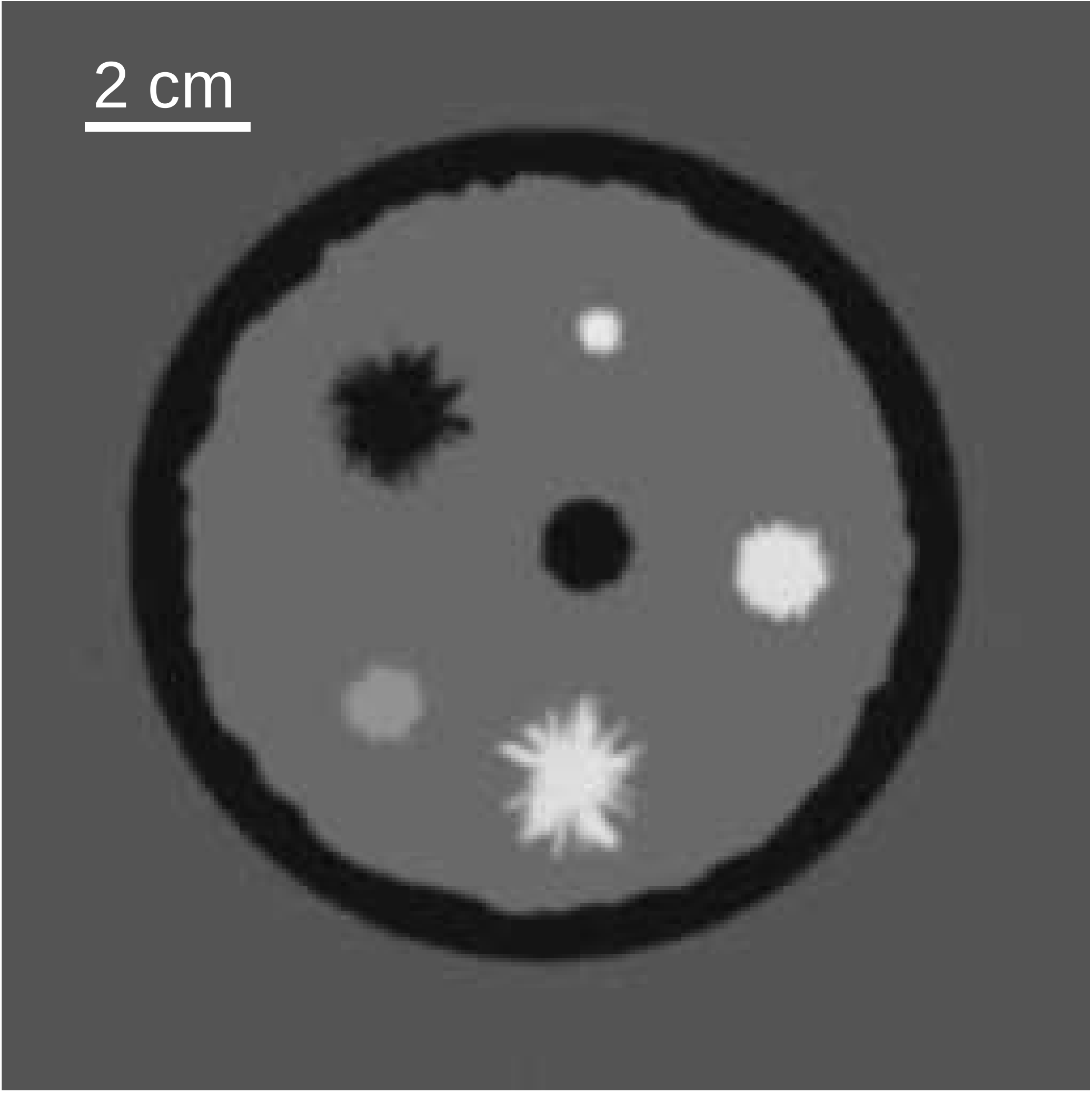}
					\caption{}
				\end{subfigure}%
				\begin{subfigure}[b]{0.25\textwidth}
					\includegraphics[width=\textwidth]{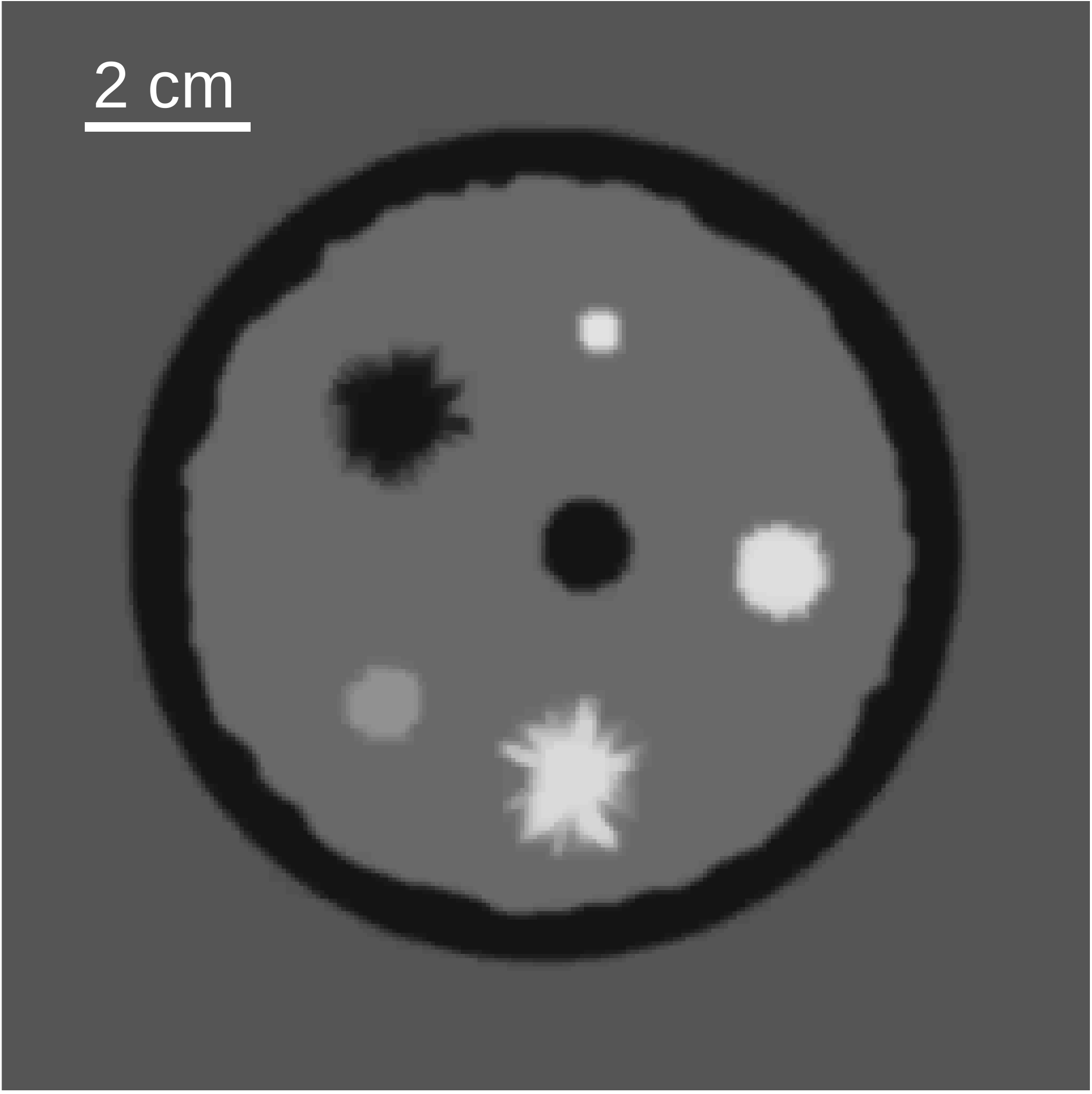}
					\caption{}
				\end{subfigure}%
				\caption{Images reconstructed by the unweighted RDA method with a fixed step size of 0.1 for regularization parameter values of (a) $1 \times 10^{-5}$, (b) $5 \times 10^{-5}$, (c) $1 \times 10^{-4}$, and (d) $5 \times 10^{-4}$, shown after 300 iterations. All images are shown  in a grayscale window of $[1.47, 1.58]~ \text{mm}/\mu \text{s}$.} \label{fig:rda_regparam_img}
			\end{figure*}

\begin{figure}[htbp!]
				\centering
				\begin{subfigure}[b]{0.25\textwidth}
					\includegraphics[width=\textwidth]{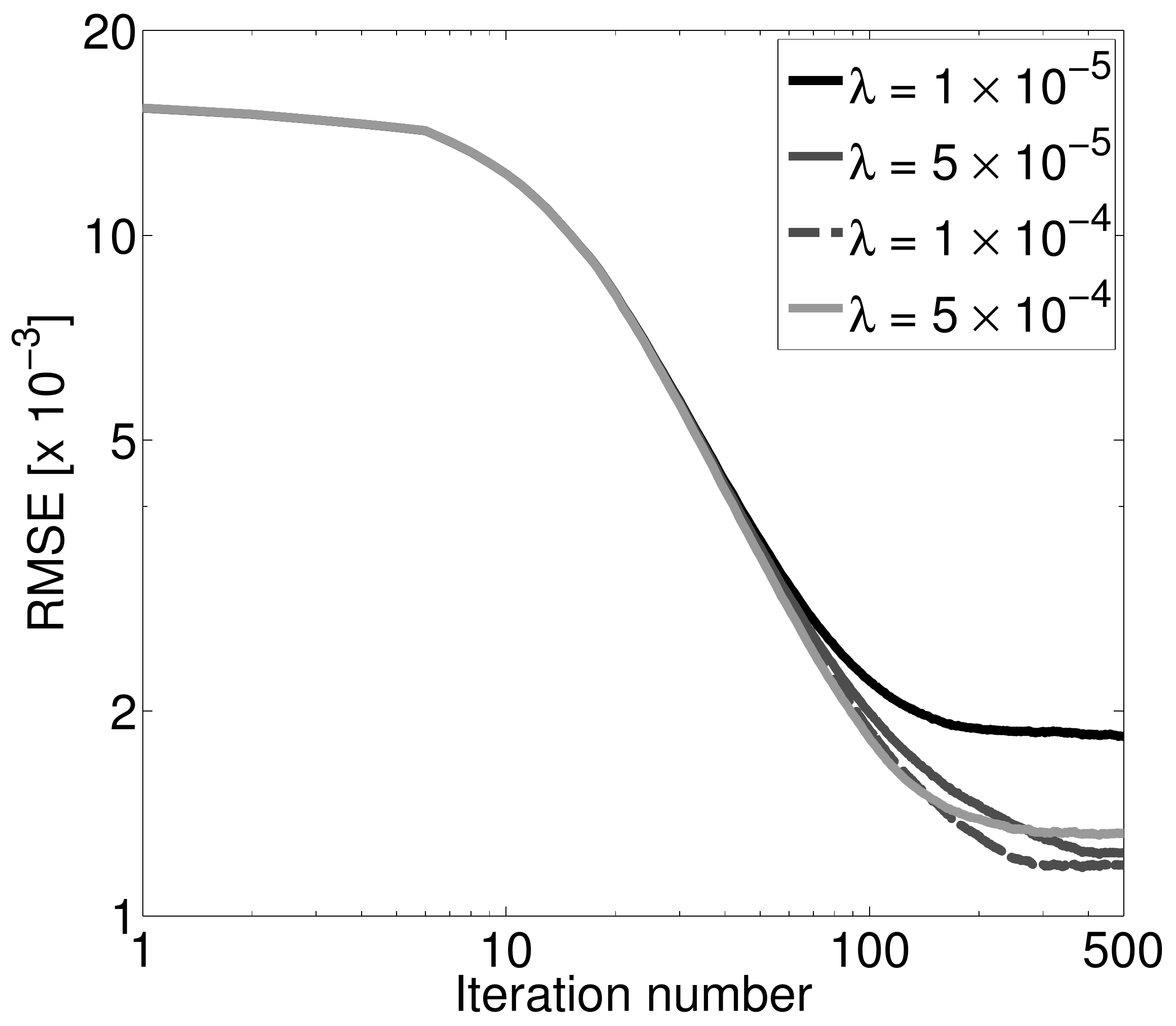}
					\caption{} \label{fig:rda_regparam_conv}
				\end{subfigure}%
				\begin{subfigure}[b]{0.25\textwidth}
					\includegraphics[width=\textwidth]{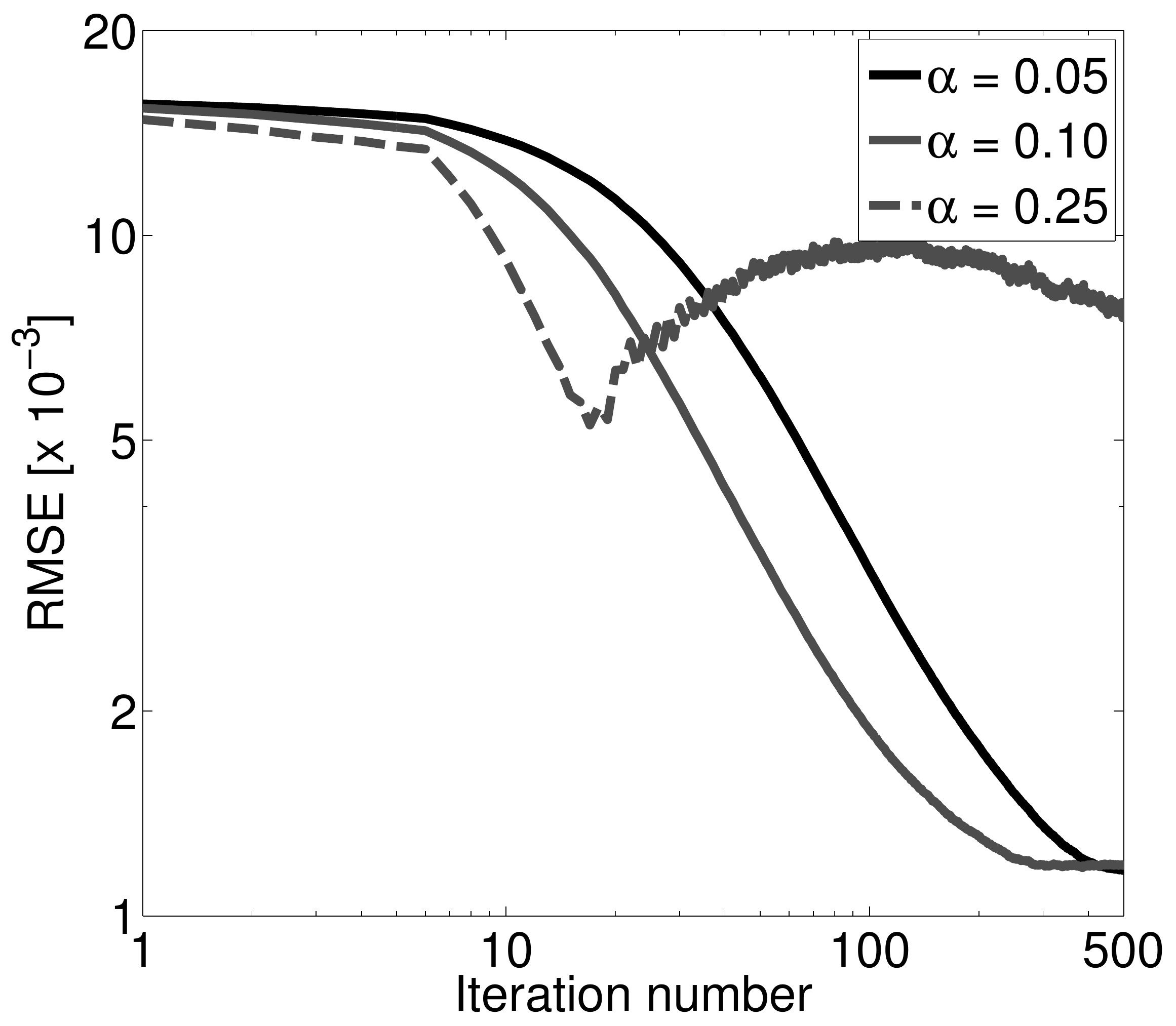}
					\caption{} \label{fig:rda_stepsize_conv}
				\end{subfigure}
				\caption{Plot of RMSE versus the number of iterations for (a) several regularization parameter values and a fixed step size of 0.1 and (b) several constant step size values and a fixed regularization parameter value of $1 \times 10^{-4}$ for images reconstructed by use of the unweighted RDA method.}
			\end{figure}

			The optimal step size (or, equivalently, value of $\gamma$ in line 9 of Algorithm~\ref{alg:rda}) and regularization parameter value for the RDA method will be determined in the same manner as employed for the SGD method. First, the regularization parameter value that resulted in the most accurate reconstructed image was determined. Example images reconstructed by RDA for several regularization parameter values are shown in Fig.~\ref{fig:rda_regparam_img}.
			From Fig.~\ref{fig:rda_regparam_conv}, it can be seen that a regularization parameter value of $1 \times 10^{-4}$ results in the most accurate reconstructed image. 
			This is smaller than the value obtained for SGD. From Fig.~\ref{fig:rda_stepsize_conv}, the optimal step size value is 0.1, the same value obtained for SGD.	
			
\begin{figure*}[htbp!]
				\begin{subfigure}[b]{0.25\textwidth}
					\includegraphics[width=\textwidth]{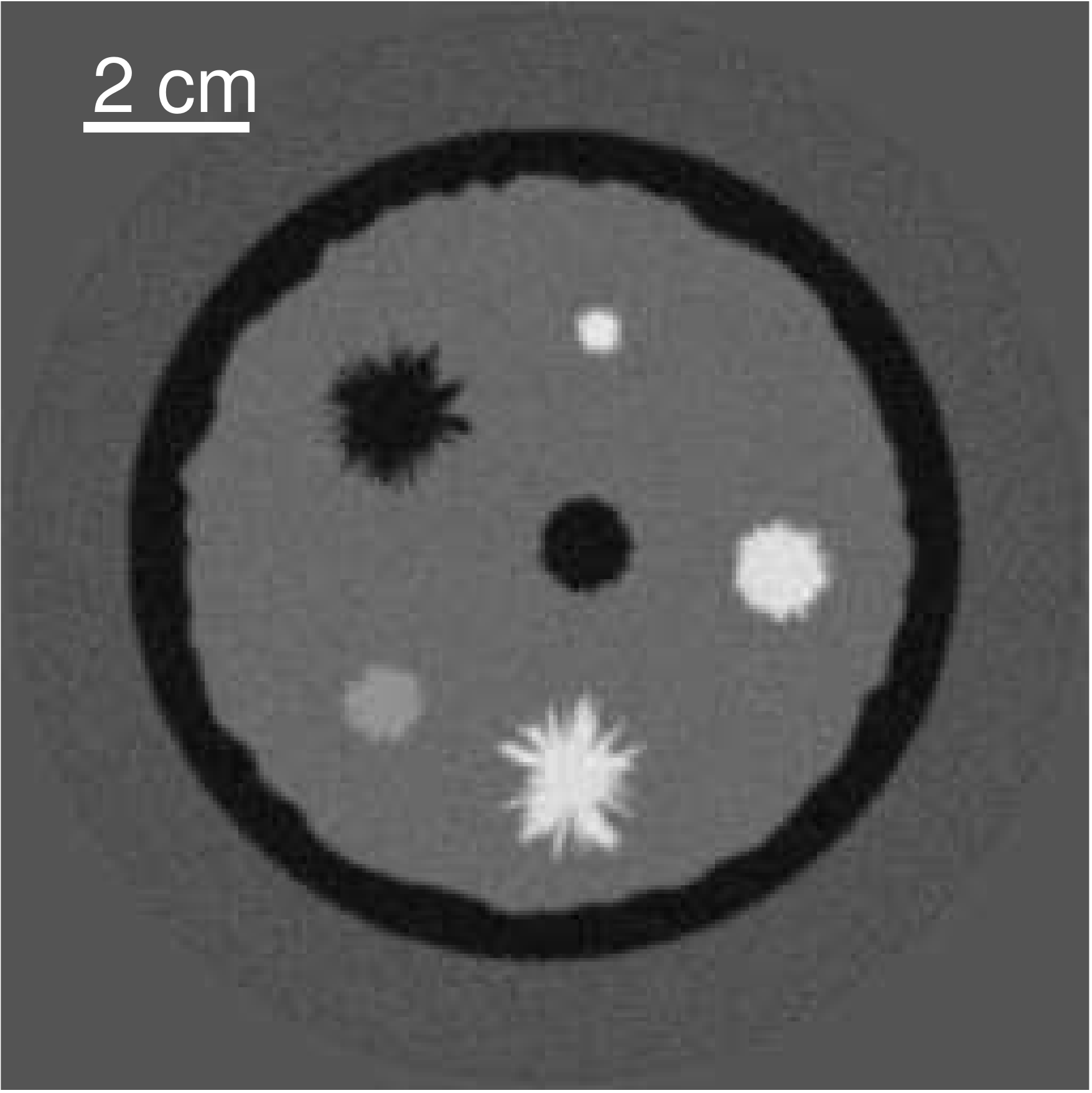}
				\end{subfigure}%
				\begin{subfigure}[b]{0.25\textwidth}
					\includegraphics[width=\textwidth]{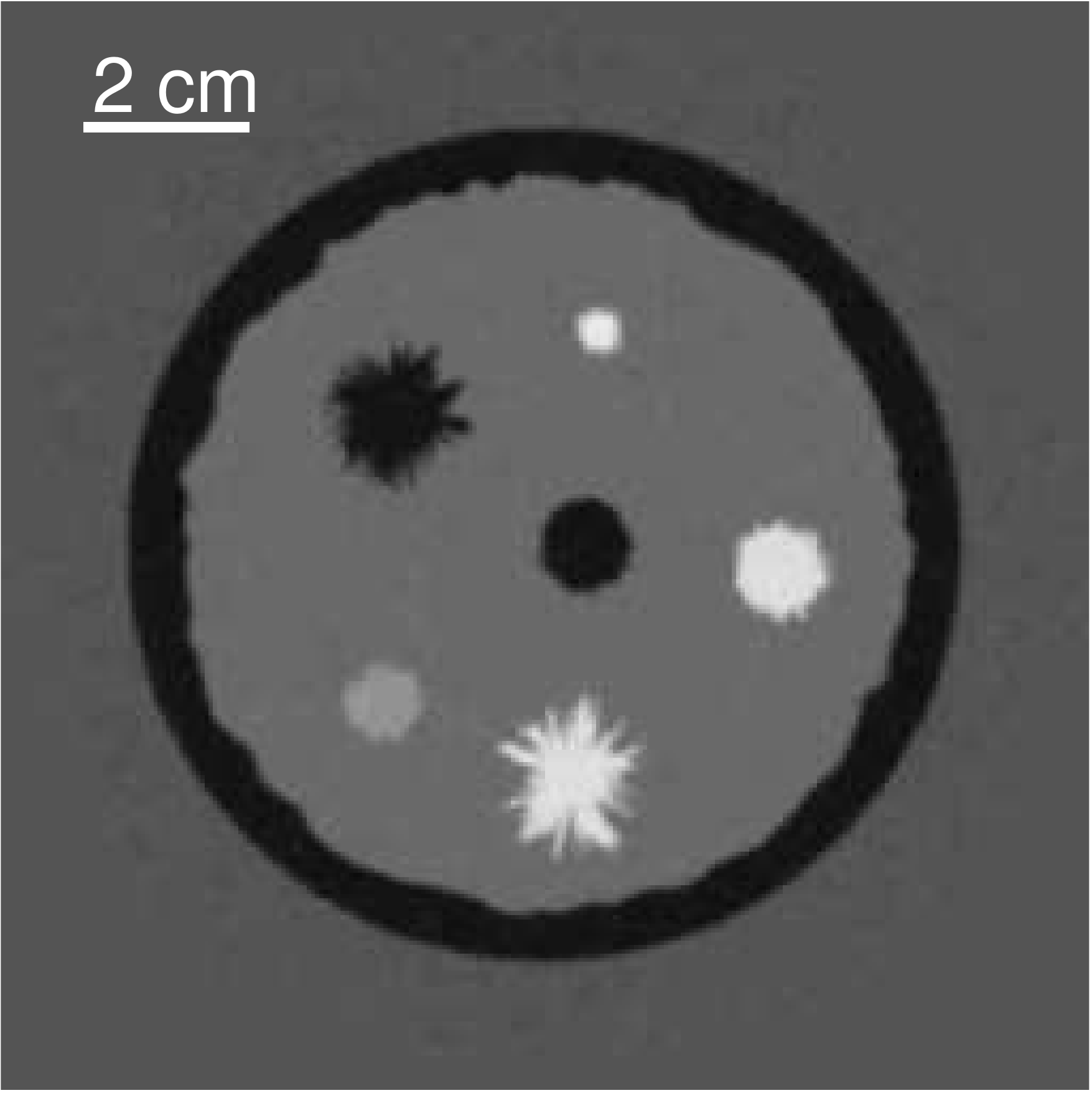}
				\end{subfigure}%
				\begin{subfigure}[b]{0.25\textwidth}
					\includegraphics[width=\textwidth]{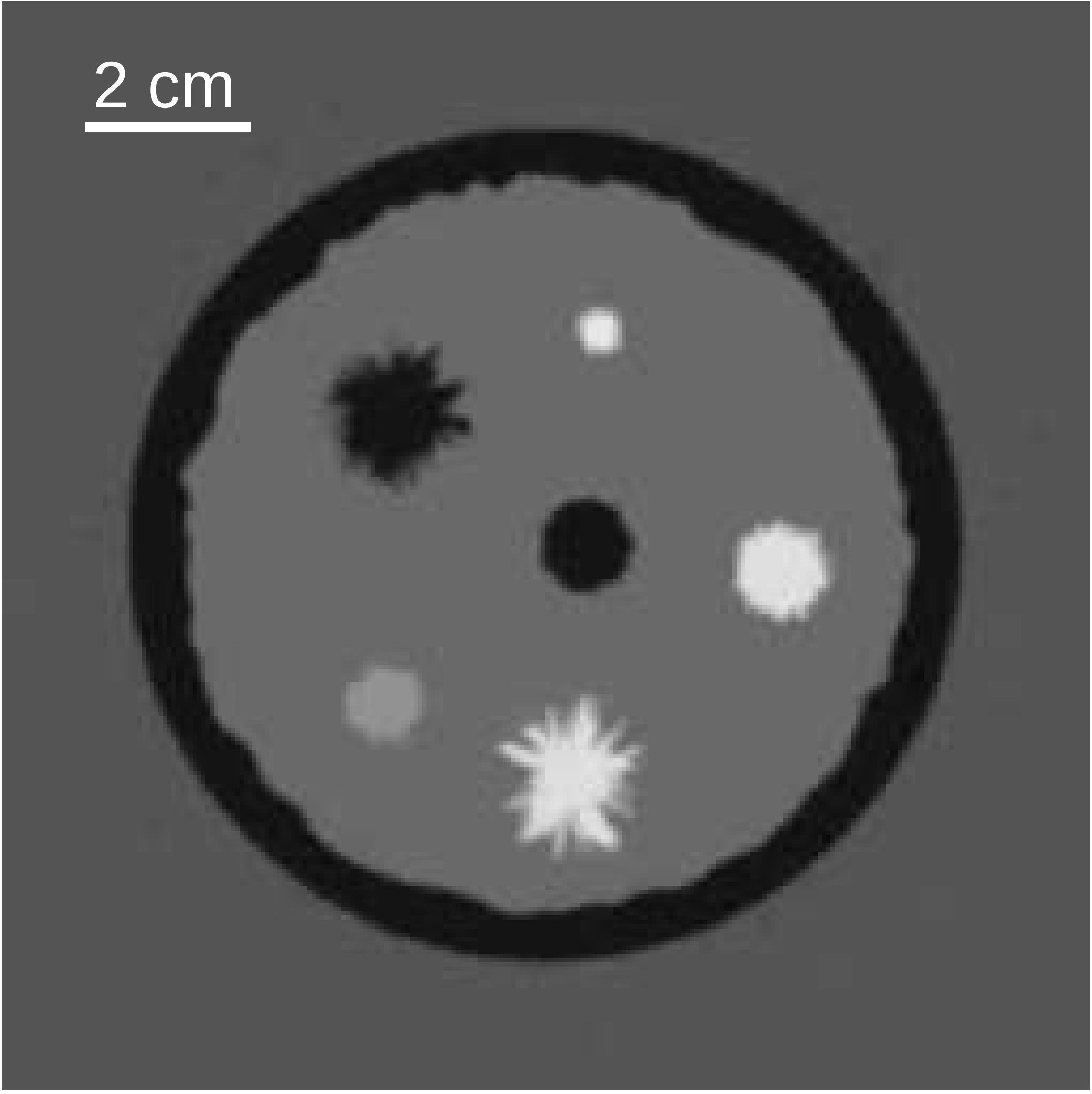}
				\end{subfigure}%
				\begin{subfigure}[b]{0.25\textwidth}
					\includegraphics[width=\textwidth]{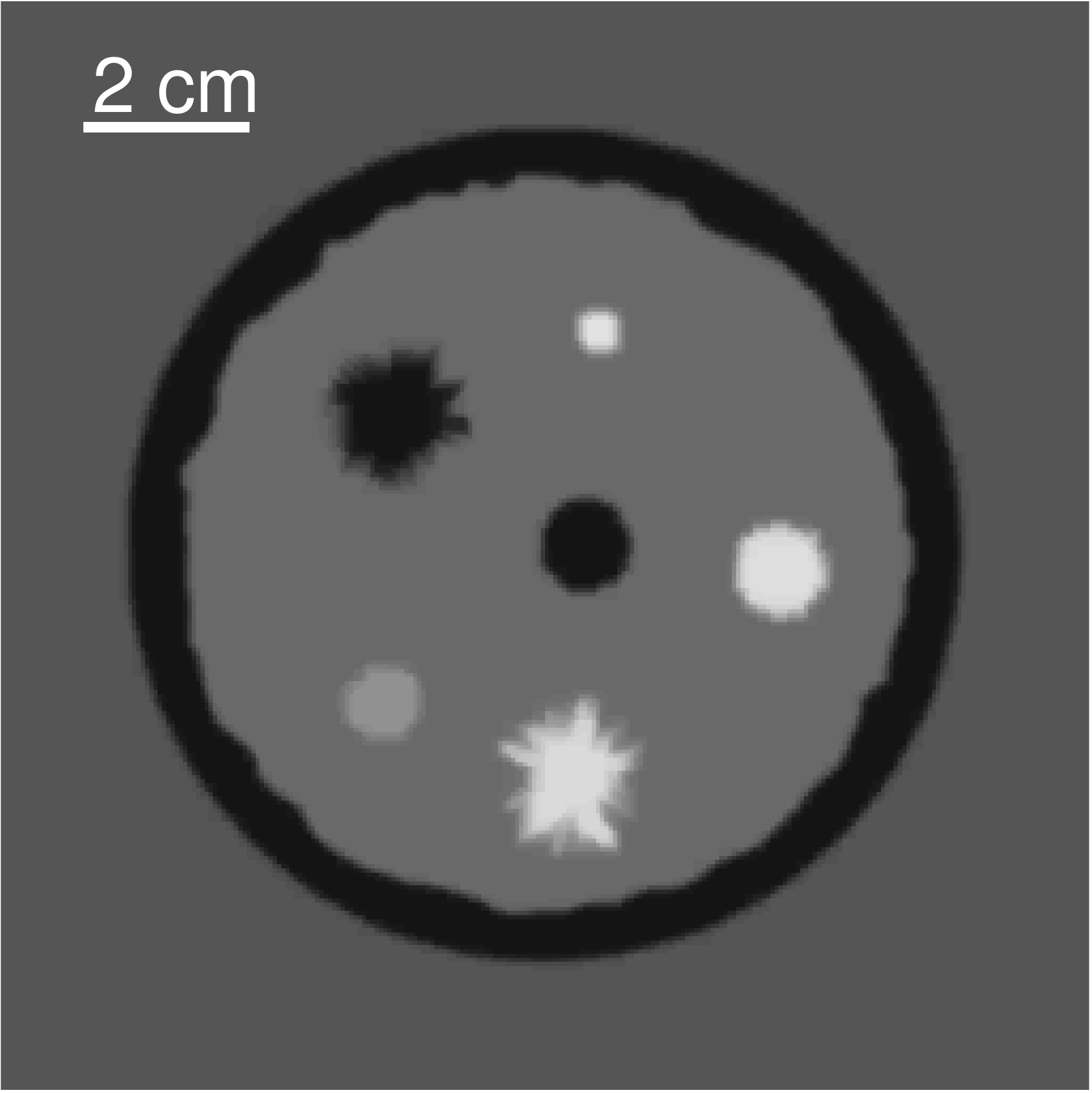}
				\end{subfigure}%
				\caption{Images reconstructed by the weighted RDA method for regularization parameter values of (a) $1 \times 10^{-5}$, (b) $5 \times 10^{-5}$, (c) $1 \times 10^{-4}$, and (d) $5 \times 10^{-4}$, shown after 300 iterations. All images are shown in a grayscale window of $[1.47, 1.58]~ \text{mm}/\mu \text{s}$.}
				%\textcolor{red}{These figures were not generated using the latest weighted RDA  implementation, but I expect the results to be similar.}}
				\label{fig:wrda_regparam_imgs}
			\end{figure*}

\begin{figure*}[htbp!]
				\begin{subfigure}[b]{0.25\textwidth}
					\includegraphics[width=\textwidth]{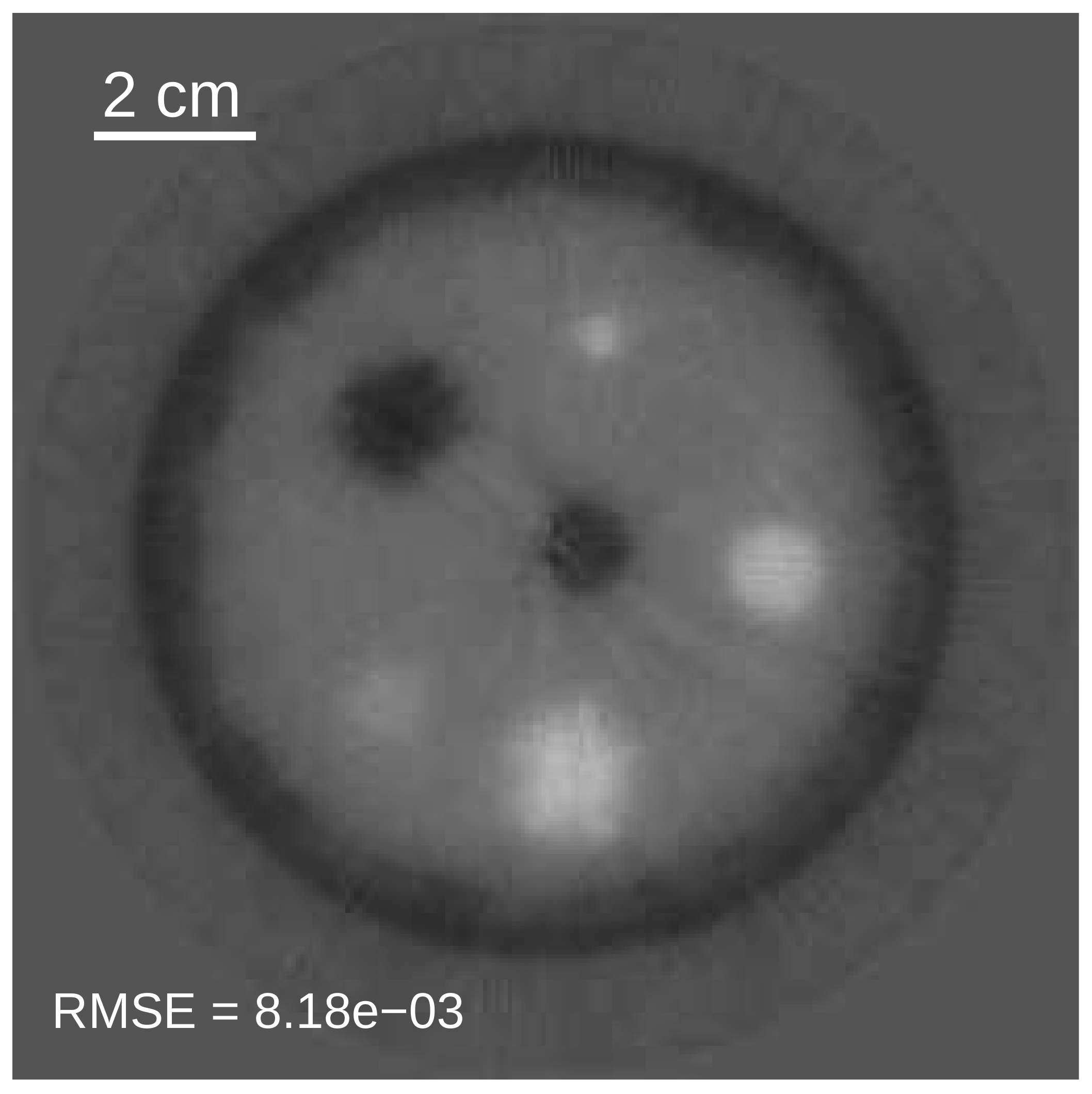}
					\caption{}
				\end{subfigure}%
				\begin{subfigure}[b]{0.25\textwidth}
					\includegraphics[width=\textwidth]{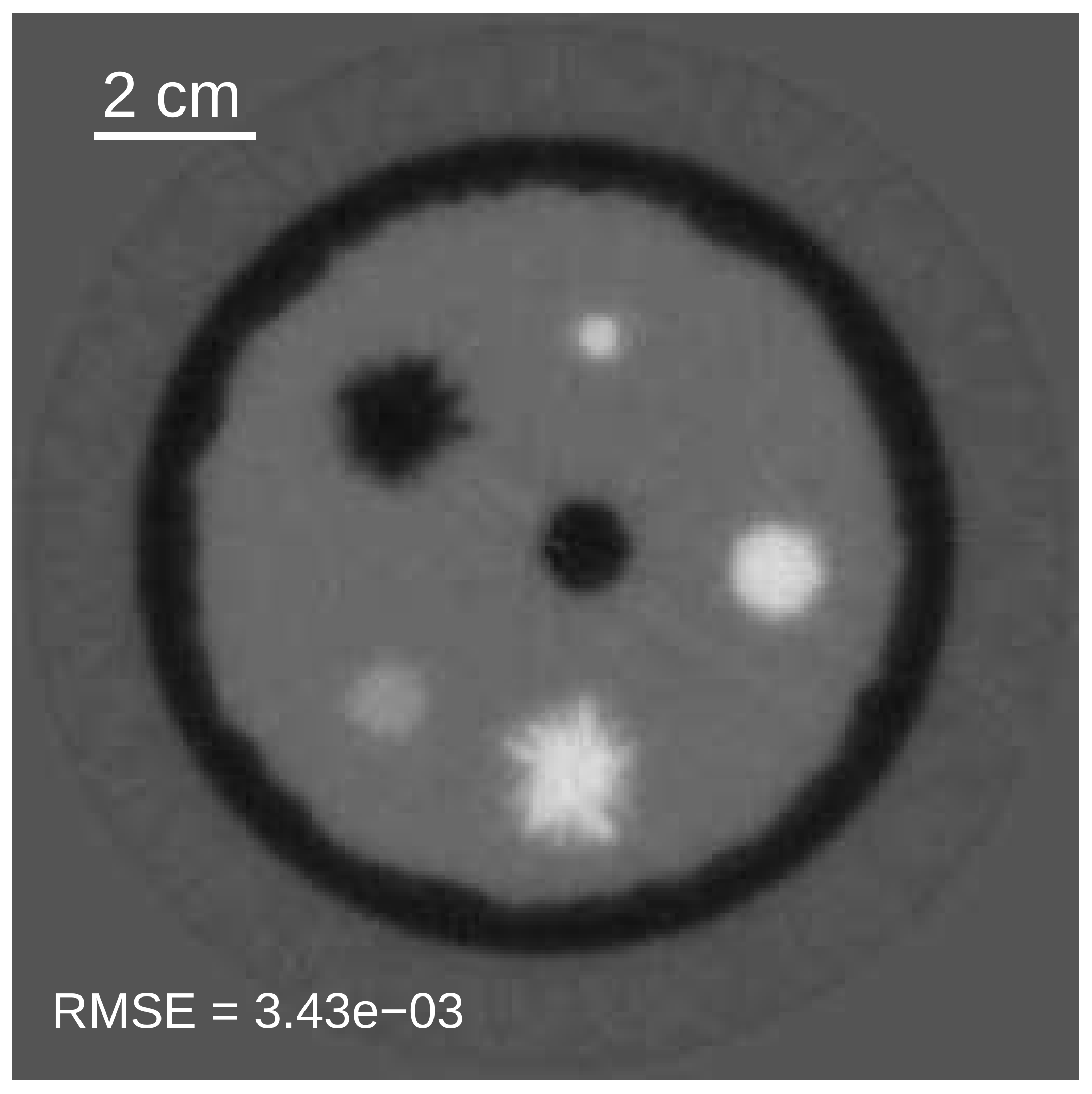}
					\caption{}
				\end{subfigure}%
				\begin{subfigure}[b]{0.25\textwidth}
					\includegraphics[width=\textwidth]{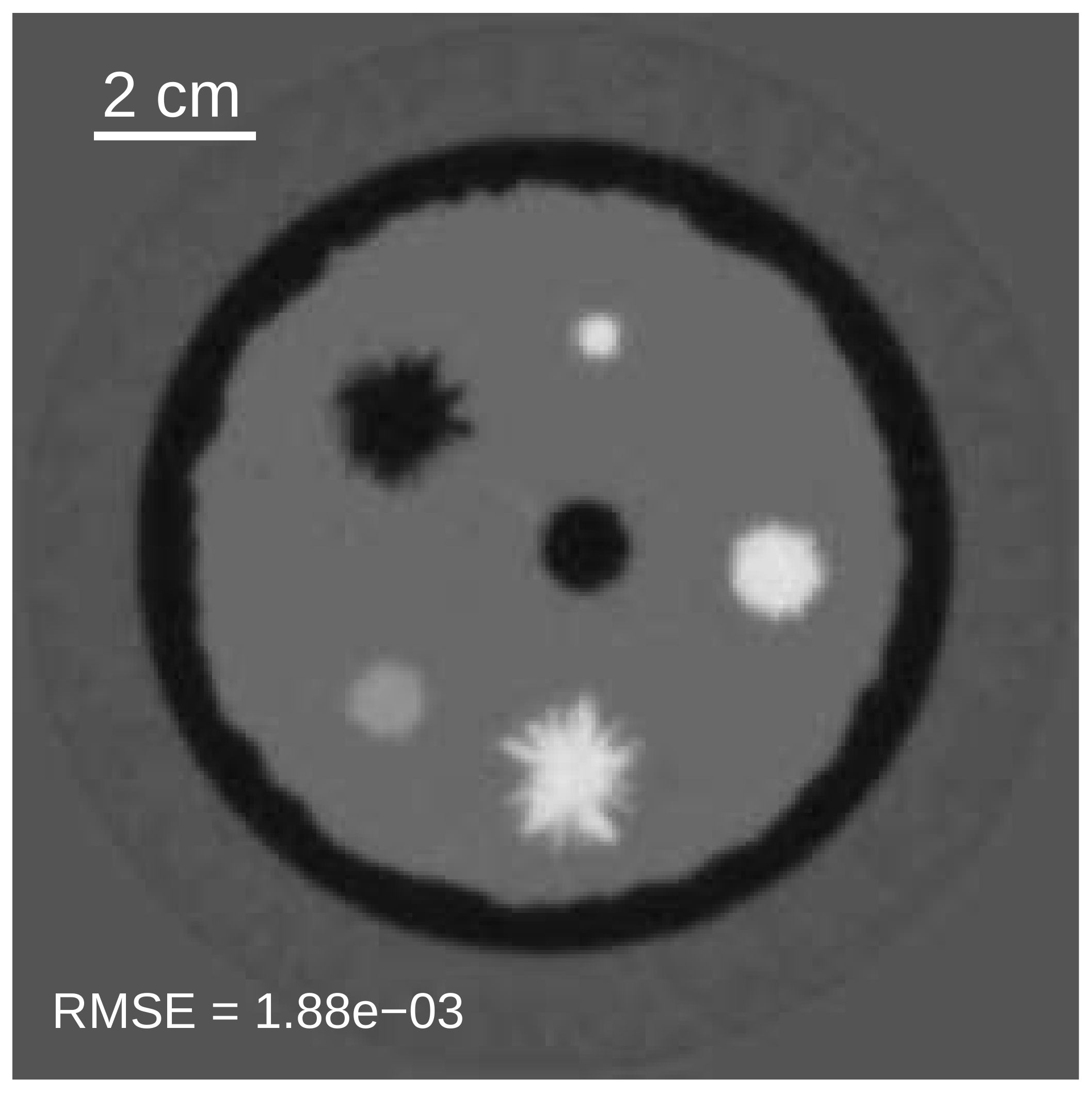}
					\caption{}
				\end{subfigure}%
				\begin{subfigure}[b]{0.25\textwidth}
					\includegraphics[width=\textwidth]{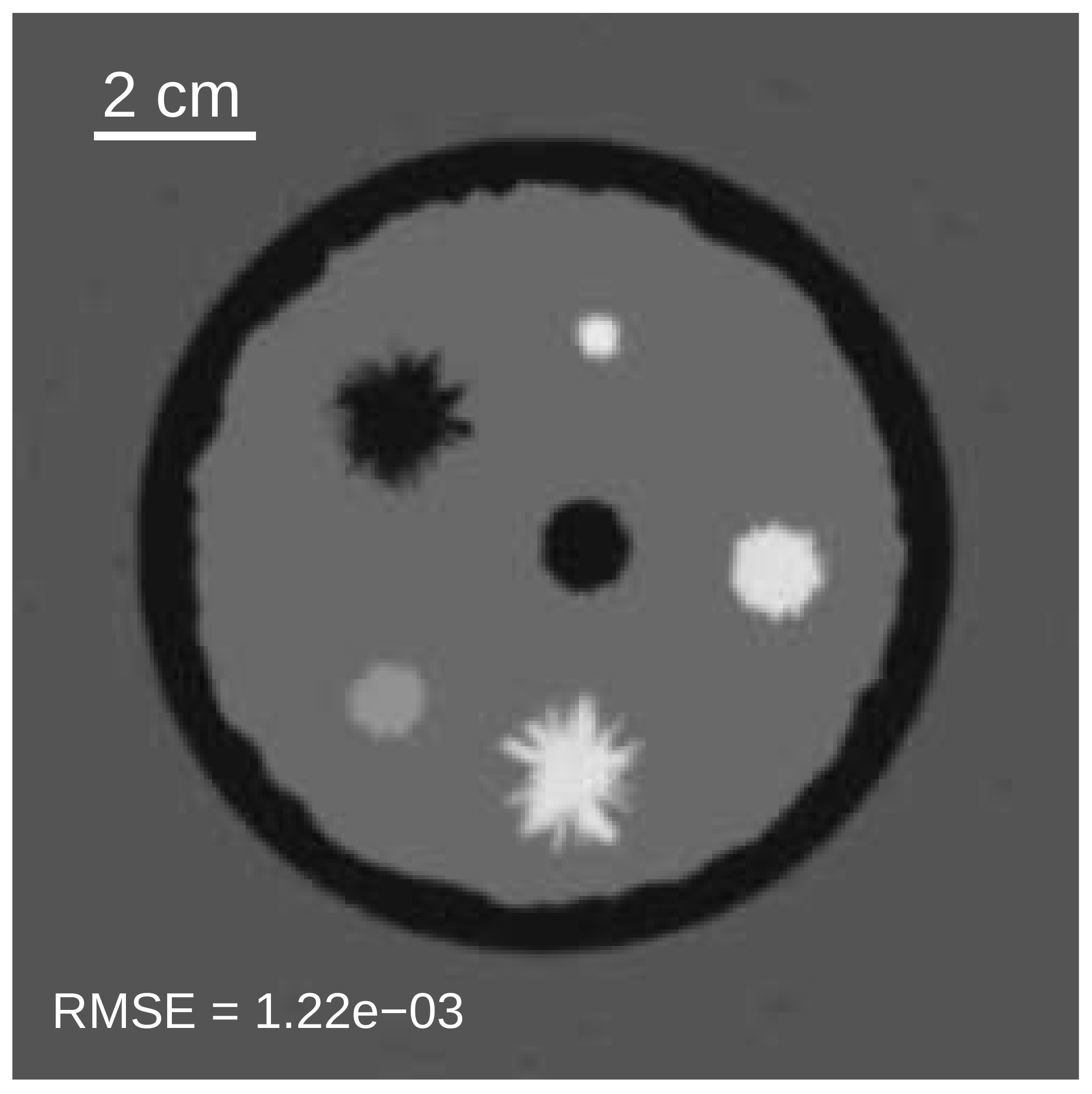}
					\caption{}
				\end{subfigure} \\ 
				\begin{subfigure}[b]{0.25\textwidth}
					\includegraphics[width=\textwidth]{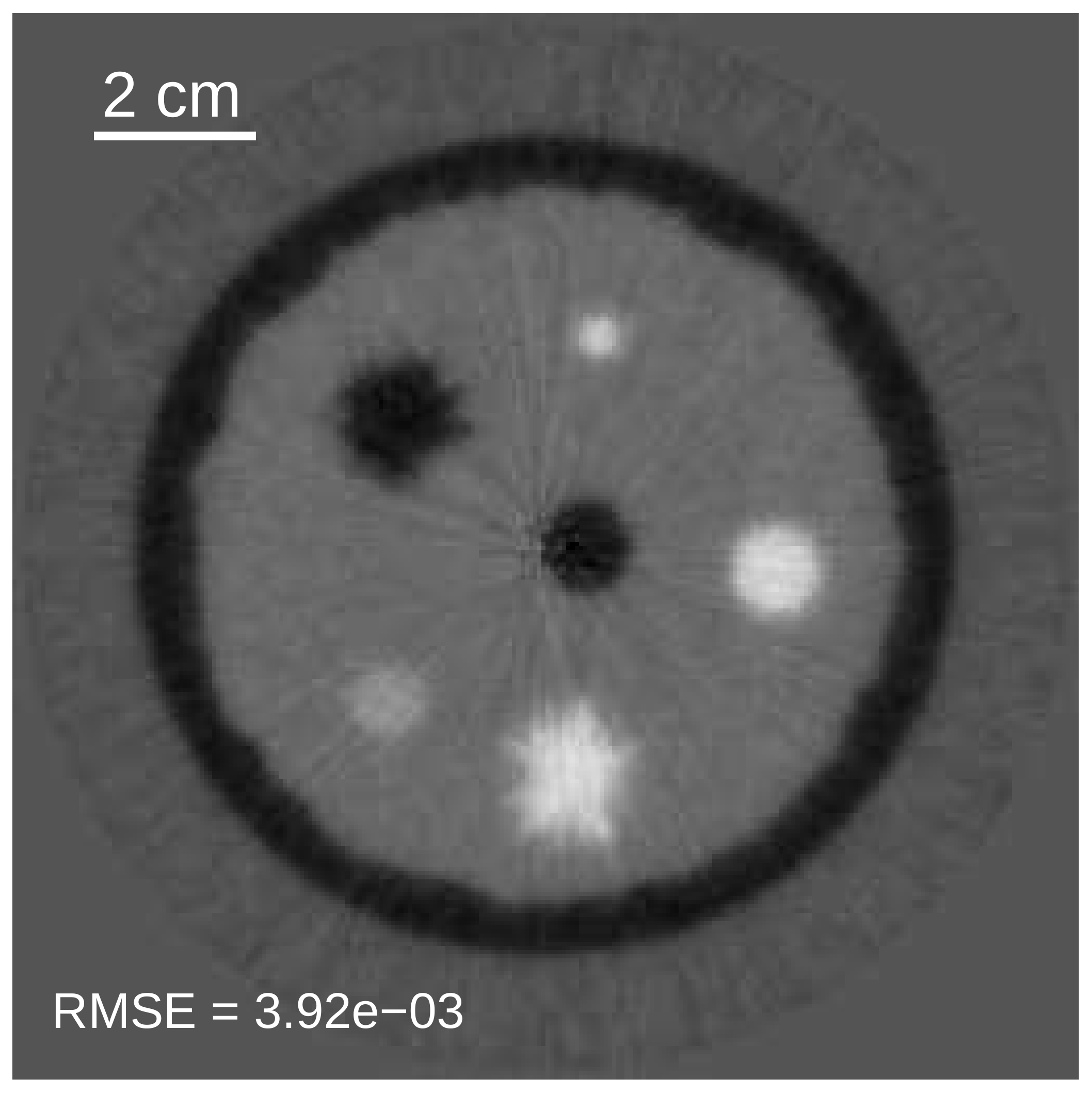}
					\caption{}
				\end{subfigure}%
				\begin{subfigure}[b]{0.25\textwidth}
					\includegraphics[width=\textwidth]{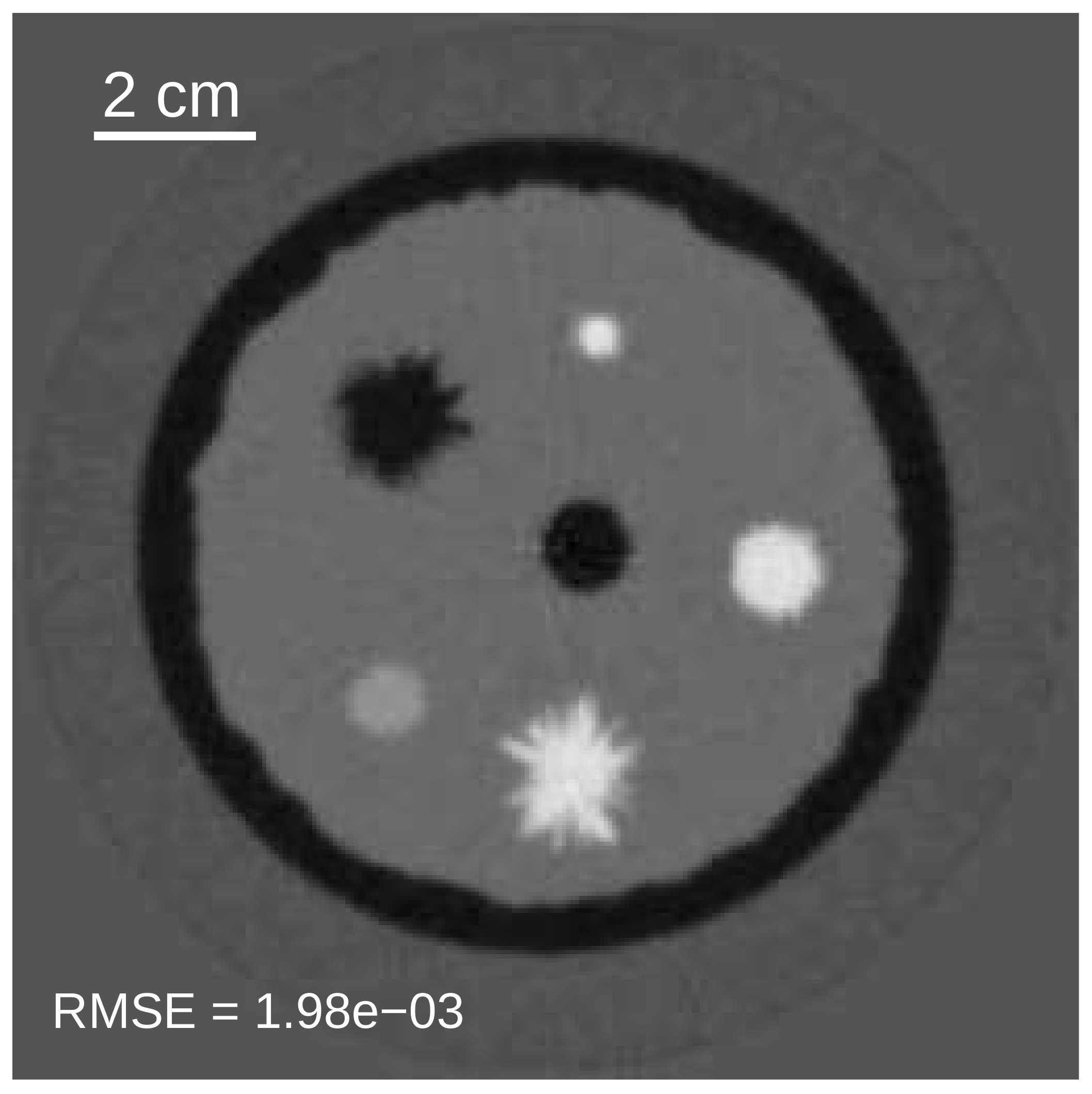}
					\caption{}
				\end{subfigure}%
				\begin{subfigure}[b]{0.25\textwidth}
					\includegraphics[width=\textwidth]{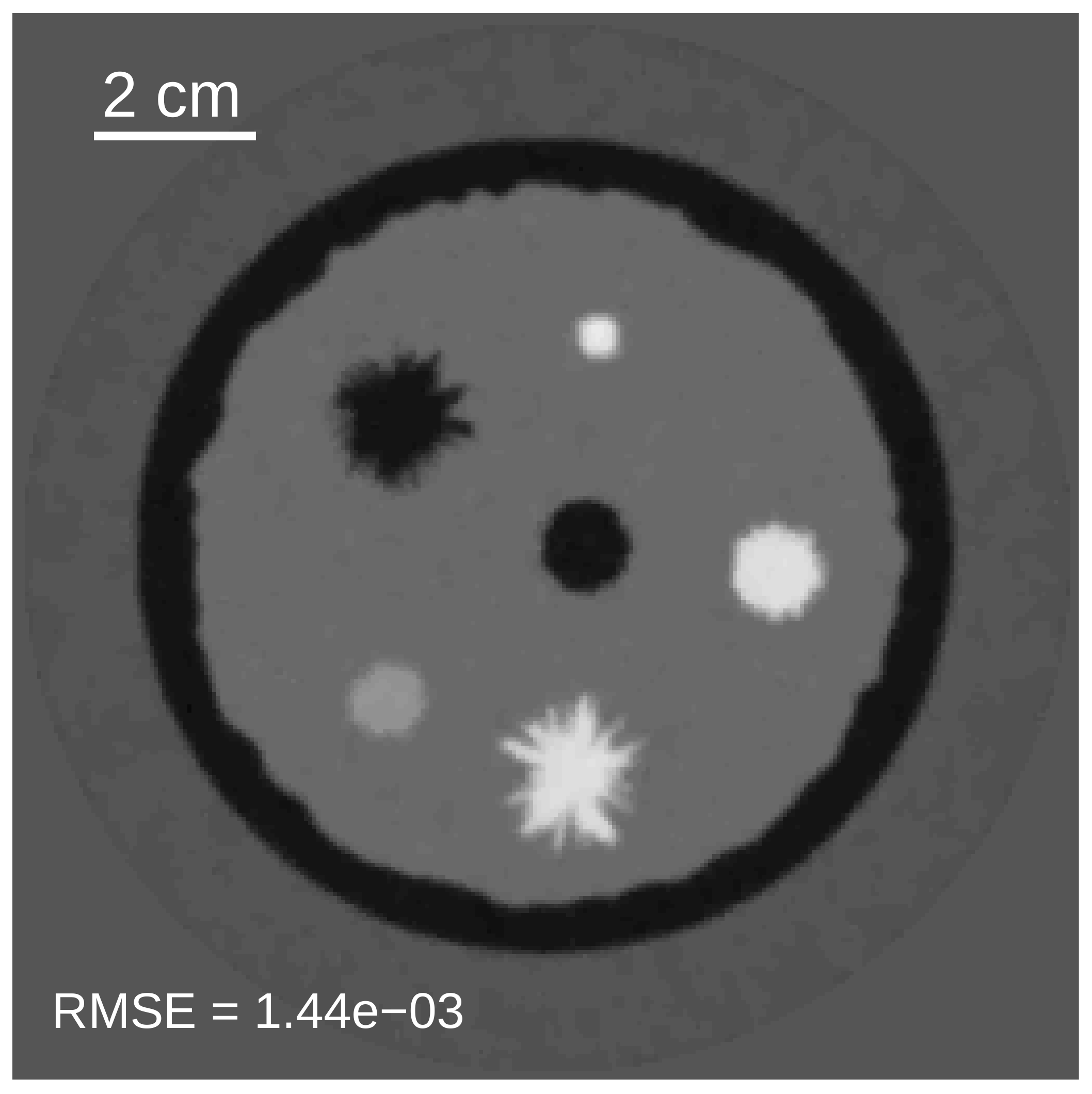}
					\caption{}
				\end{subfigure}%
				\begin{subfigure}[b]{0.25\textwidth}
					\includegraphics[width=\textwidth]{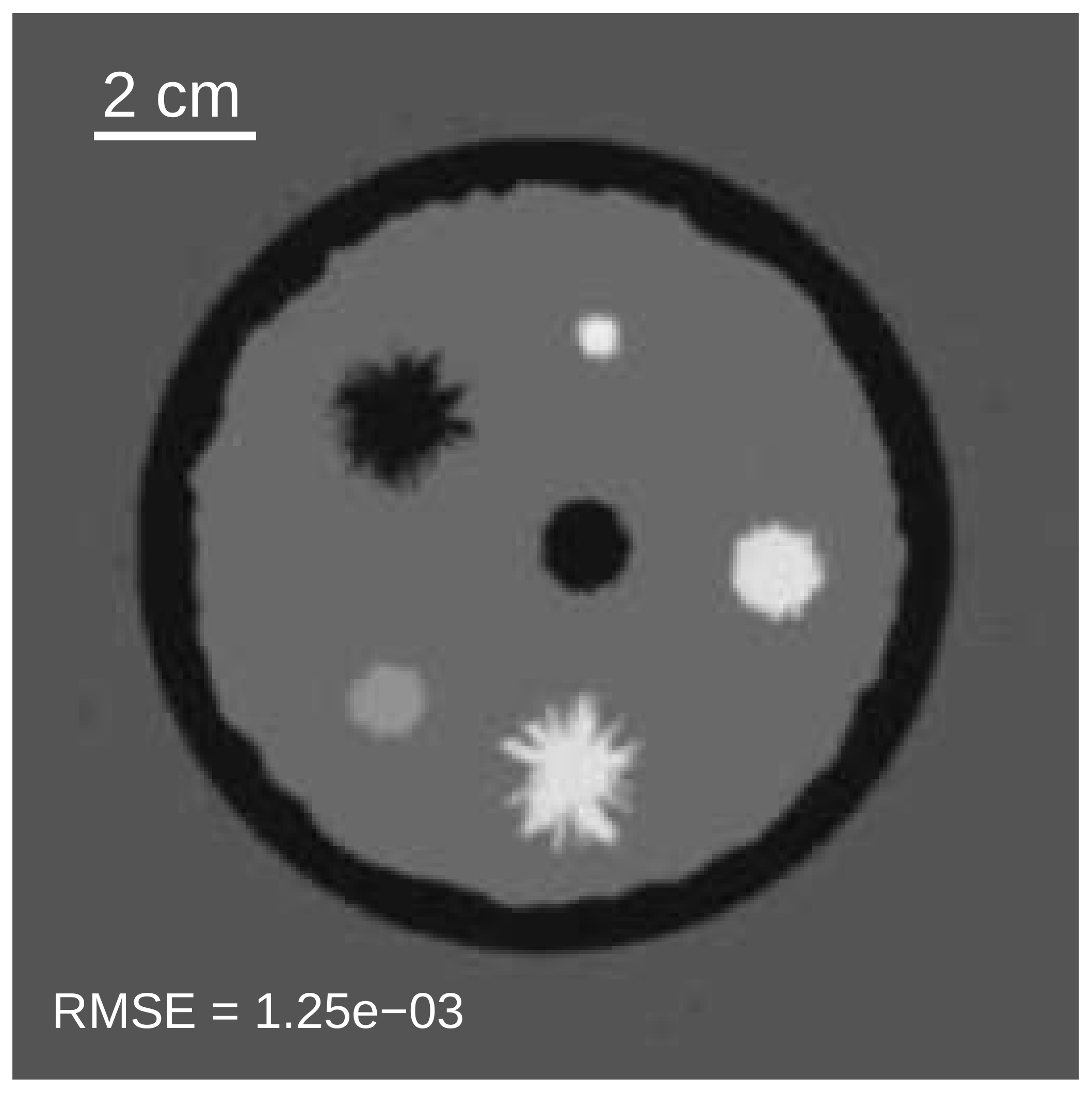}
					\caption{}
				\end{subfigure}%
				\caption{Images reconstructed by use of the unweighted dual averaging method with a fixed step size of 0.1 after (a) 20, (b) 50, (c) 100, and (d) 250 iterations. Images reconstructed by use of the weighted dual averaging method after (e) 20, (f) 50, (g) 100, and (h) 250 iterations. All results are shown for a regularization parameter value of $1 \times 10^{-4}$ and in a grayscale window of $[1.47, 1.58]~ \text{mm}/\mu \text{s}$. The RMSEs for each reconstructed image are displayed in the bottom left of each subfigure.}
				\label{fig:rda_comp_iter_imgs}
			\end{figure*}

\begin{figure}[htbp!]
				\centering
				\begin{subfigure}[b]{0.25\textwidth}
					\includegraphics[width=\textwidth]{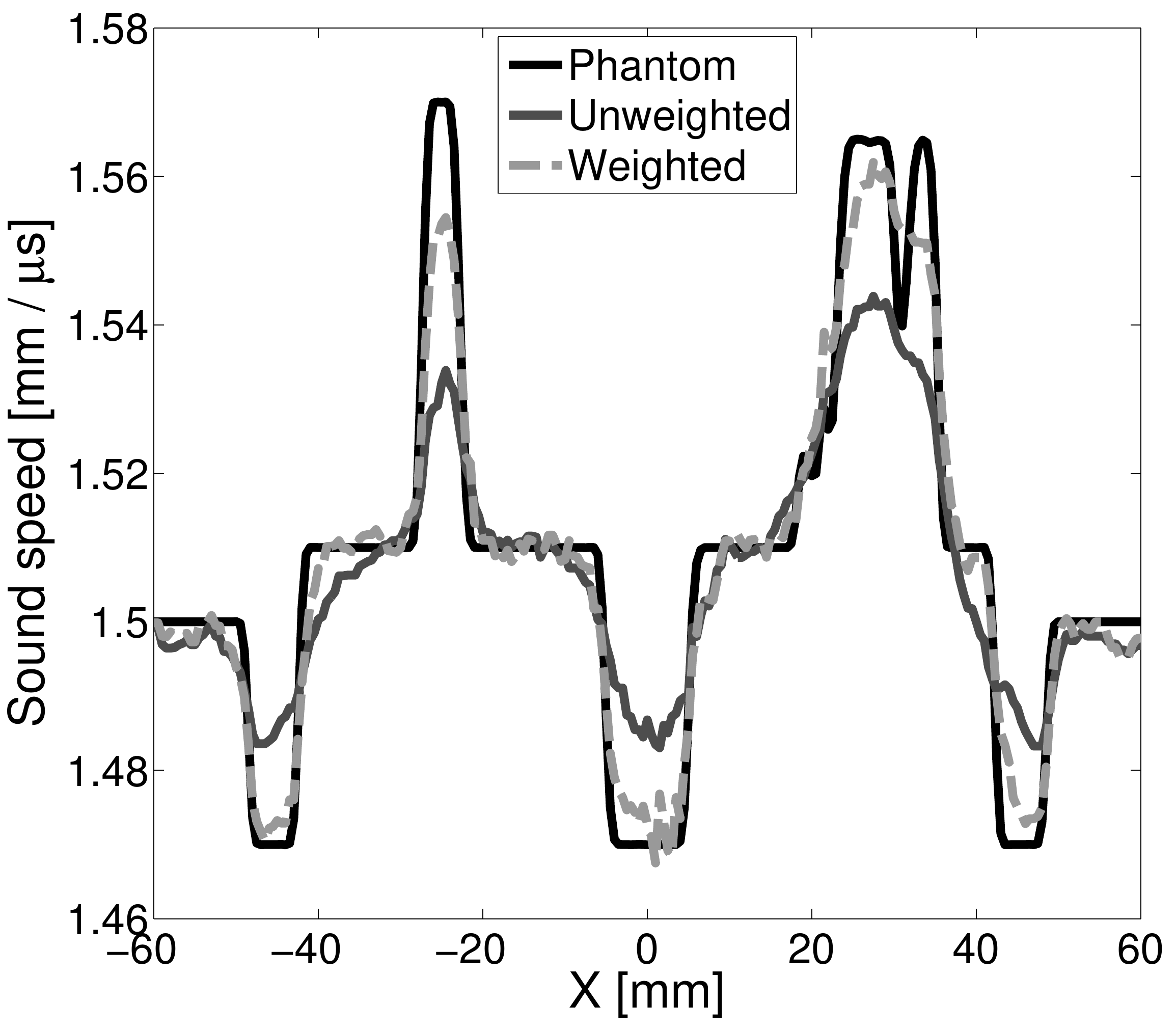}
					\caption{}
				\end{subfigure}%
				\begin{subfigure}[b]{0.25\textwidth}
					\includegraphics[width=\textwidth]{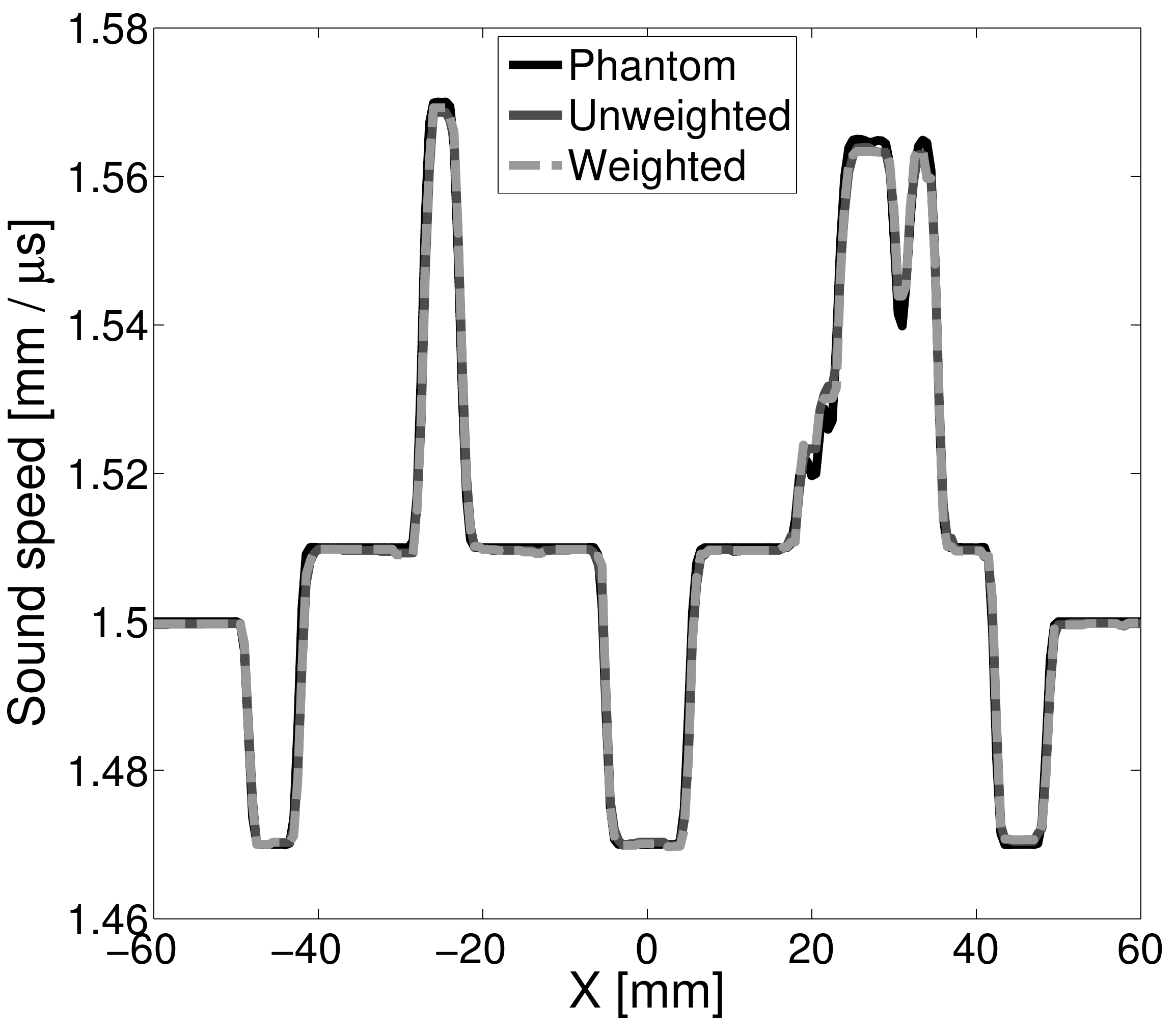}
					\caption{}
				\end{subfigure}
				\caption{(a) Profiles through y = -6.5 mm for reconstructed images obtained by use of the weighted RDA method and the unweighted RDA method with a fixed step size of 0.1, shown after 20 iterations. (b) Profiles through y = -6.5 mm for reconstructed images obtained by use of the weighted RDA method and the unweighted RDA method with a fixed step size of 0.1, shown after 250 iterations.}
				\label{fig:rda_comp_iter_prof}
			\end{figure}

\begin{figure}[htbp!]
				\centering
				\begin{subfigure}[b]{0.25\textwidth}
					\includegraphics[width=\textwidth]{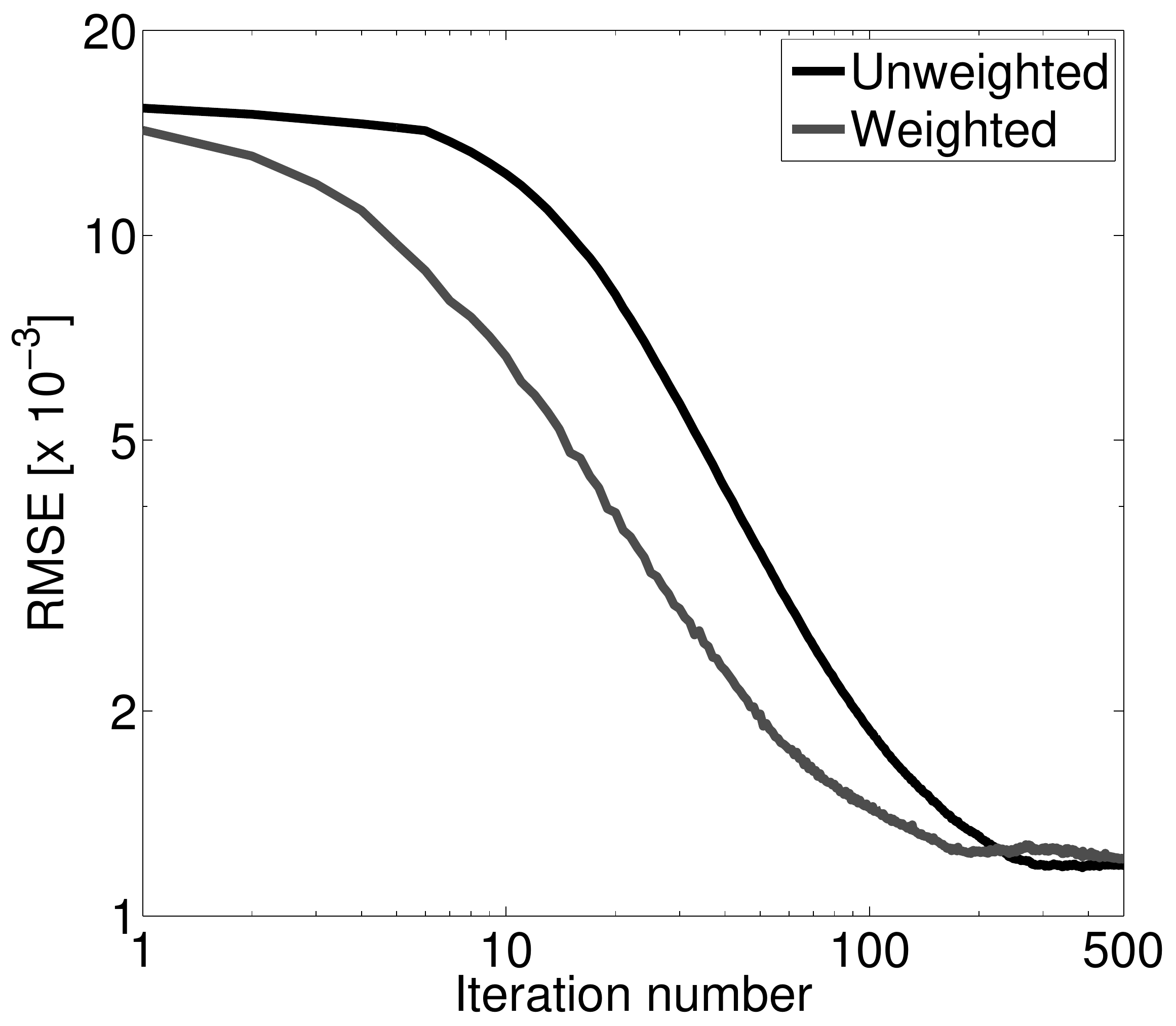}
					\caption{}
				\end{subfigure}%
				\begin{subfigure}[b]{0.25\textwidth}
					\includegraphics[width=\textwidth]{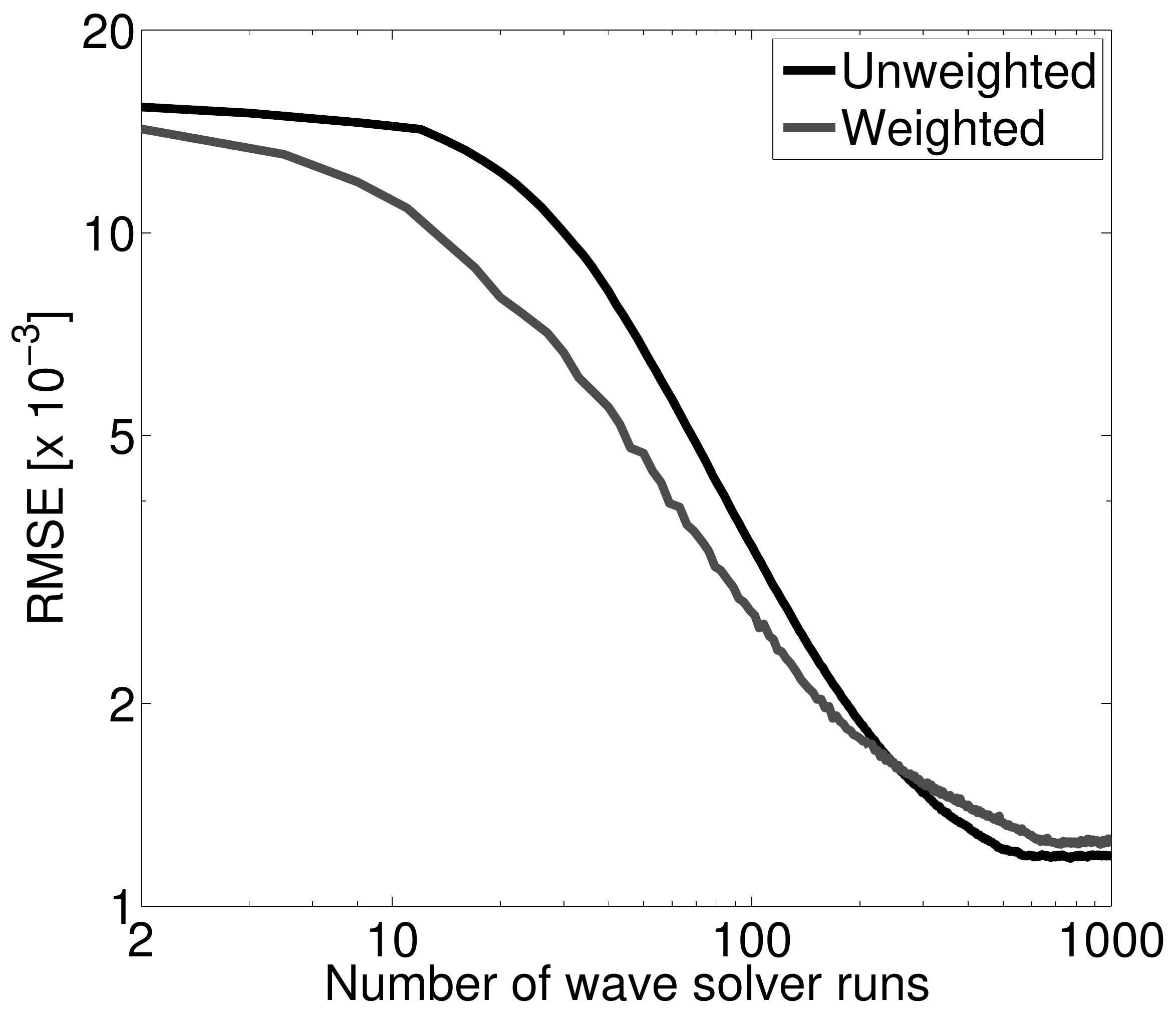}
					\caption{}
				\end{subfigure}
				\caption{Plot of RMSE vs. (a) the number of iterations and (b) the number of wave solver runs for the weighted and unweighted RDA methods.}
				\label{fig:rda_comp_conv}
			\end{figure}

			The weighted RDA method can be used to accelerate the convergence of the RDA method. As was done for the unweighted implementation, images were reconstructed for several regularization parameter values (see Fig.~\ref{fig:wrda_regparam_imgs}). 
			The impact of regularization appears unchanged by the weighting strategy. Once again, a regularization parameter value of $1 \times 10^{-4}$ results in the smallest RMSE. While the ultimate image obtained after many iterations is largely unchanged by the weighting strategy, reconstructed images obtained at early iterations can be greatly improved. As seen in Fig.~\ref{fig:rda_comp_iter_imgs}, the accuracy of the reconstructed images after 20, 50, and even 100 iterations is improved by use of the weighted RDA method. This is seen in both the RMSE of the reconstructed images and in the apparent visual quality of the images. 
			This improvement is reflected in the profiles through the reconstructed images shown in Fig.~\ref{fig:rda_comp_iter_prof}. 
			This improvement is maintained even when the convergence of the reconstruction methods is viewed in terms of the number of wave solver runs as opposed to the number of iterations (see Fig.~\ref{fig:rda_comp_conv}). After approximately 250 wave solver runs (or 250 iterations for the unweighted method), the weighted and unweighted approaches produce images of similar accuracy.				
			
			\subsection{Comparison of images reconstructed by use of SGD and RDA}
\begin{figure*}[htbp!]
				\begin{subfigure}[b]{0.25\textwidth}
					\includegraphics[width=\textwidth]{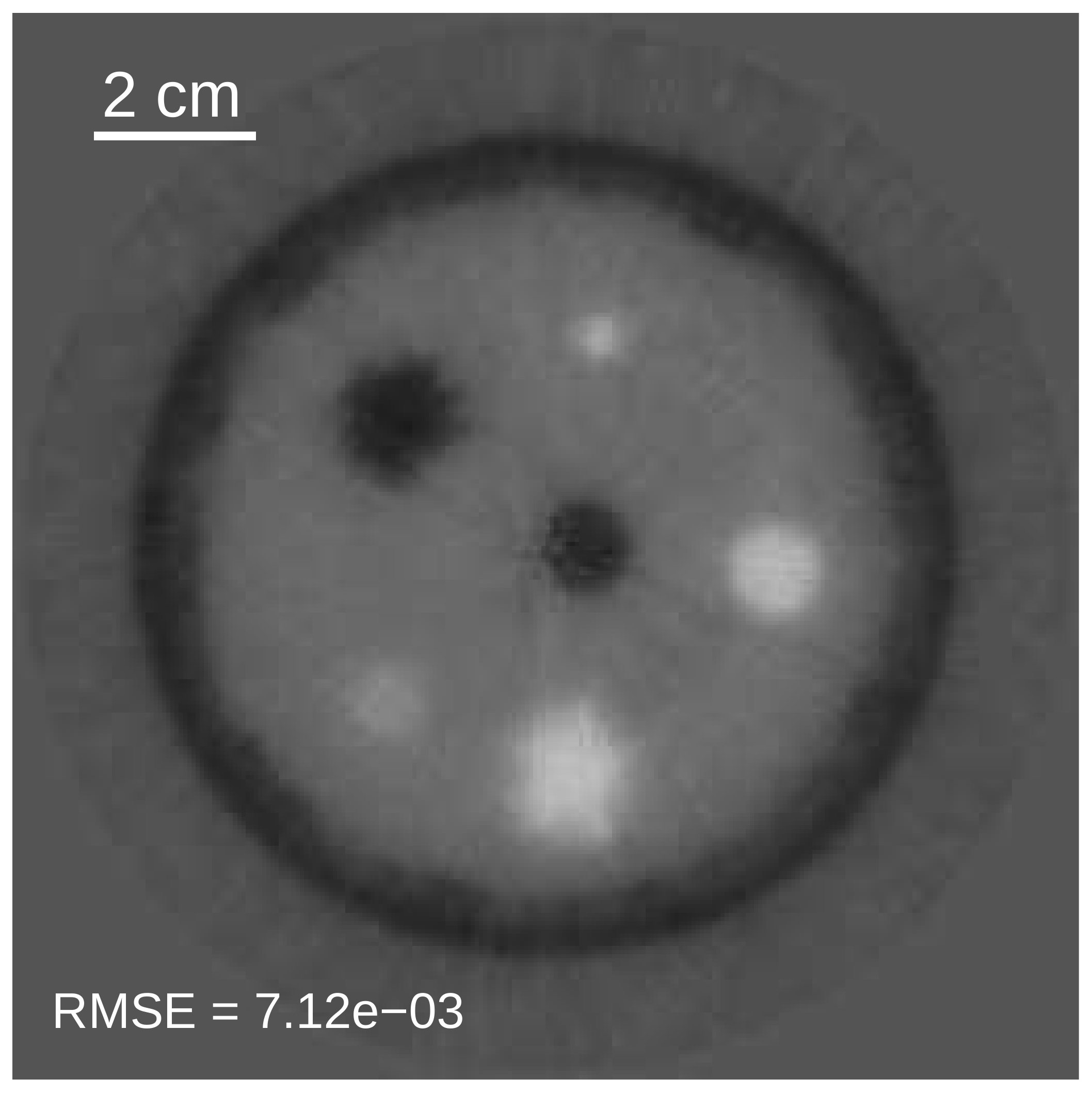}
					\caption{}
				\end{subfigure}%
				\begin{subfigure}[b]{0.25\textwidth}
					\includegraphics[width=\textwidth]{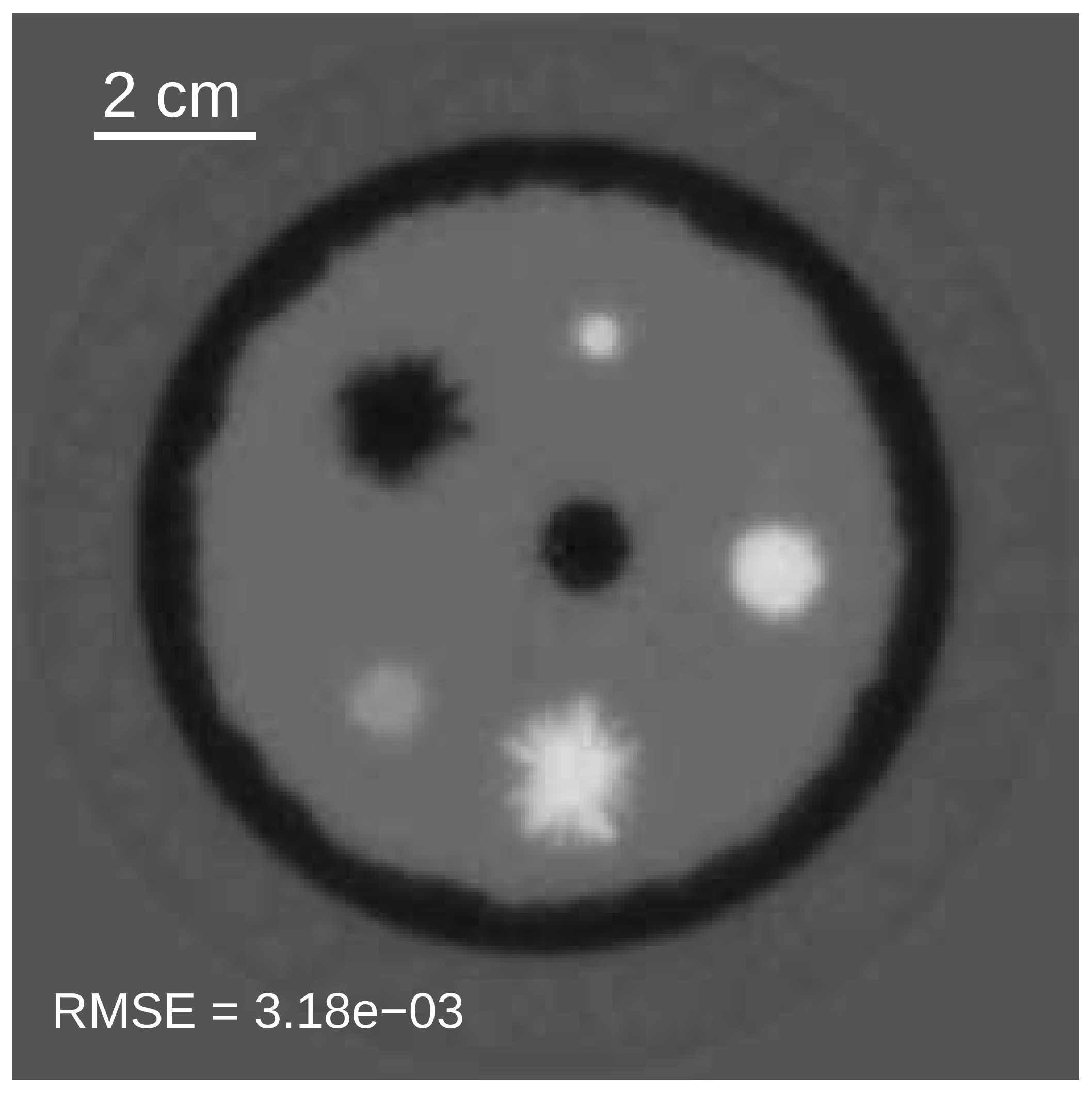}
					\caption{}
				\end{subfigure}%
				\begin{subfigure}[b]{0.25\textwidth}
					\includegraphics[width=\textwidth]{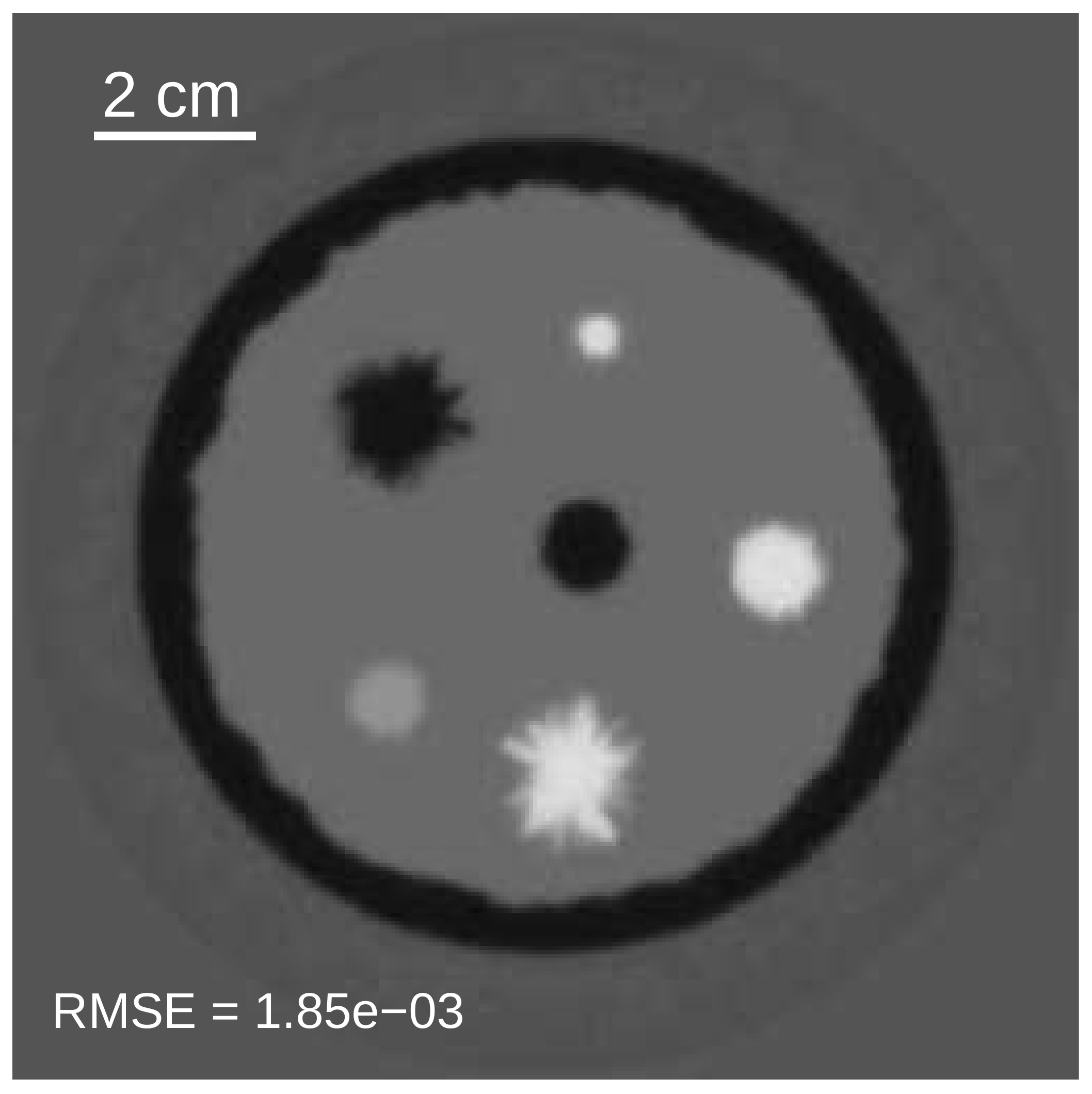}
					\caption{}
				\end{subfigure}%
				\begin{subfigure}[b]{0.25\textwidth}
					\includegraphics[width=\textwidth]{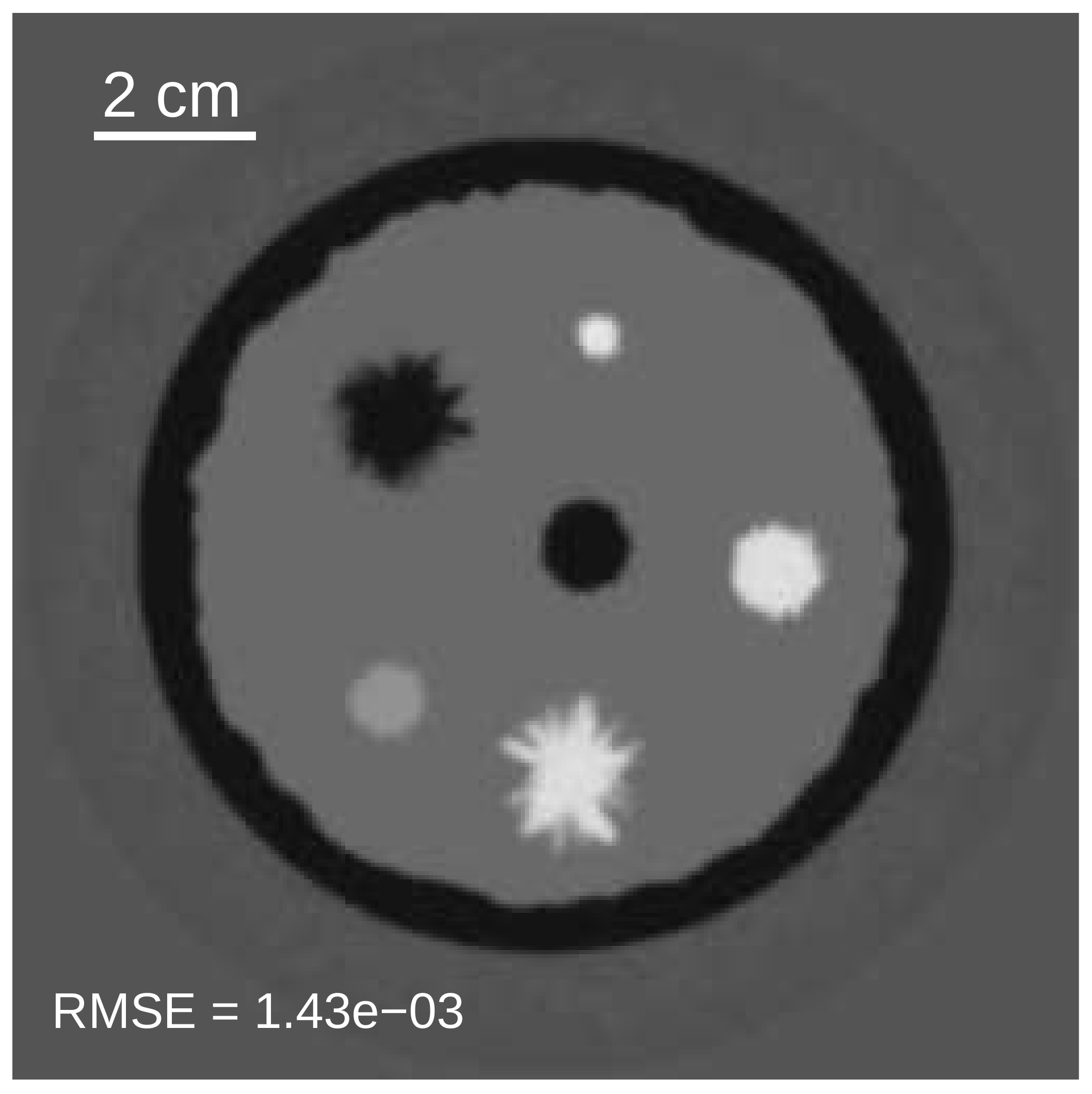}
					\caption{}
				\end{subfigure} \\ 
				\begin{subfigure}[b]{0.25\textwidth}
					\includegraphics[width=\textwidth]{recon19_daconst01tv1e_4}
					\caption{}
				\end{subfigure}%
				\begin{subfigure}[b]{0.25\textwidth}
					\includegraphics[width=\textwidth]{recon49_daconst01tv1e_4}
					\caption{}
				\end{subfigure}%
				\begin{subfigure}[b]{0.25\textwidth}
					\includegraphics[width=\textwidth]{recon99_daconst01tv1e_4}
					\caption{}
				\end{subfigure}%
				\begin{subfigure}[b]{0.25\textwidth}
					\includegraphics[width=\textwidth]{recon249_daconst01tv1e_4}
					\caption{}
				\end{subfigure} \\
				\begin{subfigure}[b]{0.25\textwidth}
					\includegraphics[width=\textwidth]{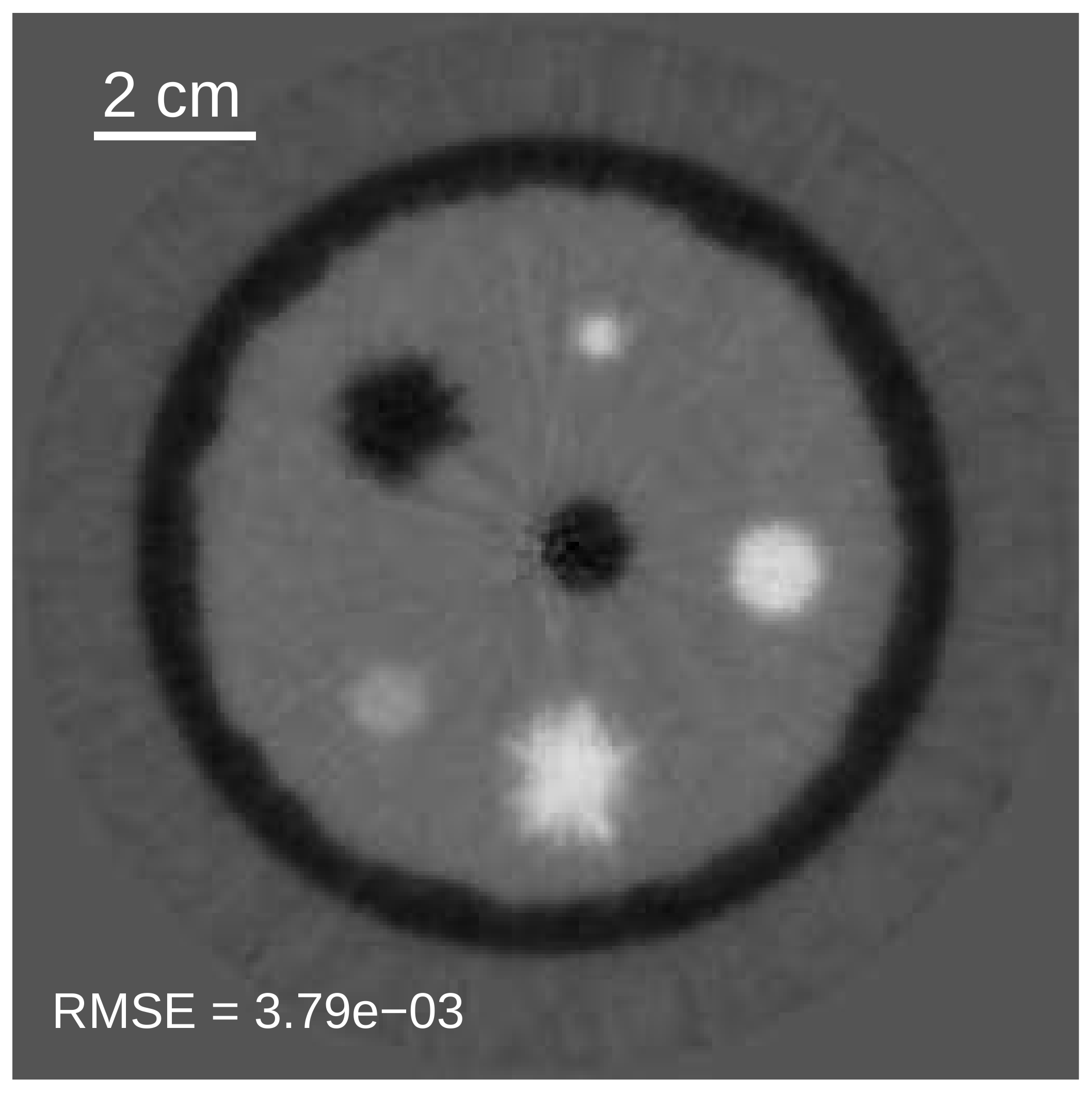}
					\caption{}
				\end{subfigure}%
				\begin{subfigure}[b]{0.25\textwidth}
					\includegraphics[width=\textwidth]{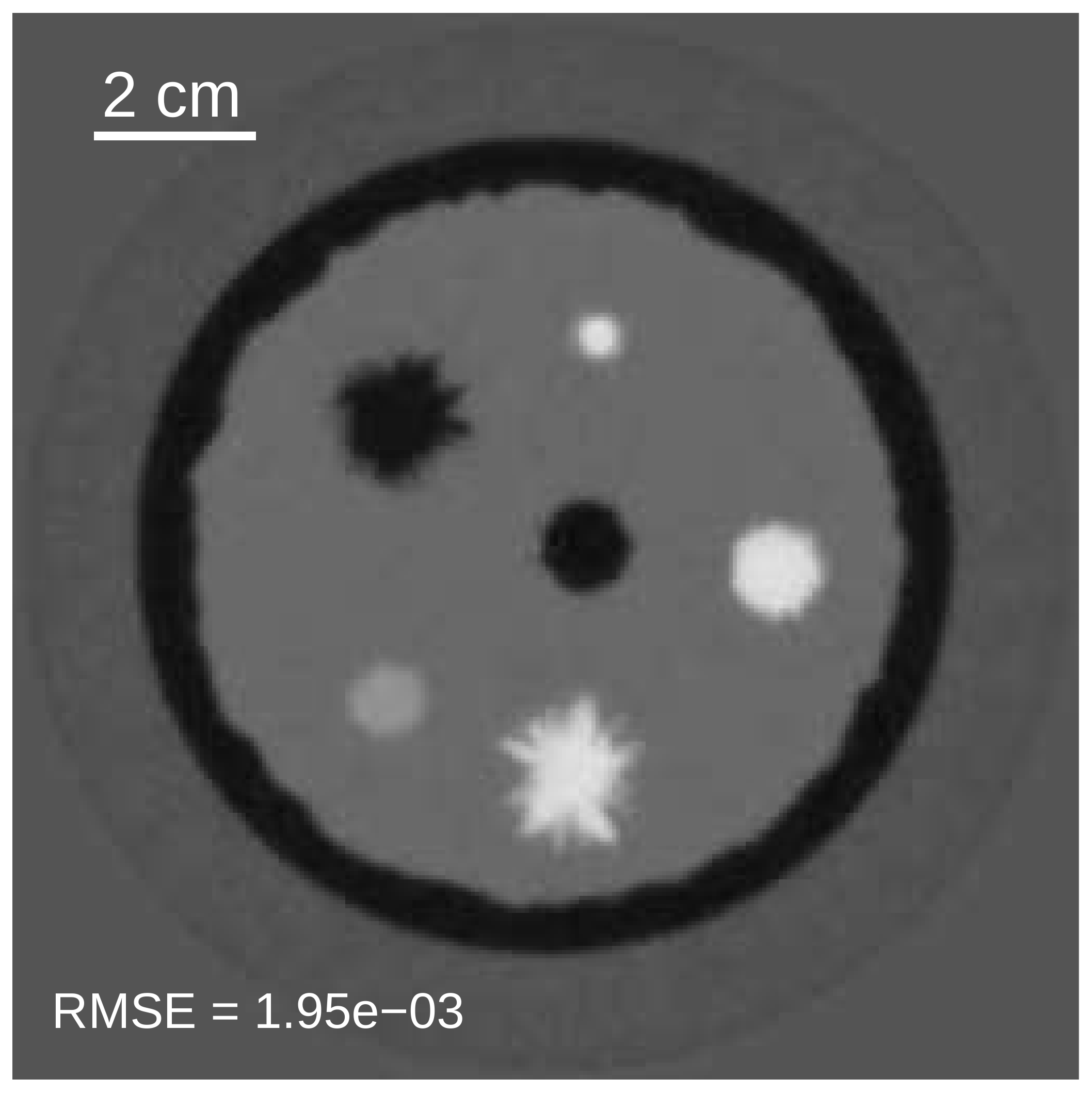}
					\caption{}
				\end{subfigure}%
				\begin{subfigure}[b]{0.25\textwidth}
					\includegraphics[width=\textwidth]{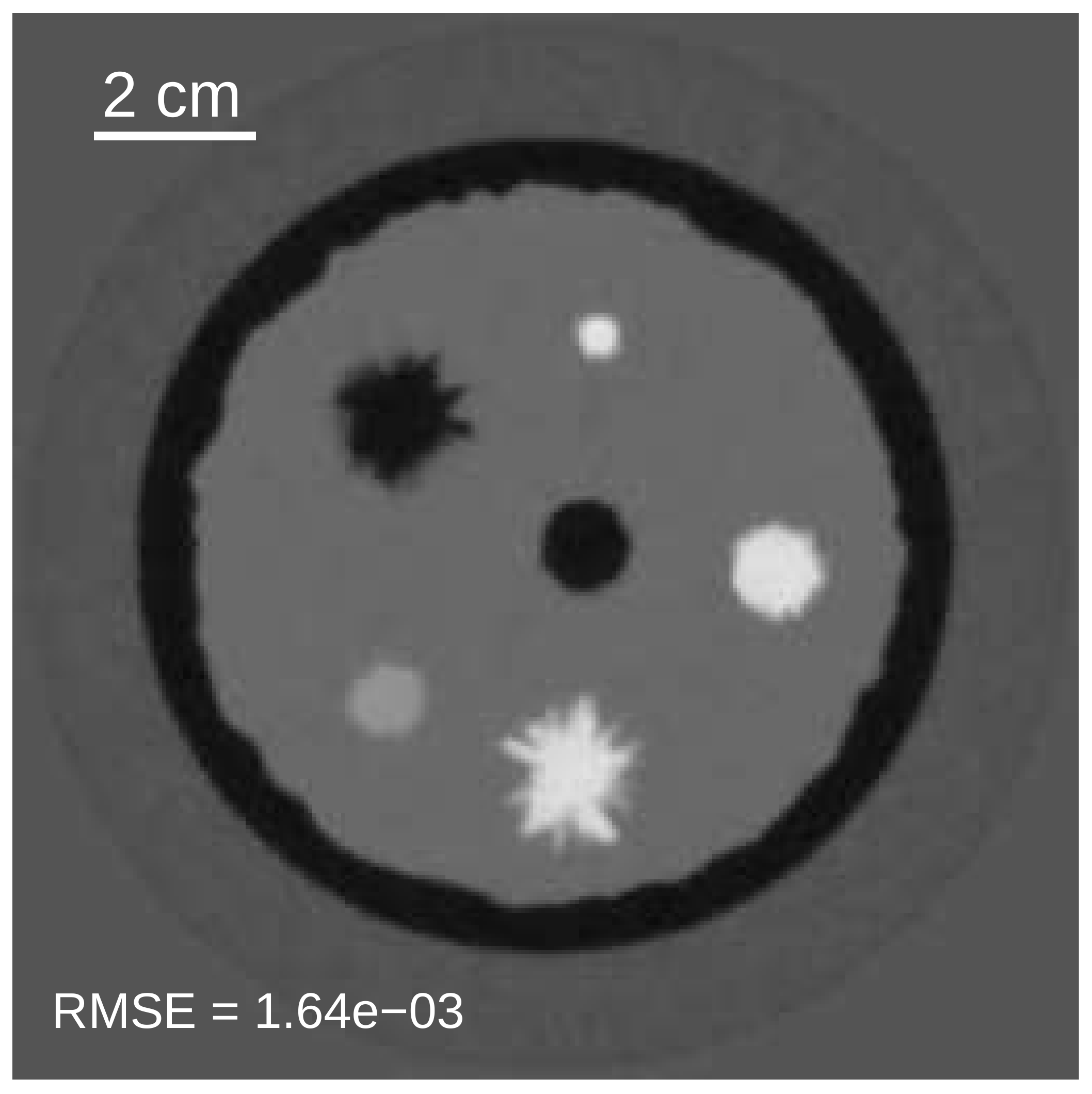}
					\caption{}
				\end{subfigure}%
				\begin{subfigure}[b]{0.25\textwidth}
					\includegraphics[width=\textwidth]{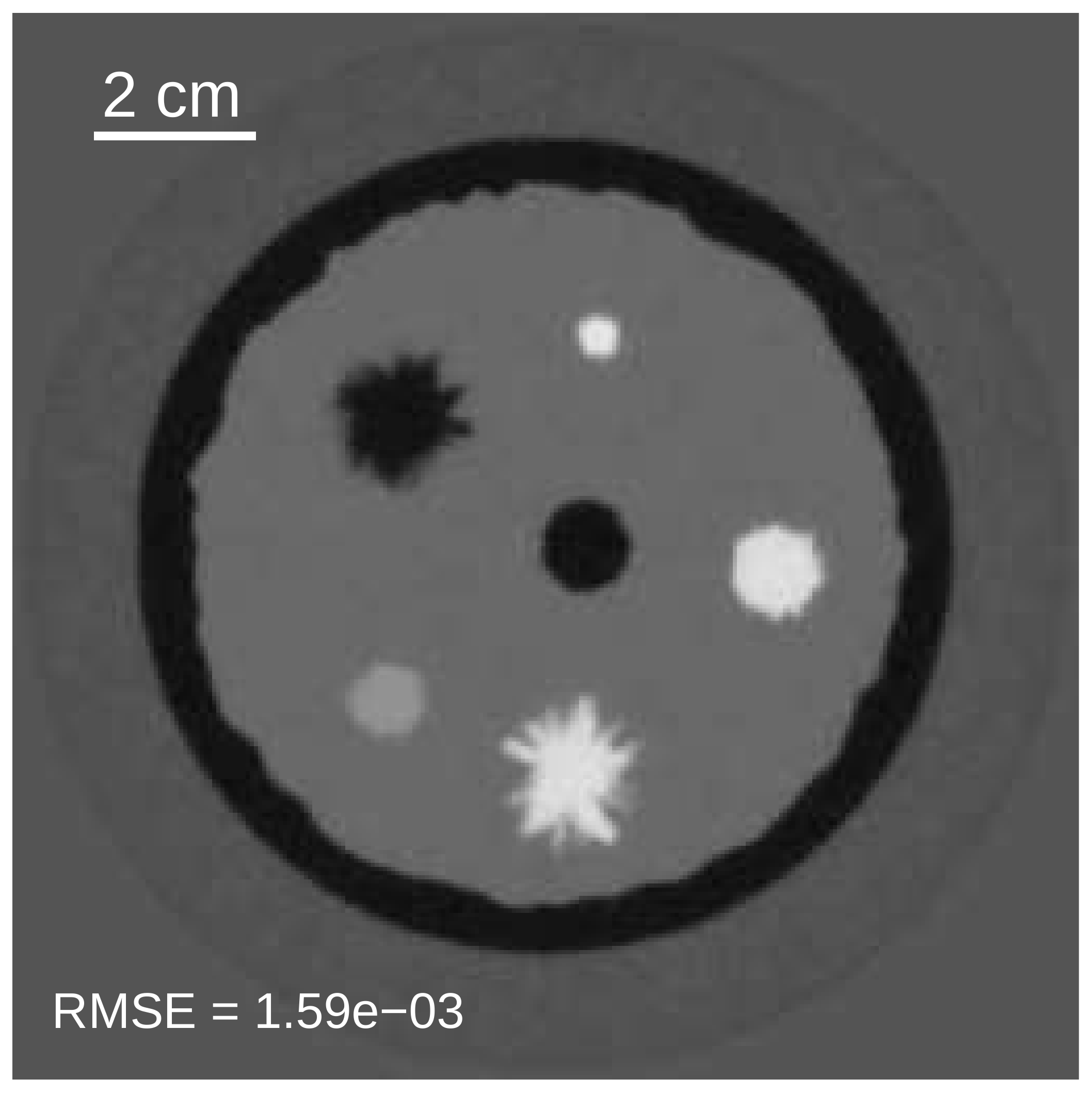}
					\caption{}
				\end{subfigure} \\ 
				\begin{subfigure}[b]{0.25\textwidth}
					\includegraphics[width=\textwidth]{recon19_dalstv1e_4}
					\caption{}
				\end{subfigure}%
				\begin{subfigure}[b]{0.25\textwidth}
					\includegraphics[width=\textwidth]{recon49_dalstv1e_4}
					\caption{}
				\end{subfigure}%
				\begin{subfigure}[b]{0.25\textwidth}
					\includegraphics[width=\textwidth]{recon99_dalstv1e_4}
					\caption{}
				\end{subfigure}%
				\begin{subfigure}[b]{0.25\textwidth}
					\includegraphics[width=\textwidth]{recon249_dalstv1e_4}
					\caption{}
				\end{subfigure}%
				\caption{(Row 1) Images reconstructed by use of SGD with a constant step size of 0.1 and a regularization parameter value of $5 \times 10^{-4}$ after (a) 20, (b) 50, (c) 100, and (d) 250 iterations. (Row 2) Images reconstructed by use of unweighted RDA with a fixed step size of 0.1 and a regularization parameter value of $1 \times 10^{-4}$ after (e) 20, (f) 50, (g) 100, and (h) 250 iterations. (Row 3) Images reconstructed by use of SGD with a line search and a regularization parameter value of $5 \times 10^{-4}$ after (i) 20, (j) 50, (k) 100, and (l) 250 iterations. (Row 4) Images reconstructed by use of weighted RDA with a regularization parameter value of $1 \times 10^{-4}$ after (m) 20, (n) 50, (o) 100, and (p) 250 iterations. All images are shown in a grayscale window of $[1.47, 1.58]~ \text{mm}/\mu \text{s}$.}
				\label{fig:all_methods_comp_imgs}
			\end{figure*}

			The images produced by use of the SGD and RDA methods were compared directly. Images reconstructed by all four implementations are shown in Fig.~\ref{fig:all_methods_comp_imgs}: (1) SGD with a constant step size, (2) unweighted RDA, (3) SGD with a line search, and (4) weighted RDA. As indicated by the RMSEs noted in the bottom left of each image, the initial convergence rates of SGD with a line search and the weighted RDA method are much faster than that of either SGD with a constant step size or the unweighted RDA method. However, the accuracy of the reconstructed images at later iterations is superior for the two RDA methods compared with the SGD-based methods. In fact, the accuracy of the image reconstructed by the weighted RDA method is better than that obtained by SGD with a constant step size. This demonstrates that the weighted RDA method can provide both fast convergence and more accurate images than was possible using the SGD method.
			
\begin{figure}[htbp!]
				\centering
				\includegraphics[width=0.4\textwidth]{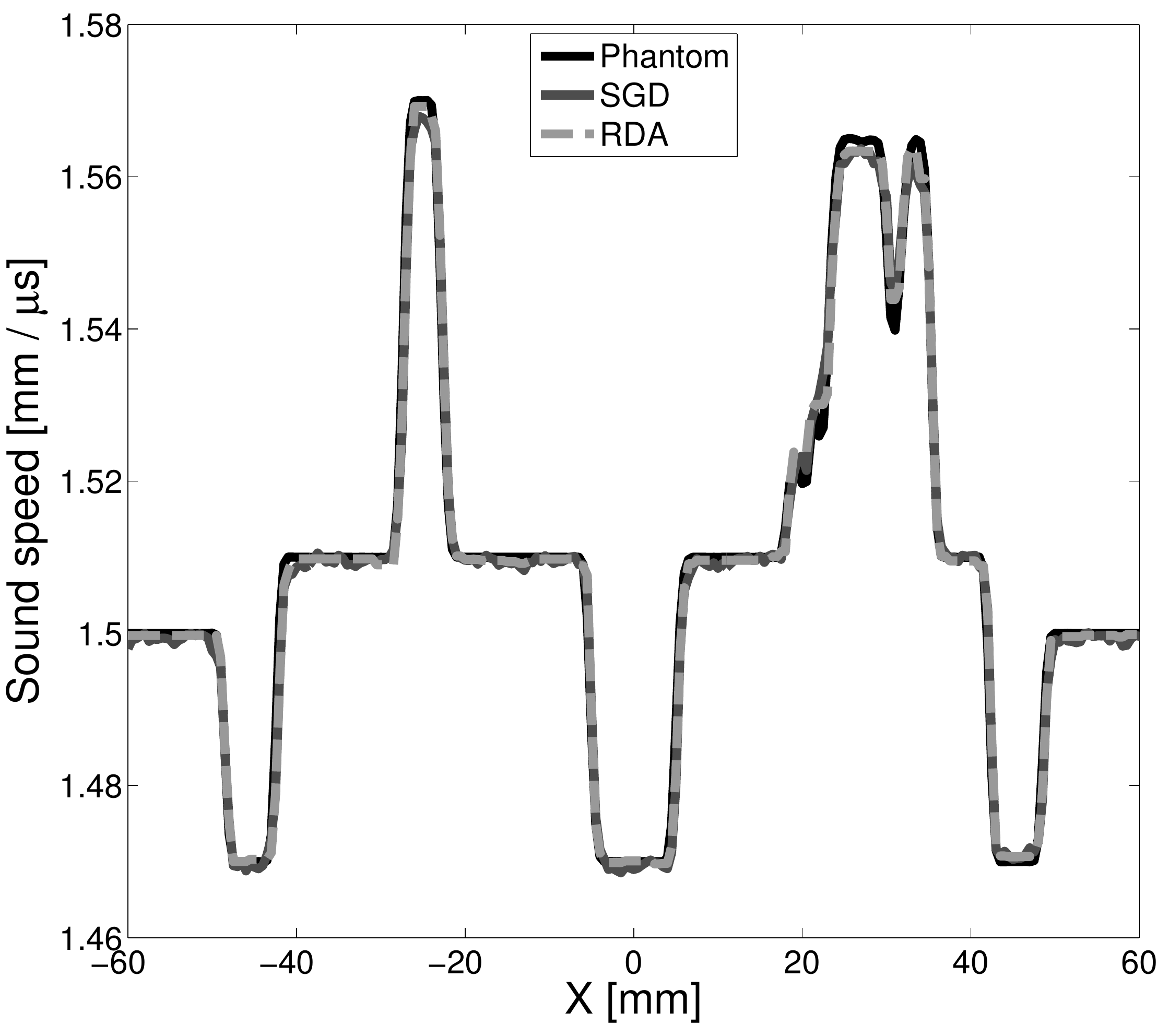}
				\caption{Profiles through y = -6.5 mm for images reconstructed by the use of SGD with a line search and weighted RDA.}
				\label{fig:rda_sgd_comp_prof}
			\end{figure}

			The improved accuracy of the weighted RDA method compared with SGD with a line search is reflected in the profiles through the reconstructed images (see Fig.~\ref{fig:rda_sgd_comp_prof}). The profile obtained by use of SGD is noticeably noisier than that obtained by use of RDA. This suggests that the RDA method may be more effective in mitigating noise than SGD. This potential benefit will be considered more closely through the use of a bias-variance analysis, detailed in Section \ref{sec:bias-variance}.
			
\begin{figure}[htbp!]
				\begin{subfigure}[b]{0.25\textwidth}
					\includegraphics[width=\textwidth]{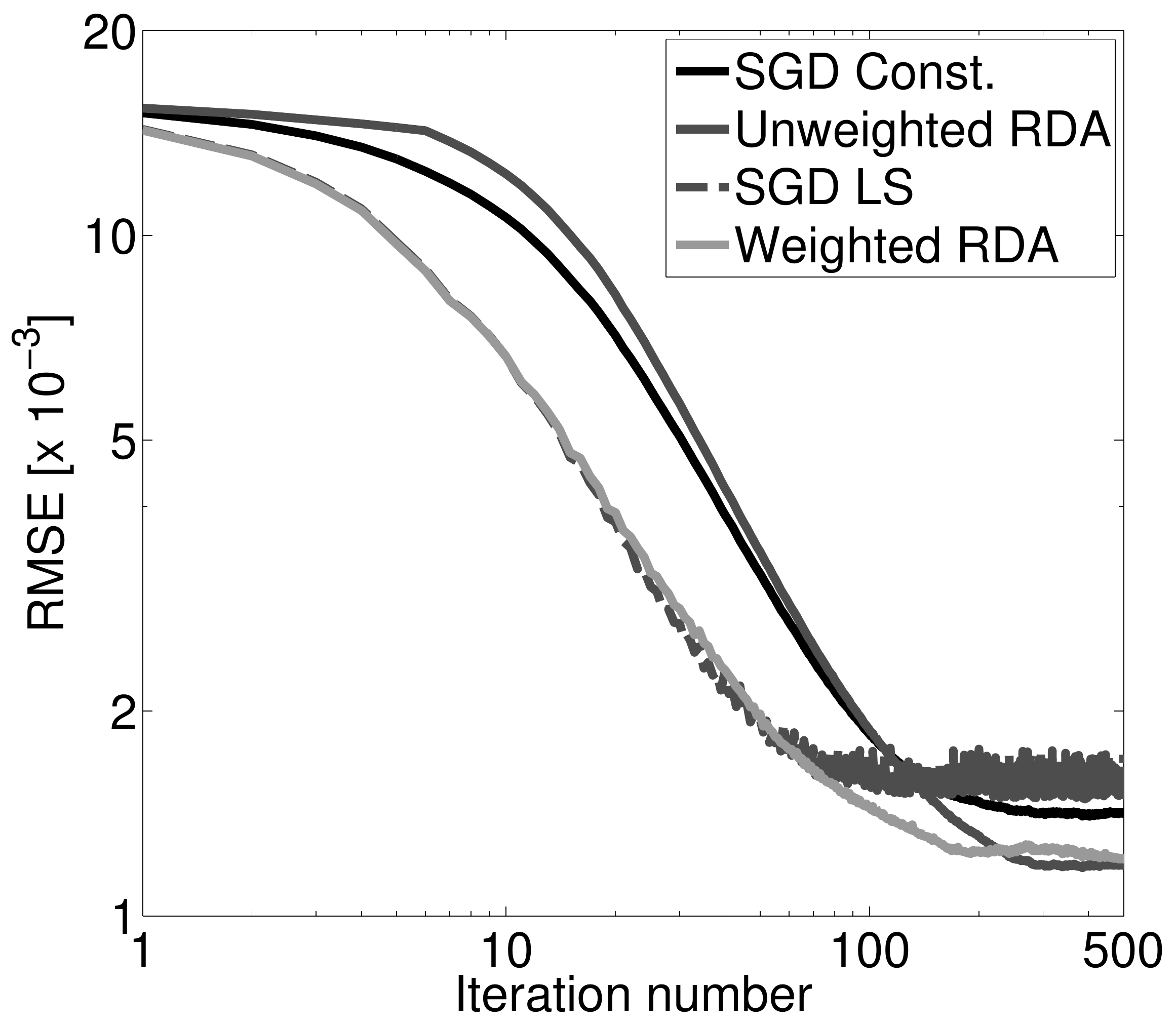}
					\caption{}
				\end{subfigure}%
				\begin{subfigure}[b]{0.25\textwidth}
					\includegraphics[width=\textwidth]{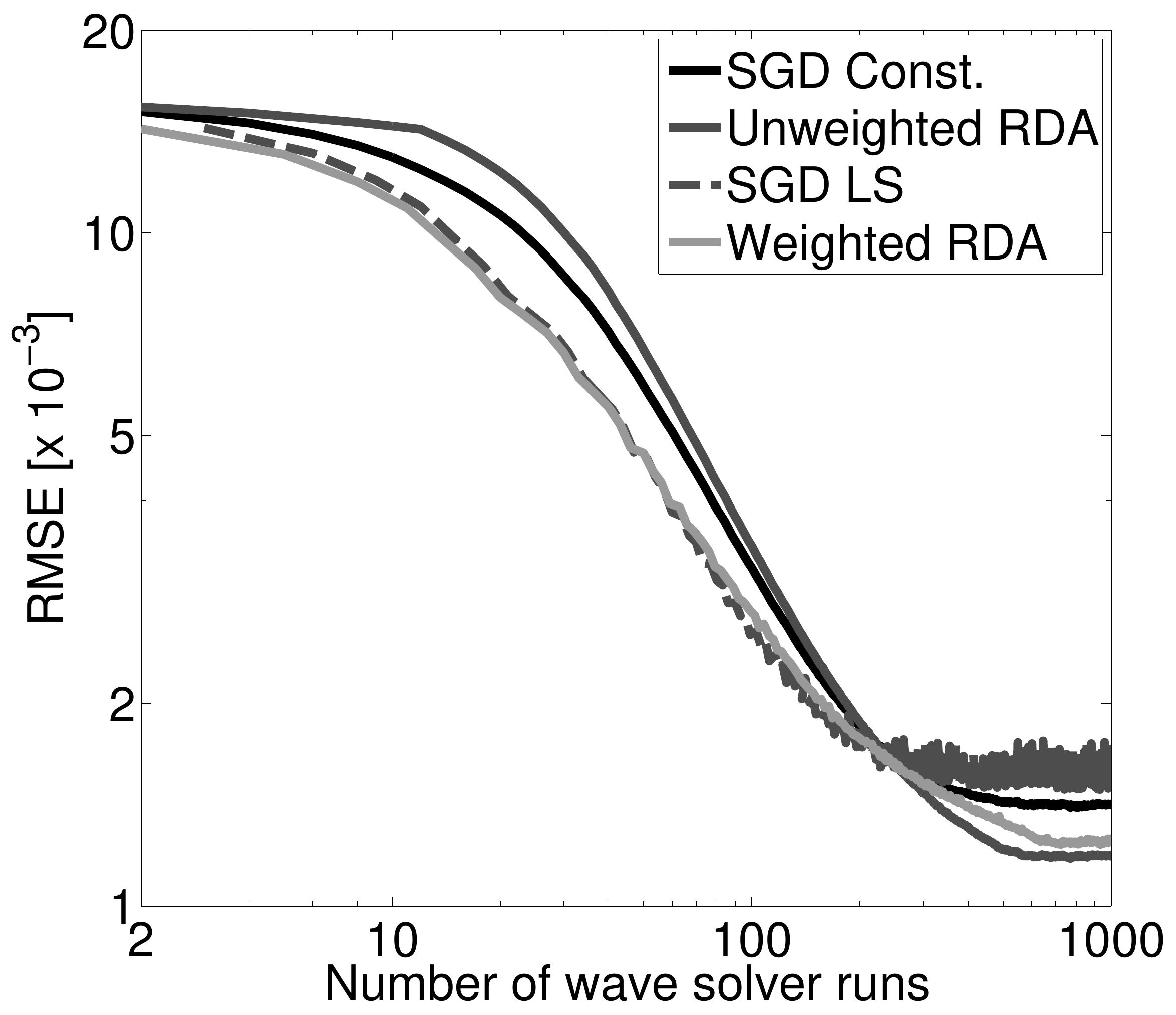}
					\caption{}
				\end{subfigure}
				\caption{Plot of RMSE versus (a) the number of iterations and (b) the number of wave solver runs for SGD with a line search, SGD with a constant step size of 0.1, unweighted RDA with a step size of 0.1, and weighted RDA.}
				\label{fig:rda_sgd_comp_conv}
			\end{figure}

			The plots of the convergence rates, shown in Fig.~\ref{fig:rda_sgd_comp_conv}, further confirm the benefits provided by the RDA method. SGD with a line search has a fast initial convergence, but results in a less accurate final image. From this plot, it is also clear that the estimates of the object provided by SGD with a line search also exhibit a high variance, even at later iterations. This is likely due to the fact that the line search only evaluates the cost function for a single realization of the encoding vector. As a result, the line search will tend to chose a larger step size that effectively minimizes the cost function evaluated for that encoding vector, but which increases the cost function when all, or a large number, of encoding vectors are considered. This behavior is not seen for the weighted RDA method. Since, for the RDA method, the search direction is given by a weighted average of the gradient estimates for all past encoding vector realizations, it does not overfit the cost function evaluated for a single realization of the encoding vector. This is true even though the weight at a given iteration is chosen only by evaluating the cost function for a single realization. Thus, the high variance of the object estimates is eliminated while the computational cost of selecting a weight for the RDA method is the same as performing the line search for SGD.
		
		\subsection{Bias-variance analysis} \label{sec:bias-variance}
		The investigations with a numerical breast phantom, described above, suggested that the RDA method could provide more effective regularization than SGD. However, care must be exhibited when evaluating this claim. Stronger regularization does not mean better image quality. It is not enough to compare two different reconstruction methods with the same regularization parameter value. While one may appear to produce a superior image, the other may produce a comparable image when another regularization value is employed. Thus, it is necessary to consider a range of regularization parameter values when comparing any two methods. Furthermore, image quality is most properly evaluated through task-based measures of image quality \cite{barrett_foundations_2004}. However, such studies are a substantial undertaking and are outside the scope of this paper. Instead, here, we use bias-variance curves as a proxy for this more complete assessment.
		
		Bias-variance curves depict the inherent trade-off between noise mitigation and close agreement with the measured data. As described above, an estimate of the sound speed is obtained by solving a minimization problem consisting of two terms, the data fidelity term and the regularization term. The relative weight of these terms is controlled by varying a scalar regularization parameter. Noise can be more severely suppressed by increasing the value of the regularization parameter, but this can result in reduced resolution or other forms of bias.
		
\begin{figure}[htbp!]
				\centering
				\includegraphics[width=0.4\textwidth]{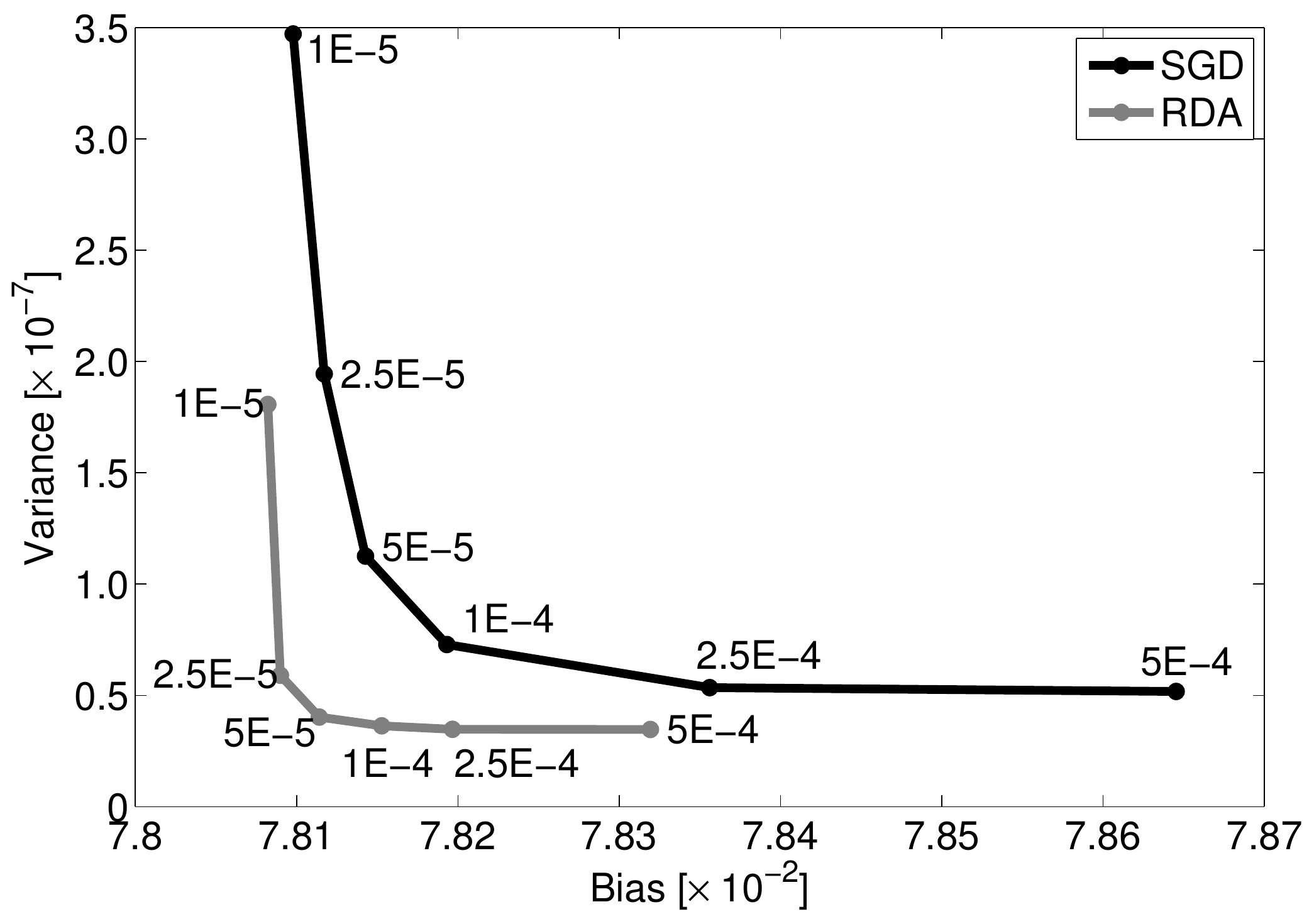}
				\caption{Bias-variance curve for SGD with constant step size and the unweighted RDA method. The corresponding regularization parameter values are given by each point.}
				\label{fig:bias_var_curve}
			\end{figure}

\begin{figure}[htbp!]
				\centering
				\begin{subfigure}[b]{0.25\textwidth}
					\includegraphics[width=\textwidth]{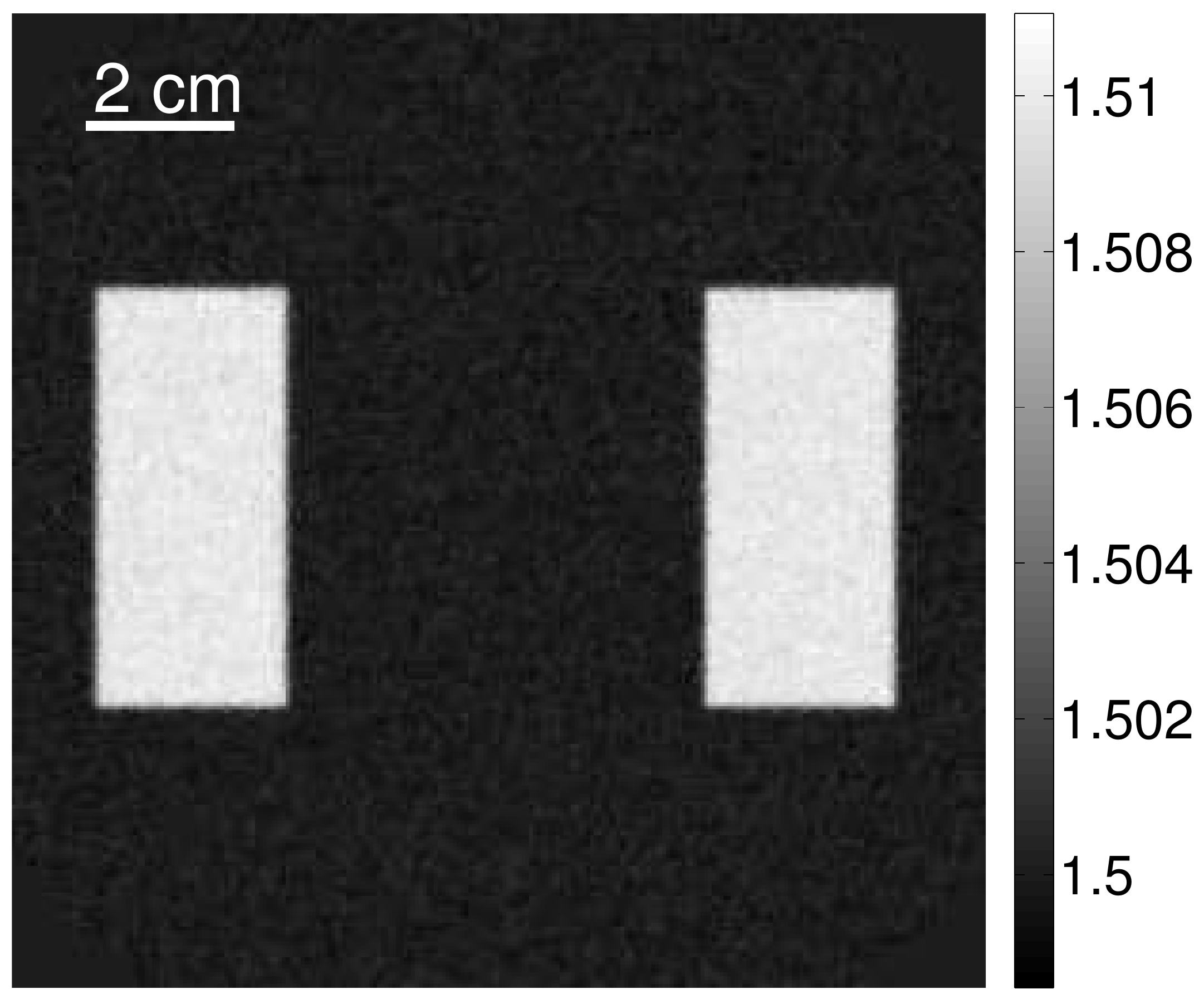}%
					\vspace*{0.5mm}%
					\caption{} 
				\end{subfigure}%
				\begin{subfigure}[b]{0.25\textwidth}
					\includegraphics[width=\textwidth]{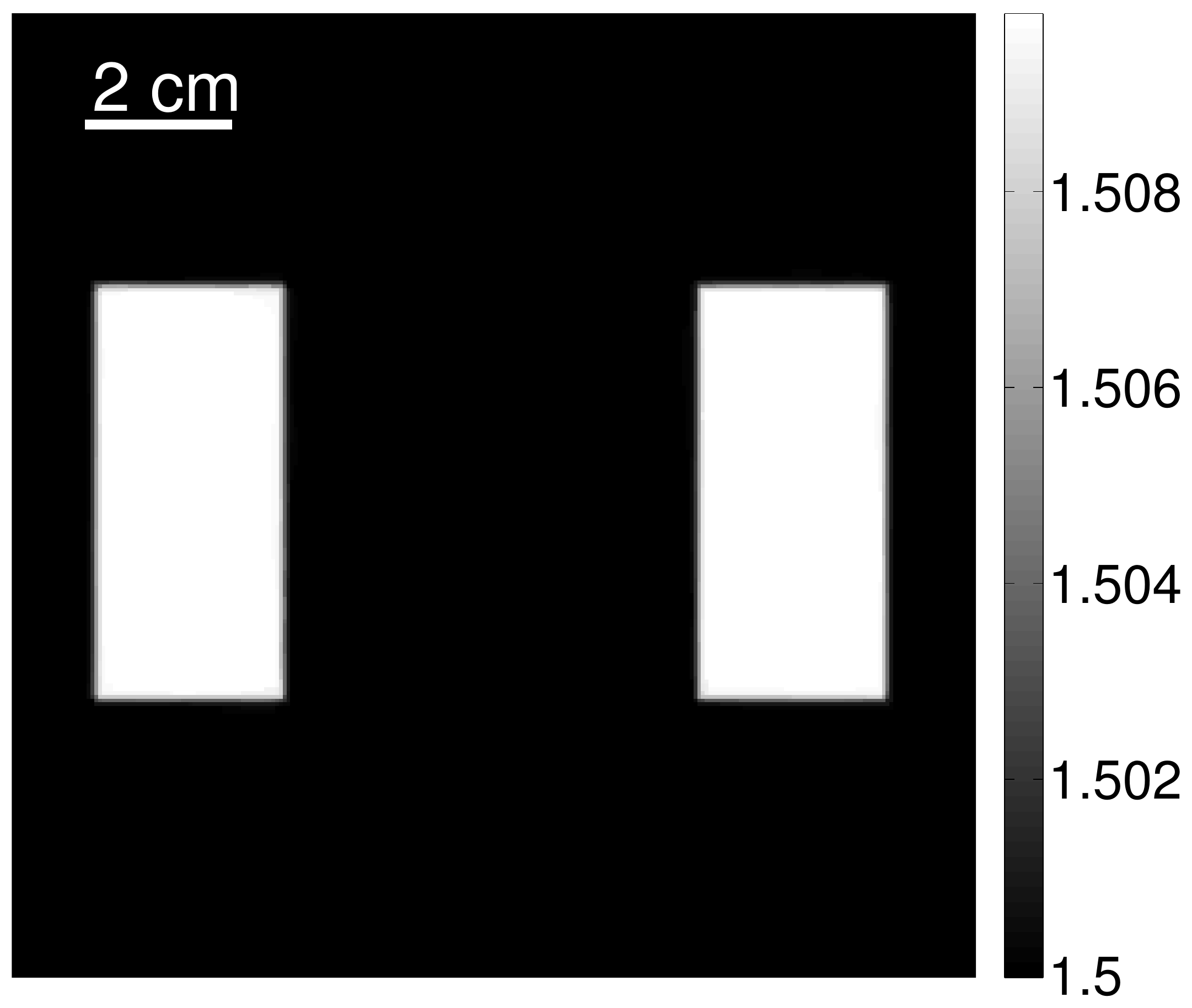}
					\caption{} 
				\end{subfigure}
				\caption{Example reconstructed images from bias-variance analysis. (a) Image reconstructed by SGD with a regularization parameter value of $5 \times 10^{-5}$. (b) Image reconstructed by RDA with a regularization parameter value of $1 \times 10^{-4}$. The two images have approximately the same bias. Both images are shown in their full dynamic ranges. The sound speed values are given in units of mm/$\mu$s.} \label{fig:bias_var_imgs}
			\end{figure}

		The bias-variance curves for SGD with a constant step size and the unweighted RDA method are shown in Fig.~\ref{fig:bias_var_curve}. The curves are generated by reconstructing a collection of images across a range of regularization parameter values. As seen in the figure, the RDA method consistently produces lower variance images (less noisy) for a given level of bias. This difference is seen in the reconstructed images. In Fig.~\ref{fig:bias_var_imgs}, reconstructed images corresponding to the same bias level are shown. The image reconstructed by use of SGD with a constant step size is noticeably noisier than the image obtained by use of the unweighted RDA method.
			
	\section{Experimental Validation}
		
		\subsection{Methods}
		Clinical data were acquired previously by use of the SoftVue USCT scanner \cite{duric_breast_2013}. The system consisted of a ring-shaped array with a radius of 110~mm, containing 2048 transducers. The transducers had a central frequency of 2.75~MHz with a pitch of 0.34~mm. Each element was elevationally focused to isolate a 3-mm-thick slice of the object. See \cite{duric_breast_2013,duric_clinical_2014} for additional information regarding the measurement system and clinical studies.
		
		%\textcolor{red}{Description of object - patient breast with tumor}
		
		Every other transducer element served as an emitter. The resulting pressure wave was then measured by the same set of 1024 transducers. The pressure was recorded with a sampling rate of 12~MHz for 2112 time points, corresponding to approximately 176~$\mu$s. This measurement process was repeated with and without the object. Forty-eight transducers were identified as bad channels following manual inspection. The data from these channels were discarded, resulting in measurements from 976 transducers. The pressure data were upsampled to a sampling rate of 20~MHz by use of linear interpolation in order to avoid the introduction of numerical errors by the numerical wave solver \cite{mast_k-space_2001}. The number of samples in each time trace was 3500. A Butterworth bandpass filter with cutoff frequencies of 0.5 and 1.0~MHz was applied to each signal.	The shape of the excitation pulse was estimated from the measured data without the object using the method described in \cite{wang_waveform_2015}.
		
\begin{figure}[htbp!]
			\centering
			\includegraphics[width=0.3\textwidth]{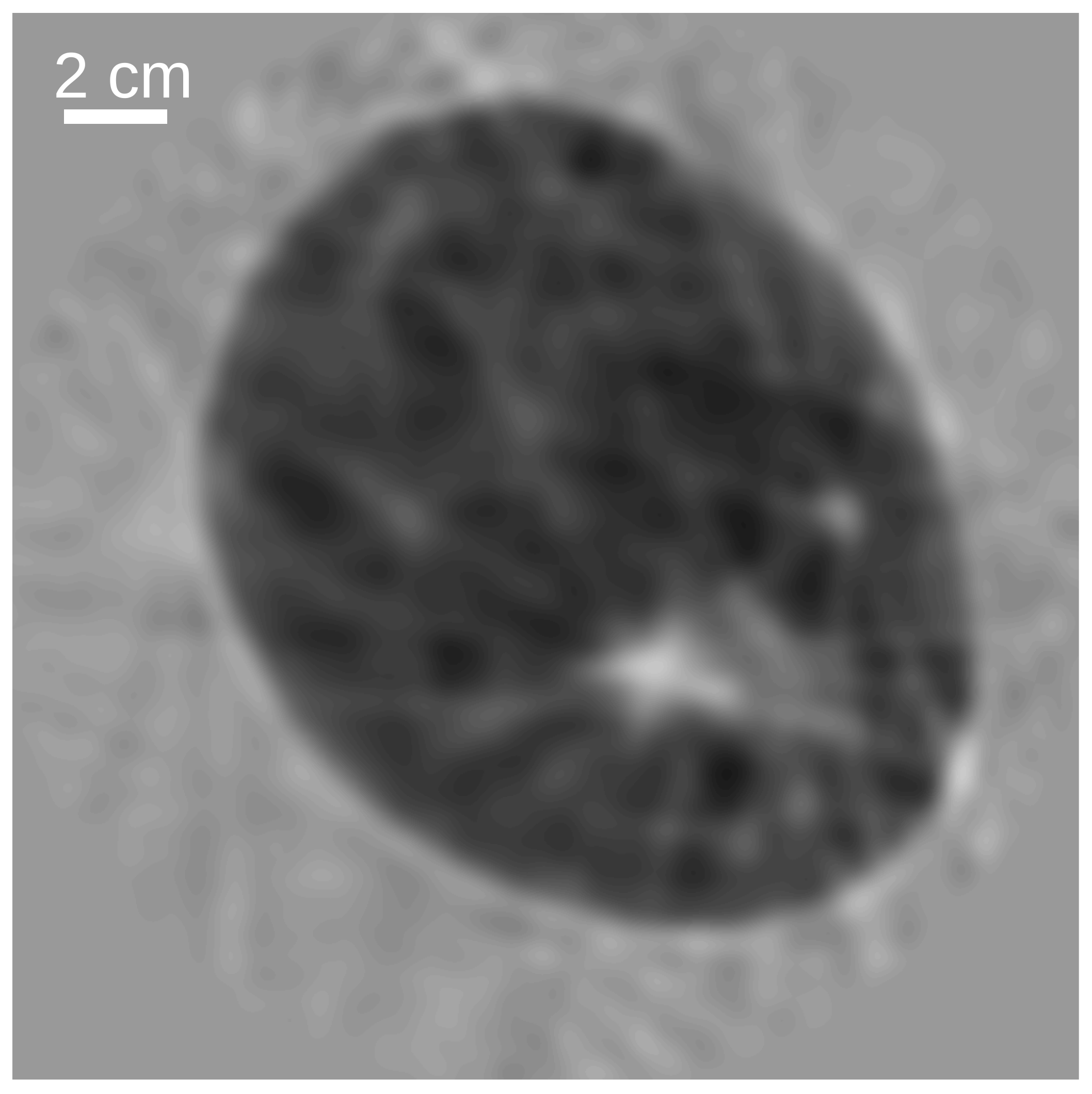}
			\caption{Initial estimate of the object reconstructed by use of an adjoint state method described in \cite{anis_investigation_2014}.} \label{fig:init_guess_clin}
		\end{figure}

		An initial estimate of the object was reconstructed by use of an adjoint state method (see Fig.~\ref{fig:init_guess_clin}) \cite{anis_investigation_2014}. This estimate was used to generate a set of synthetic data. As detailed in \cite{wang_waveform_2015}, measurements near the emitter may not contribute positively to the reconstructed image due to mechanical cross-talk, model mismatch, and measurement noise. The impact of these effects can be mitigated by replacing the measurements near the emitter with synthetic data. Unlike \cite{wang_waveform_2015}, here, we substitute pressure data corresponding to an estimate of the object, provided by an adjoint state method \cite{anis_investigation_2014}, rather than a homogeneous medium. The 512 measurements from transducers opposite the emitter were kept. The others were replaced with the synthetic data. %\hl{The differences in the reconstructed images between the two data filling strategies appear small in this case, but may be more significant when the object induces strong reflections.}
		
		The images were reconstructed by solving Eqn.~(\ref{eqn:stoc_cost_fnc}), where the operator $\mathbf{H}\left(\mathbf{c}\right)$ was calculated by use of the second-order k-space pseudo-spectral wave equation solver as described in Section~\ref{sec:pressure_data} \cite{mast_k-space_2001}. The calculation domain was $512 \times 512~\text{mm}^2$, divided into a $2560 \times 2560$ Cartesian grid with a spacing of $0.2~\text{mm}$. The sound speed was updated within a circle of radius 105~mm. Reconstruction was performed on a platform consisting of dual quad-core CPUs, 128~GB of RAM, and a NVIDIA Tesla K40 GPU. These reconstruction parameters are summarized in Table~\ref{table:recon_pars}.
		
		While image quality is most objectively assessed using task-based methods of image quality \cite{barrett_foundations_2004}, here, for reasons of expediency, the contrast-to-noise (CNR) ratio was employed as a proxy for the detectability of the tumor. The CNR of the reconstructed images was calculated by identifying three regions. The tumor was segmented manually. Regions of similar size corresponding to the parenchymal tissue and the water bath were also identified. The contrast was calculated based on the tumor and parenchymal tissue regions. The noise, however, was calculated based on the water bath to avoid mis-attributing any real variations within the parenchymal tissue to noise. The CNR was calculated as
		%\begin{empheq}[box=\colorbox{yellow}]{align}
		\begin{align}
			CNR = \frac{\bar{c}_{t} - \bar{c}_{p}}{\sigma_n} ,
		\end{align}
		%\end{empheq}
		where $\bar{c}_{t}$ is the average sound speed of the tumor, $\bar{c}_{p}$ is the average sound speed over a comparably sized region of the parenchymal tissue, and $\sigma_n$ is the standard deviation over a comparably sized region of the water bath.
			
		\subsection{Clinical results}
\begin{figure*}[htbp!]
			\centering
			\vspace*{-2mm}
			\begin{subfigure}[b]{0.25\textwidth}
				\includegraphics[width=\textwidth]{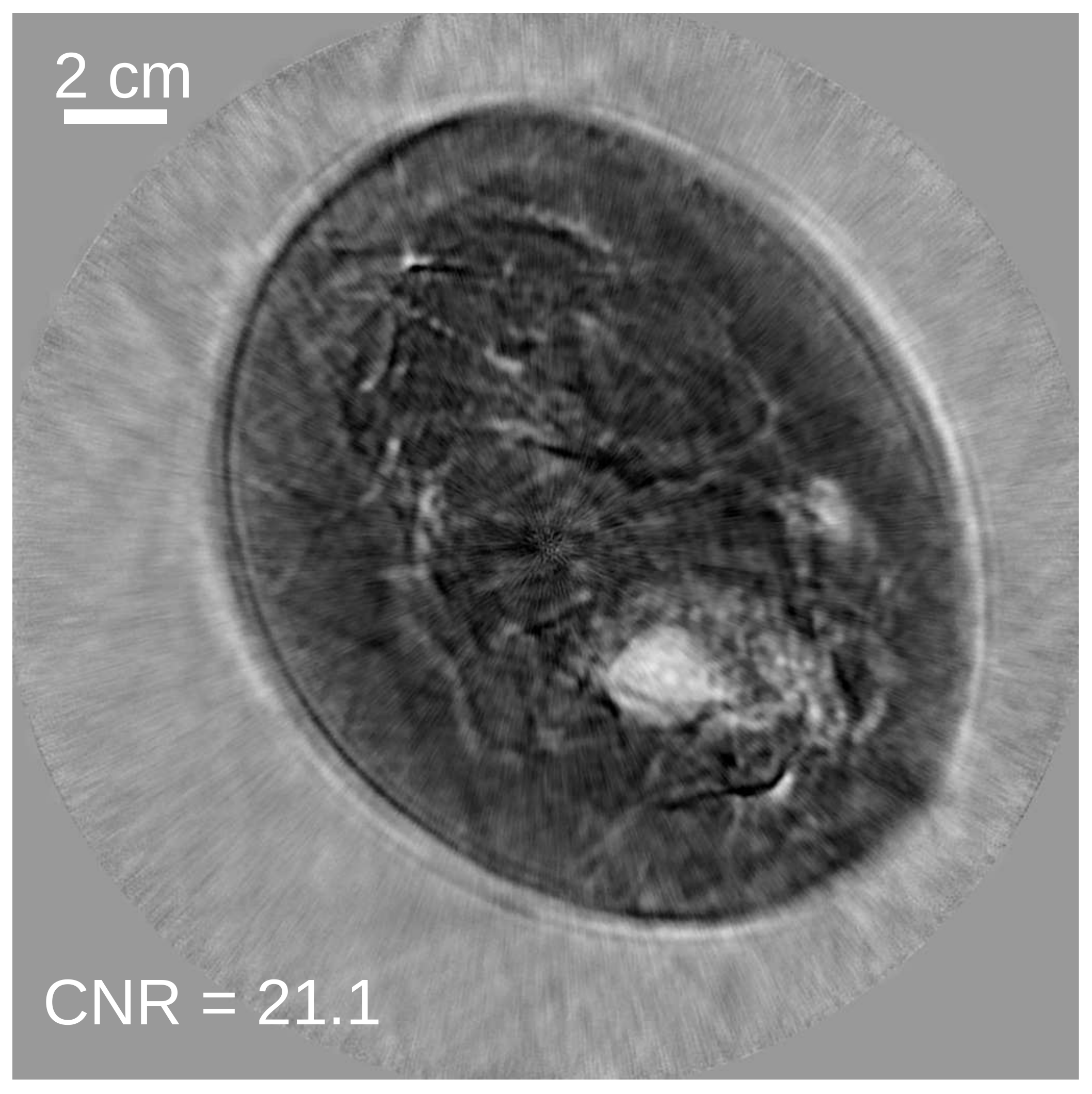}
				\caption{}
			\end{subfigure}%
			\begin{subfigure}[b]{0.25\textwidth}
				\includegraphics[width=\textwidth]{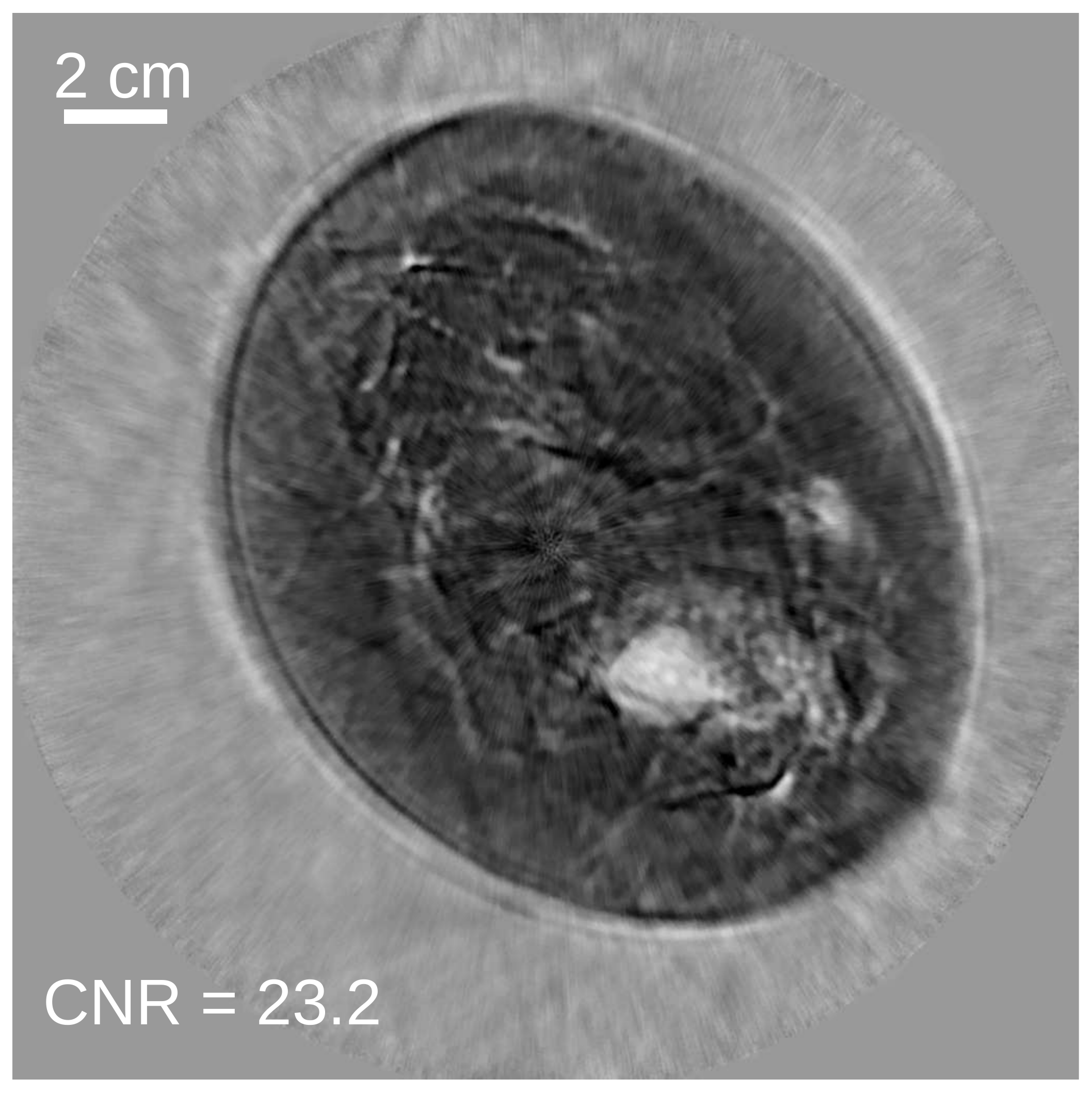}
				\caption{}
			\end{subfigure}%
			\begin{subfigure}[b]{0.25\textwidth}
				\includegraphics[width=\textwidth]{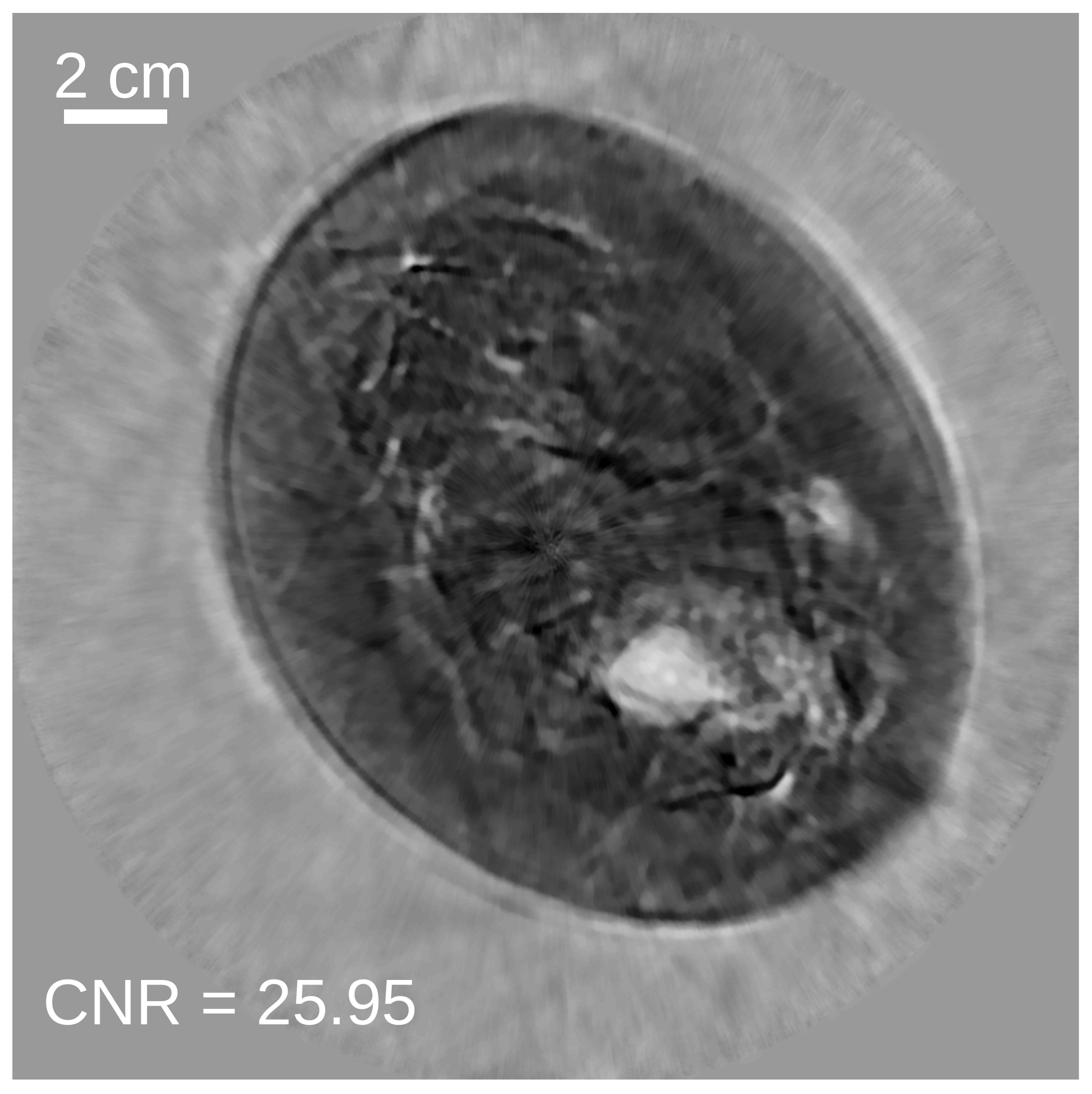}
				\caption{}
			\end{subfigure}%
			\begin{subfigure}[b]{0.25\textwidth}
				\includegraphics[width=\textwidth]{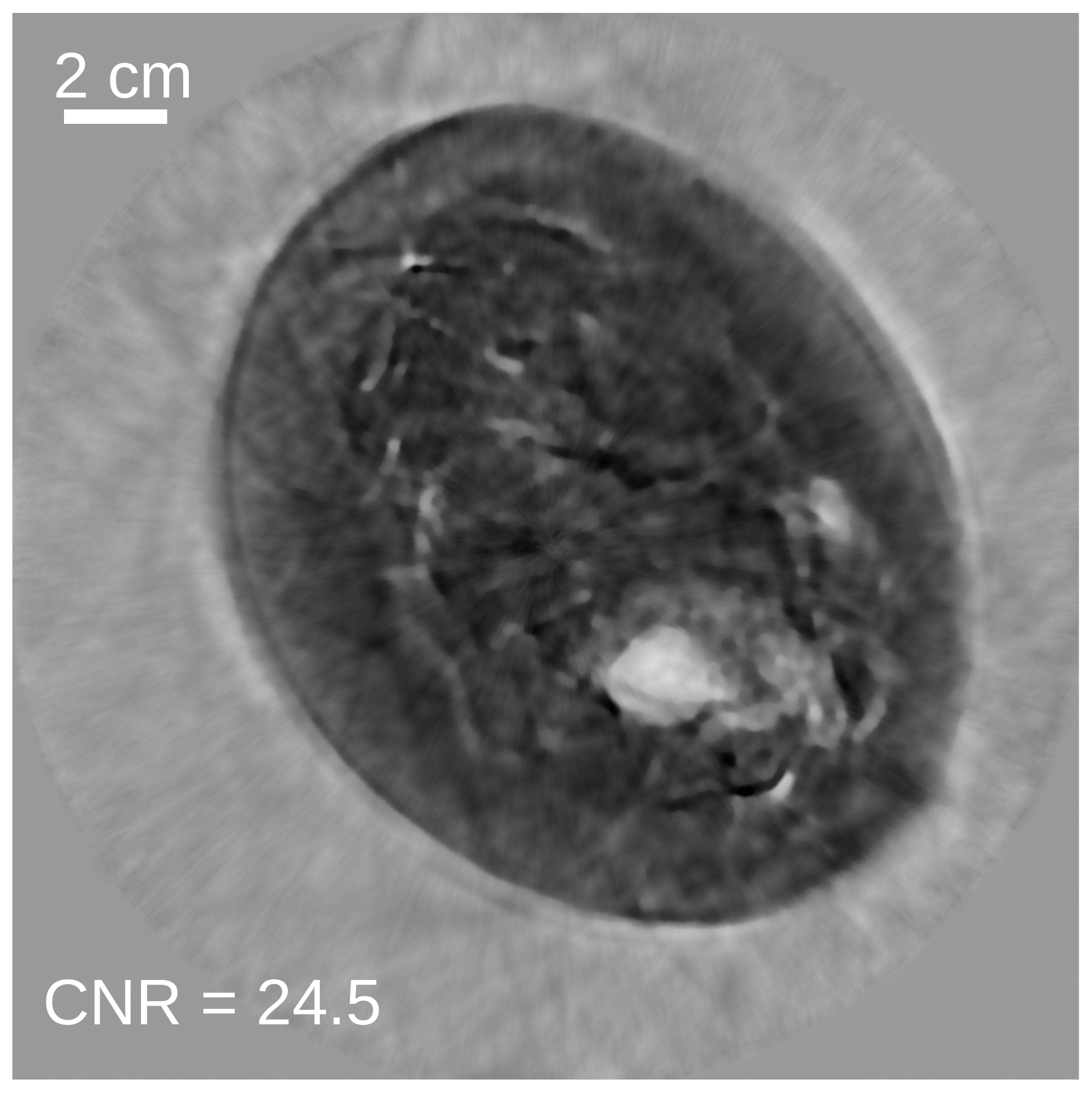}
				\caption{}
			\end{subfigure} \\
			\begin{subfigure}[b]{0.25\textwidth}
				\includegraphics[width=\textwidth]{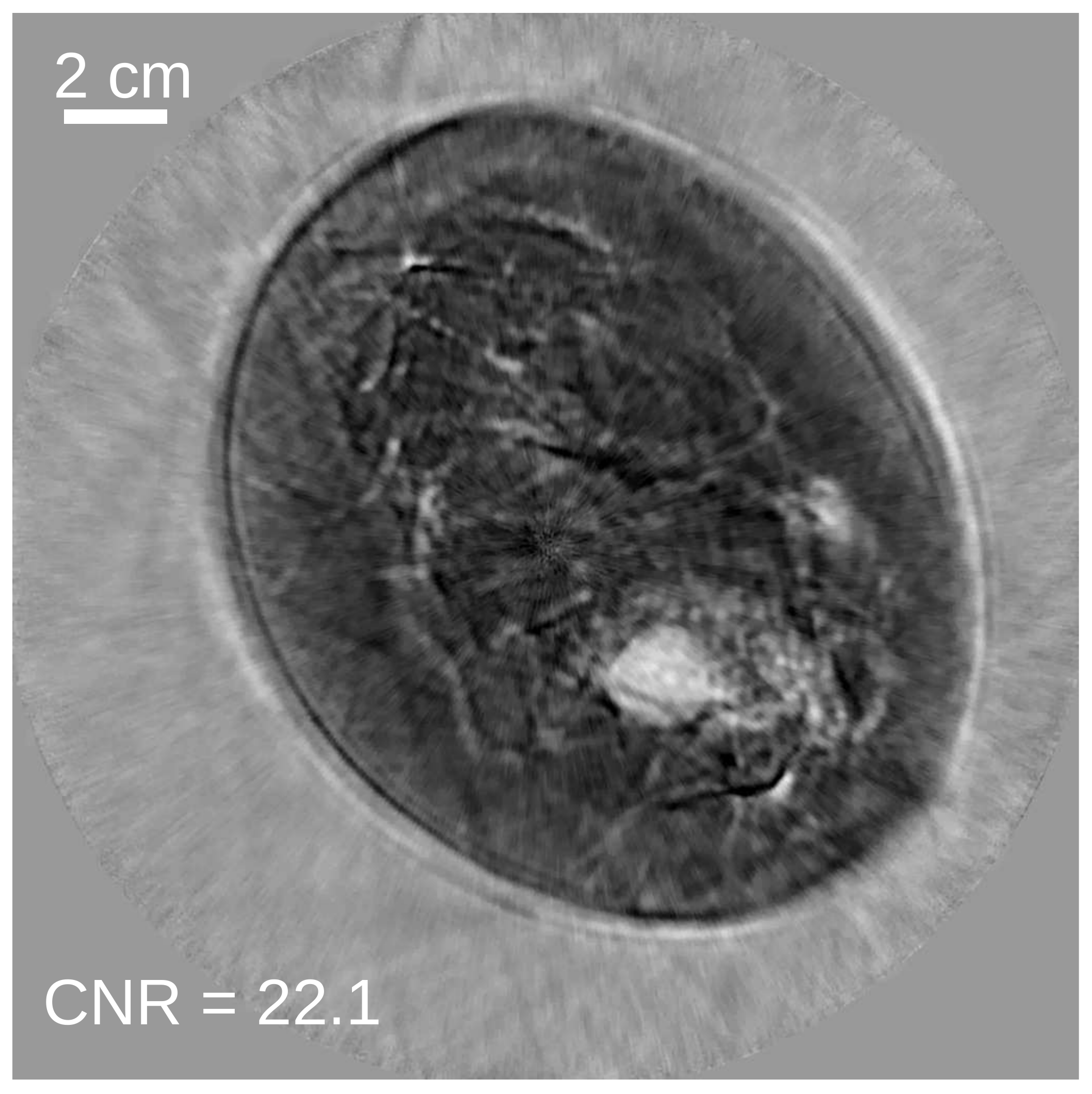}
				\caption{}
			\end{subfigure}%
			\begin{subfigure}[b]{0.25\textwidth}
				\includegraphics[width=\textwidth]{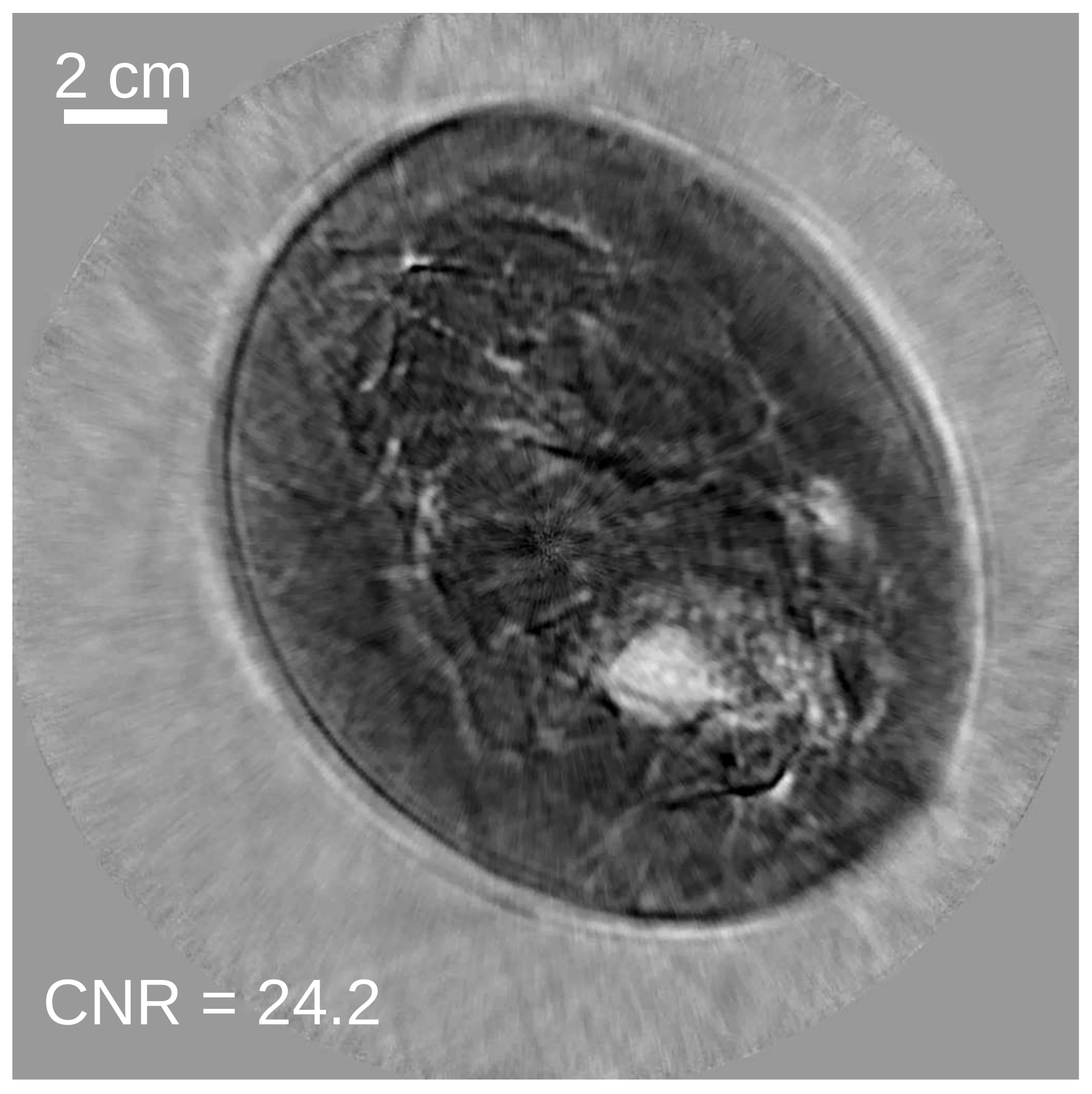}
				\caption{}
			\end{subfigure}%
			\begin{subfigure}[b]{0.25\textwidth}
				\includegraphics[width=\textwidth]{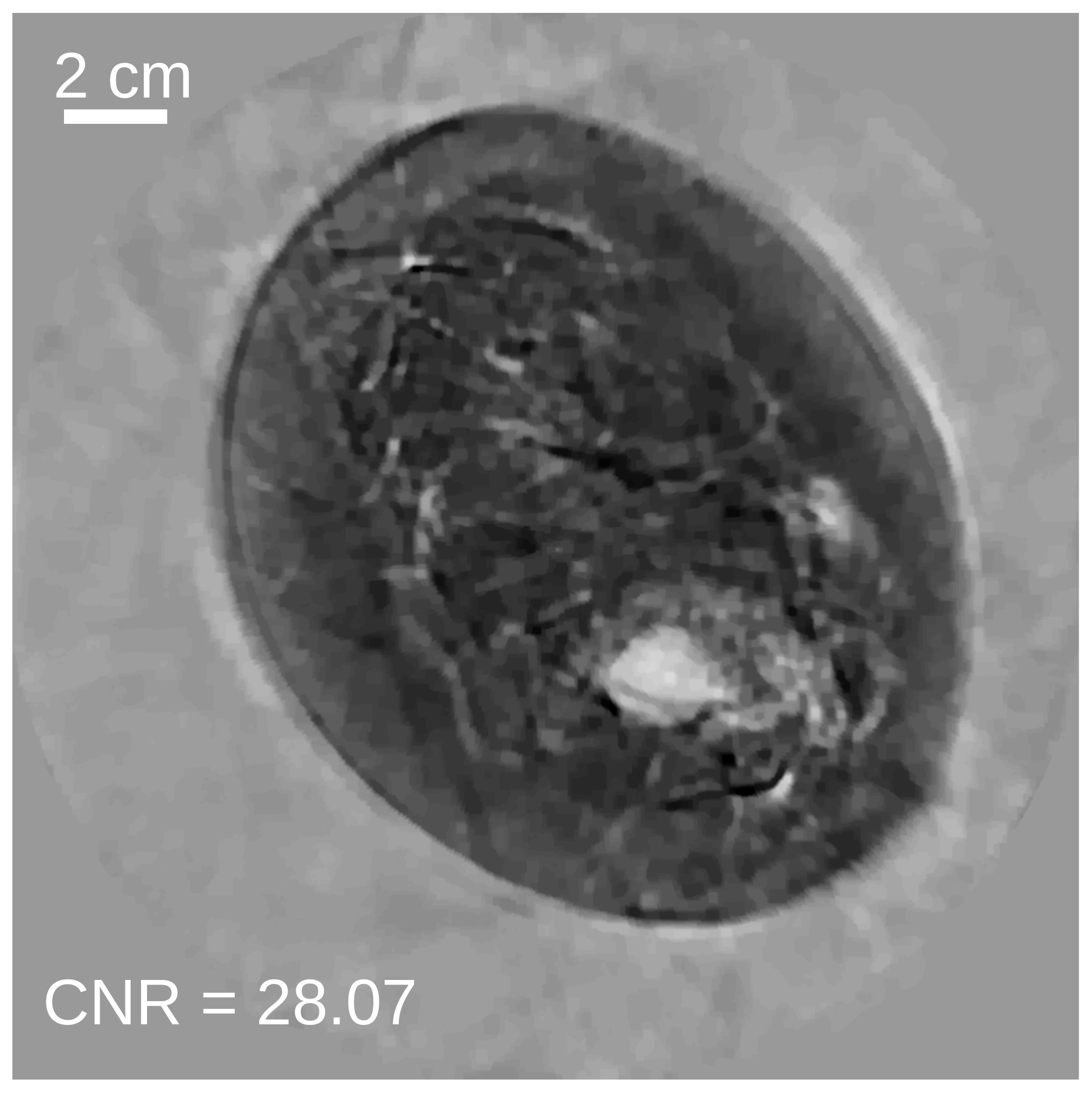}
				\caption{}
			\end{subfigure}%
			\begin{subfigure}[b]{0.25\textwidth}
				\includegraphics[width=\textwidth]{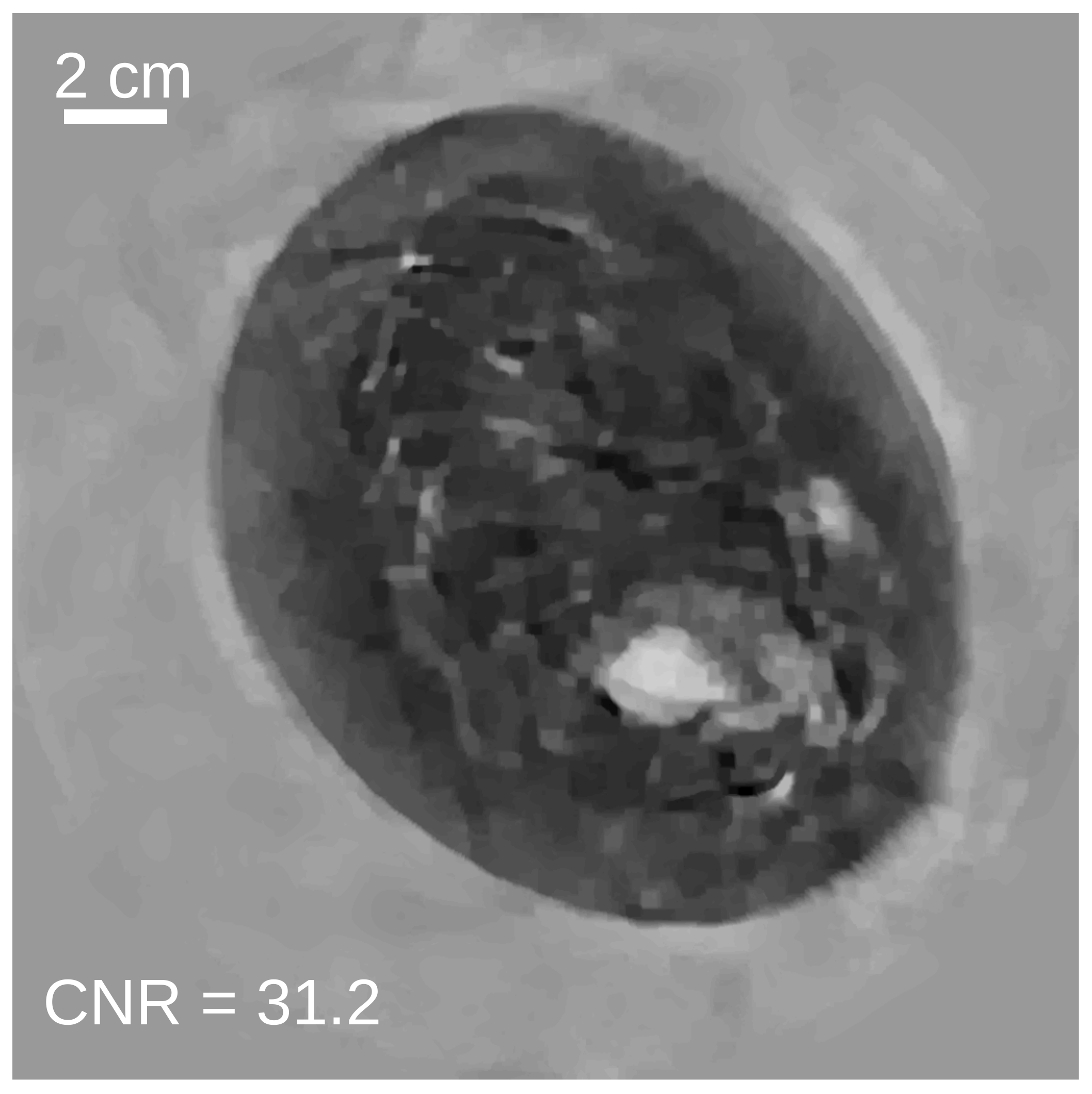}
				\caption{}
			\end{subfigure}%
			\caption{(Top row) Images reconstructed by use of SGD with a constant step size of $2.5 \times 10^5$ and regularization parameter values of (a)~$1 \times 10^{-10}$, (b)~$3 \times 10^{-10}$, (c)~$1 \times 10^{-9}$, and (d)~$3 \times 10^{-9}$. (Bottom row) Images reconstructed by use of the weighted RDA method with regularization parameter values of (e)~$1 \times 10^{-10}$, (f)~$3 \times 10^{-10}$, (g)~$1 \times 10^{-9}$, and (h)~$3 \times 10^{-9}$. Images are shown after 100 iterations and in a grayscale window of [1.38, 1.60] mm/$\mu$s. }
			\label{fig:clin_reg_param}
		\end{figure*}

		As seen in Fig.~\ref{fig:clin_reg_param}, the weighted RDA method consistently produces reconstructed images with higher CNRs than SGD with a constant step size, as indicated by the CNR values that label each image. This is shown across a range of regularization parameter values. Further, the maximum CNR obtained by SGD is lower even when the regularization parameter value is optimized. This improvement in the CNR is likely due to the favorable noise mitigation properties of the RDA method observed in the computer-simulation studies. While not shown, the CNRs for both methods do not continue to increase beyond 100 iterations. The CNR can serve as a proxy of detectability in cases where task-based measures of image quality cannot be performed \cite{barrett_foundations_2004}. While the CNRs of all the images shown in Fig.~\ref{fig:clin_reg_param} are quite high, the improvement in CNR could be more impactful for small or low-contrast tumors.
		
\begin{figure*}[htbp!]
				\centering
				\vspace*{-2mm}
				\begin{subfigure}[b]{0.25\textwidth}
					\includegraphics[width=\textwidth]{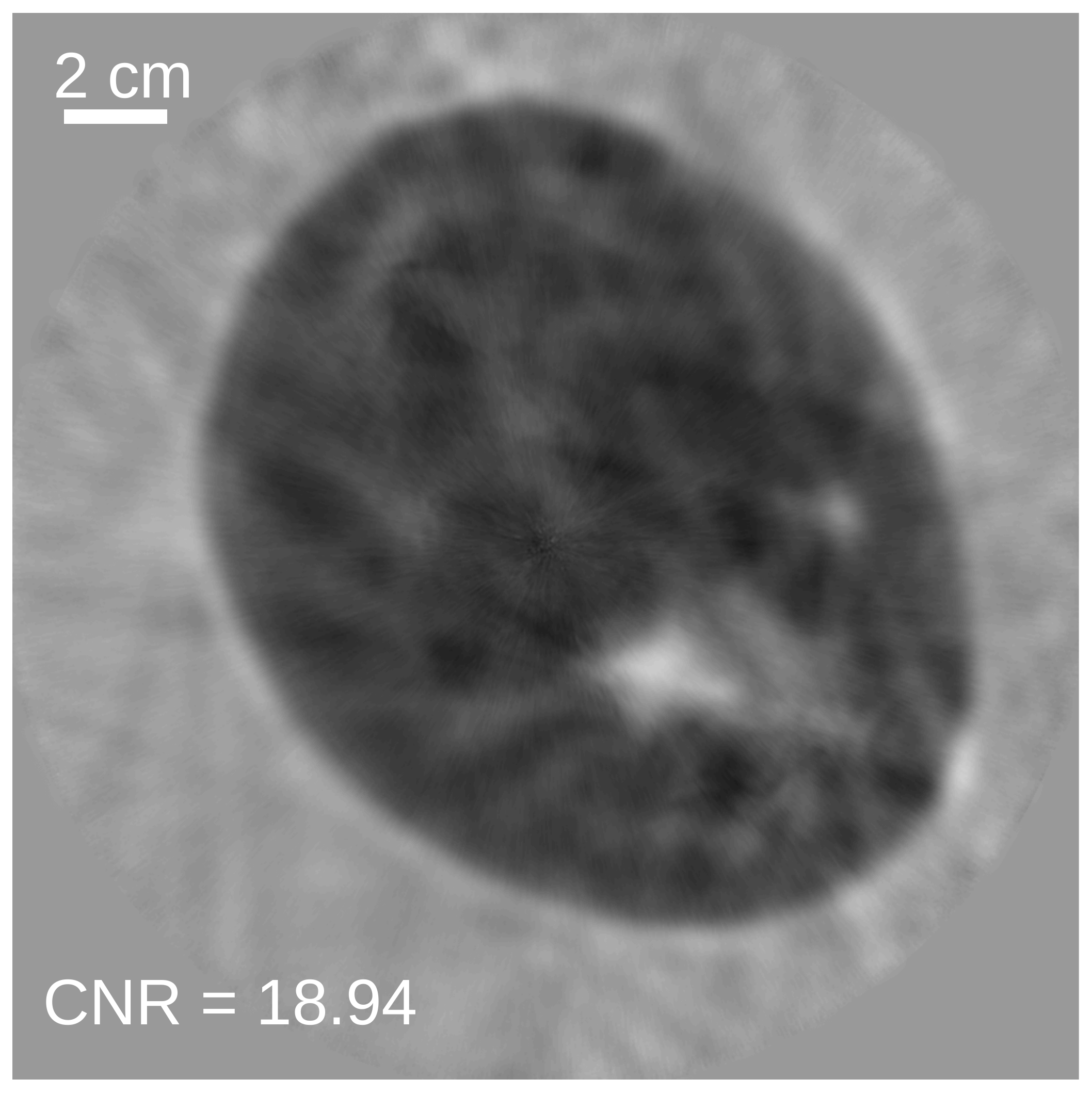}
					\caption{}
				\end{subfigure}%
				\begin{subfigure}[b]{0.25\textwidth}
					\includegraphics[width=\textwidth]{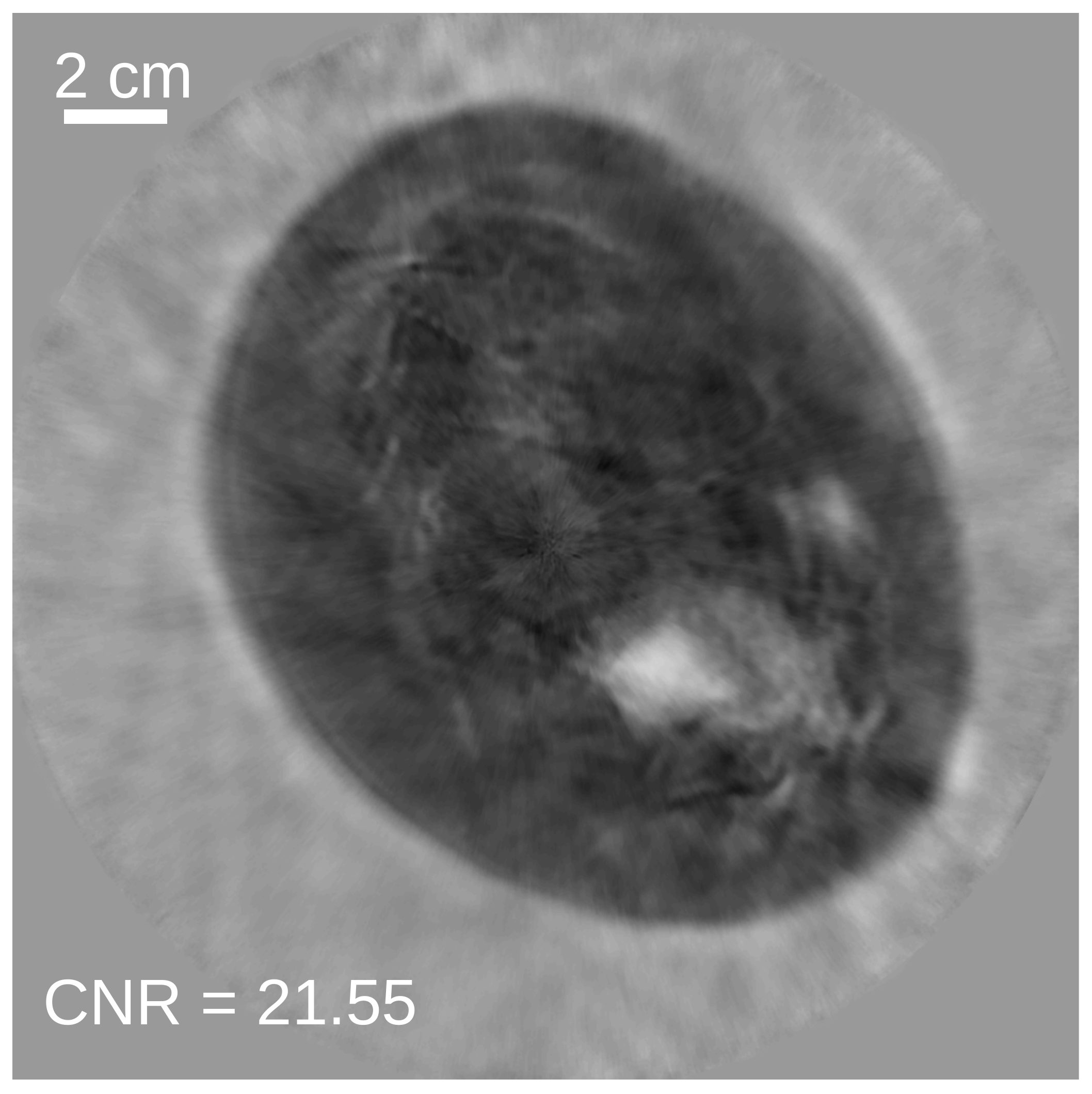}
					\caption{}
				\end{subfigure}%
				\begin{subfigure}[b]{0.25\textwidth}
					\includegraphics[width=\textwidth]{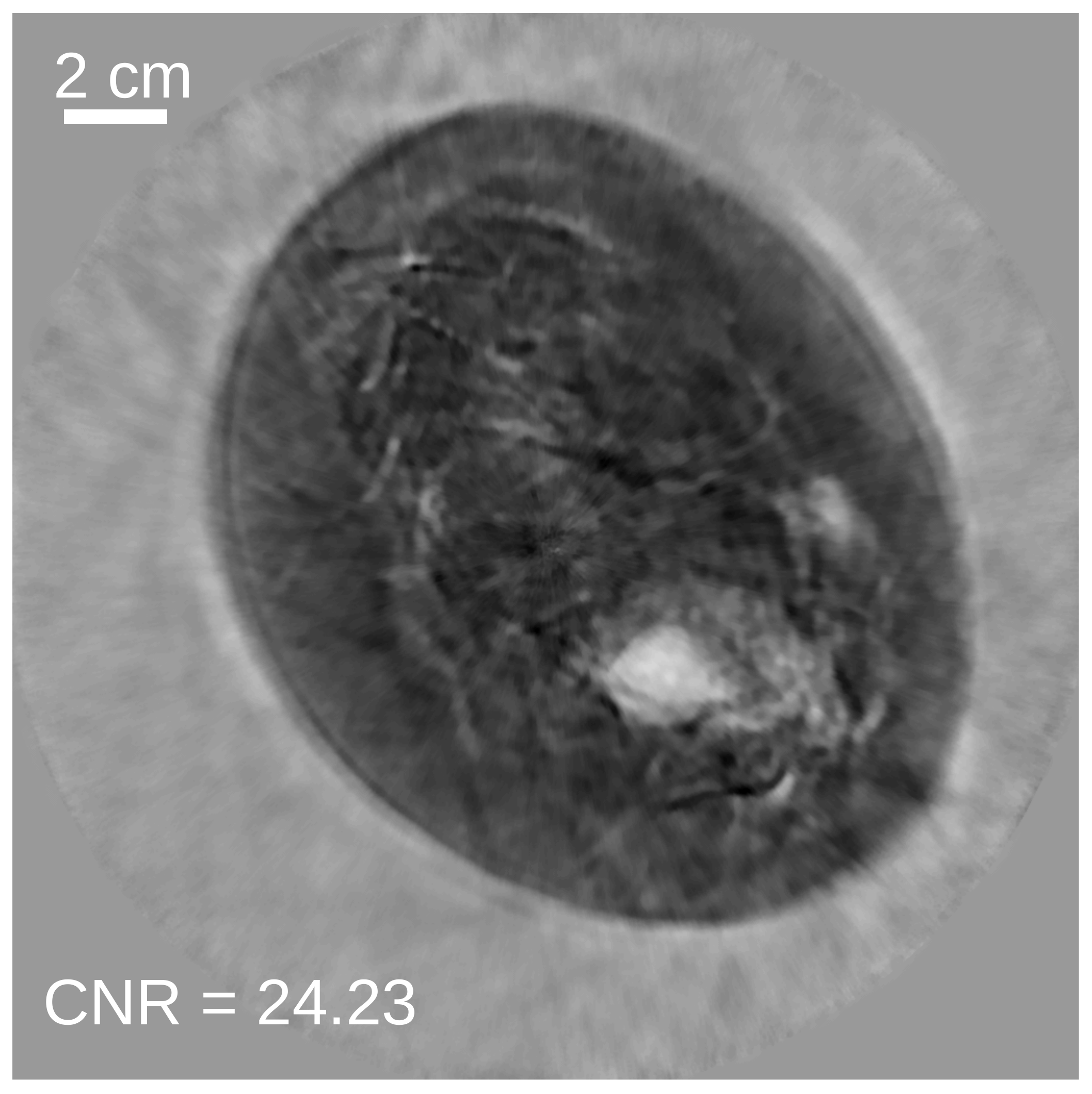}
					\caption{}
				\end{subfigure}%
				\begin{subfigure}[b]{0.25\textwidth}
					\includegraphics[width=\textwidth]{recon99_clinsgdtv1e_9const}
					\caption{}
				\end{subfigure} \\
				\begin{subfigure}[b]{0.25\textwidth}
					\includegraphics[width=\textwidth]{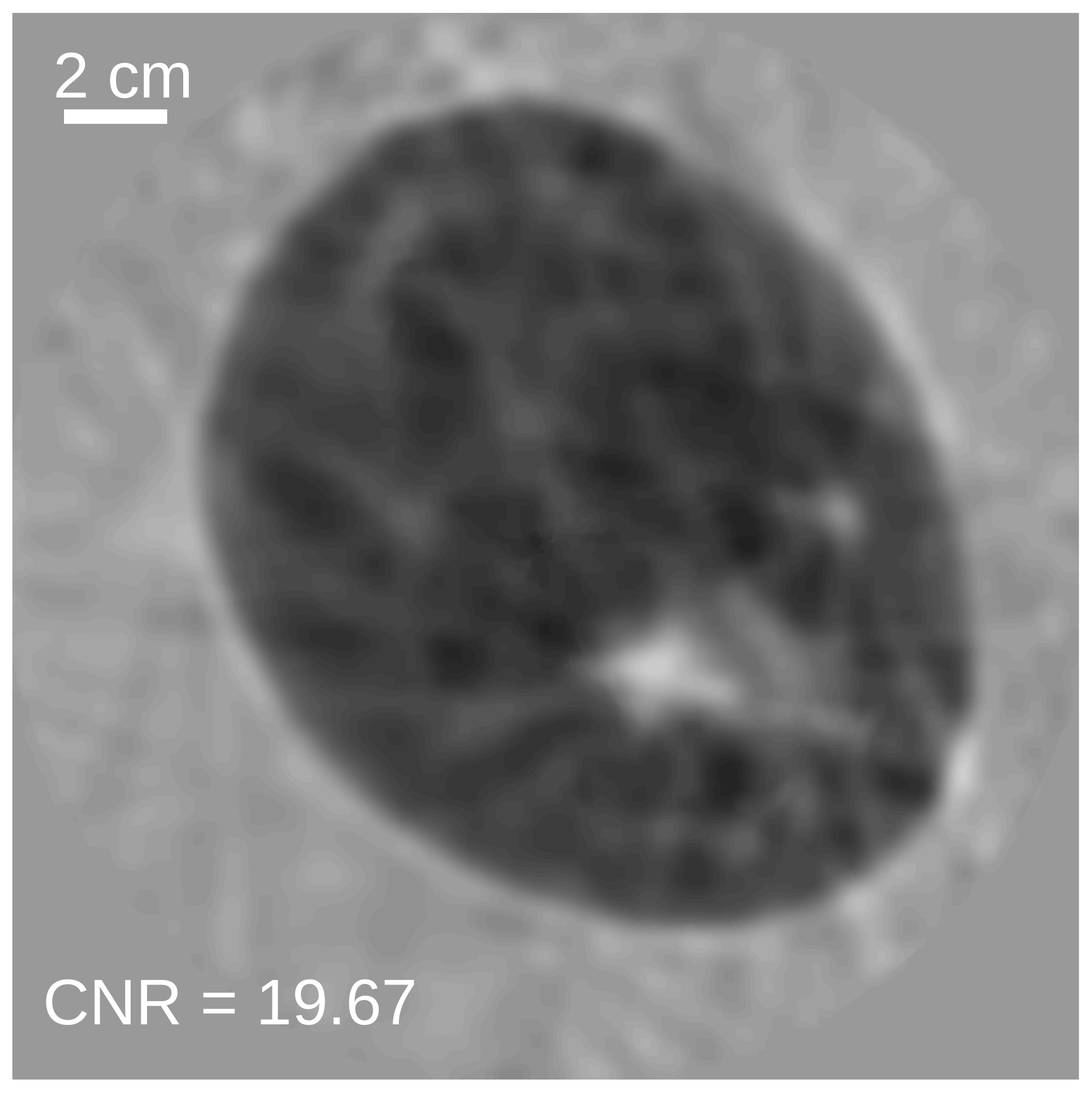}
					\caption{}
				\end{subfigure}%
				\begin{subfigure}[b]{0.25\textwidth}
					\includegraphics[width=\textwidth]{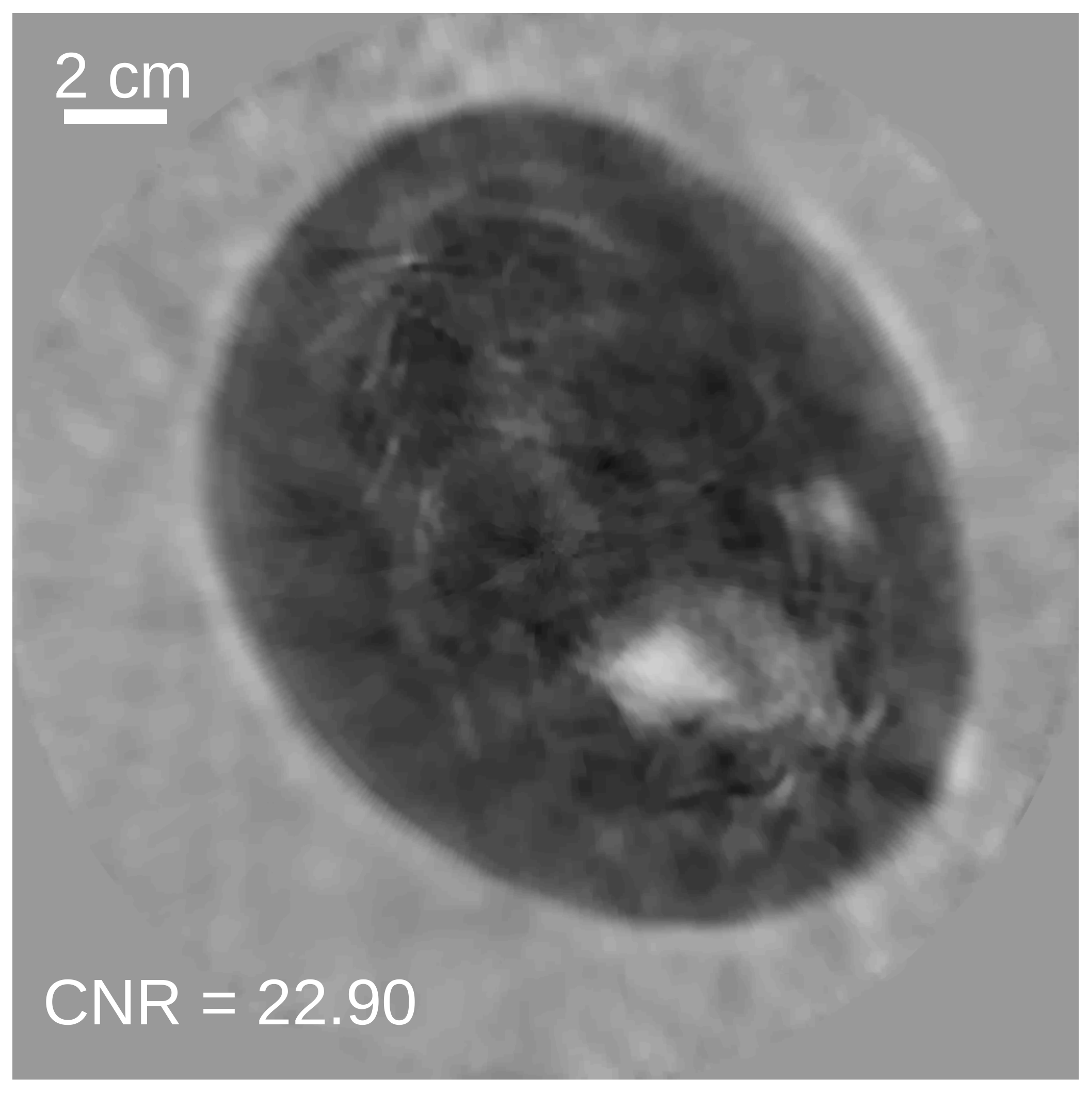}
					\caption{}
				\end{subfigure}%
				\begin{subfigure}[b]{0.25\textwidth}
					\includegraphics[width=\textwidth]{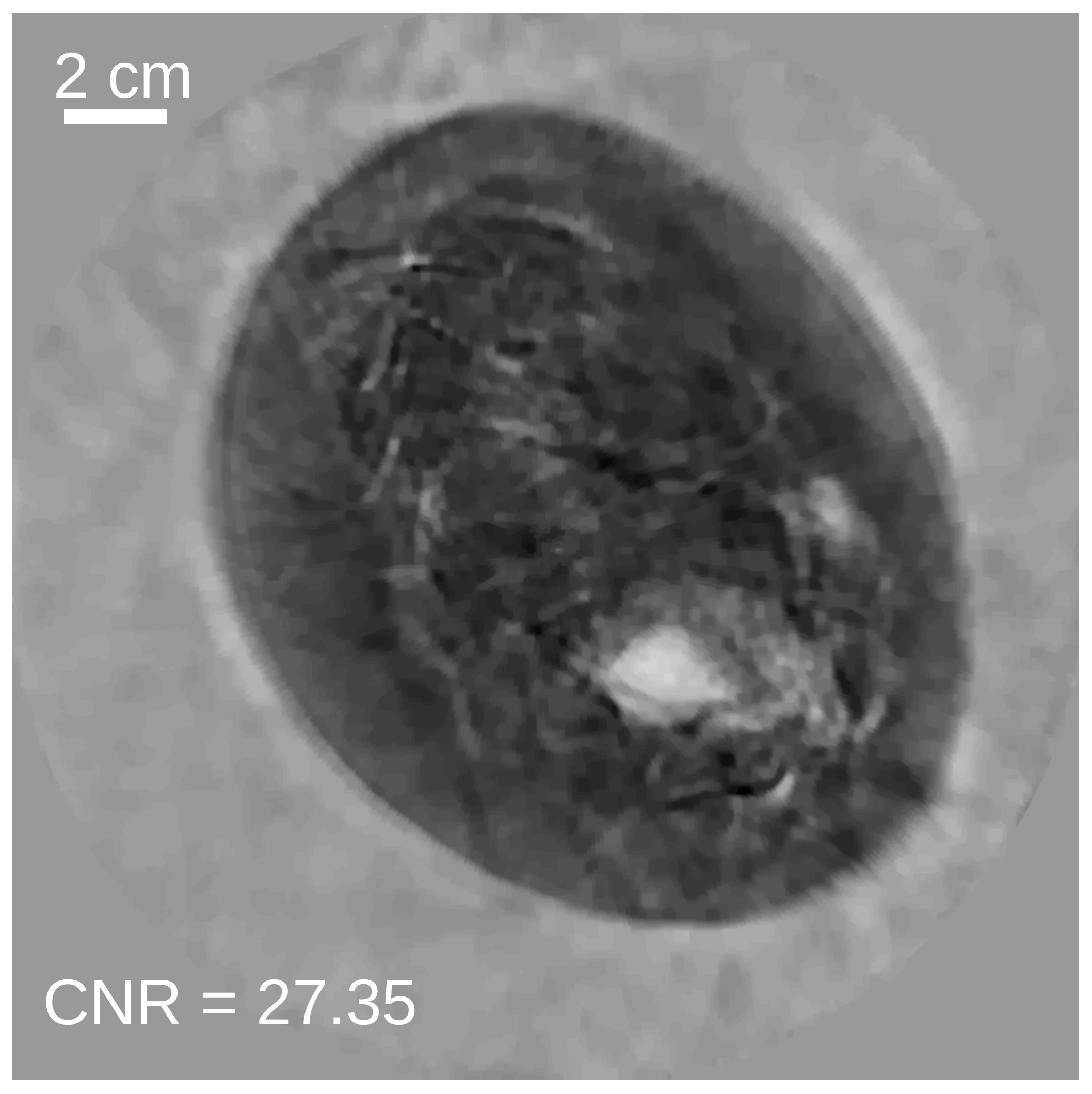}
					\caption{}
				\end{subfigure}%
				\begin{subfigure}[b]{0.25\textwidth}
					\includegraphics[width=\textwidth]{recon99_clinwrdatv1e_9}
					\caption{}
				\end{subfigure}%
				\caption{(Top row) Images reconstructed by use of SGD with a constant step size of $2.5 \times 10^{5}$ after (a) 5, (b) 20, (c) 50, and (d) 100 iterations with a regularization parameter value of $1 \times 10^{-9}$. (Bottom row) Images reconstructed by use of weighted RDA after (e) 5, (f) 20, (g) 50, and (h) 100 iterations with a regularization parameter value of $1 \times 10^{-9}$. All images are shown in a grayscale window of [1.38, 1.60] mm/$\mu$s.}
				\label{fig:clin_iter_num}
			\end{figure*}

		The reconstructed images as a function of iteration number are shown in Fig.~\ref{fig:clin_iter_num}. Since a non-constant initial guess was provided, the differences in the convergence rates of SGD with a constant step size and the weighted RDA method are less pronounced. However, a good initial guess is needed to avoid local minima since the data fidelity term is non-convex. Still, the weighted RDA method produces a higher CNR at each iteration. The difference between the CNRs of the two methods continues to grow over the first 50 iterations. This gap is eventually decreased at later iterations. This suggests that the weighted RDA method is able to provide some initial improvement in the convergence rate. This is consistent with the computer-simulation studies.			
			
\begin{figure*}[htbp!]
				\begin{subfigure}[b]{0.25\textwidth}
					\includegraphics[width=\textwidth]{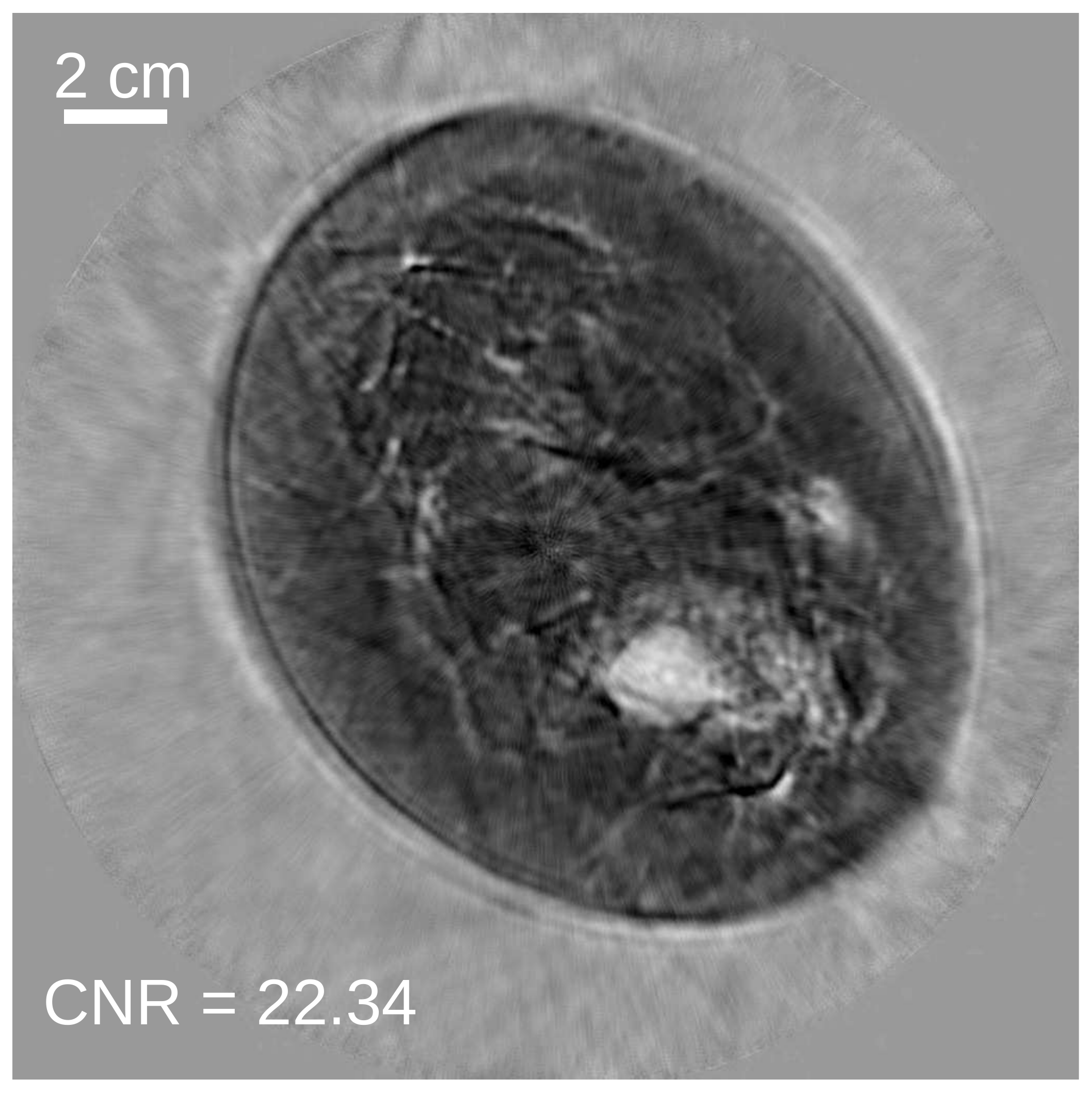}
					\caption{}
				\end{subfigure}%
				\begin{subfigure}[b]{0.25\textwidth}
					\includegraphics[width=\textwidth]{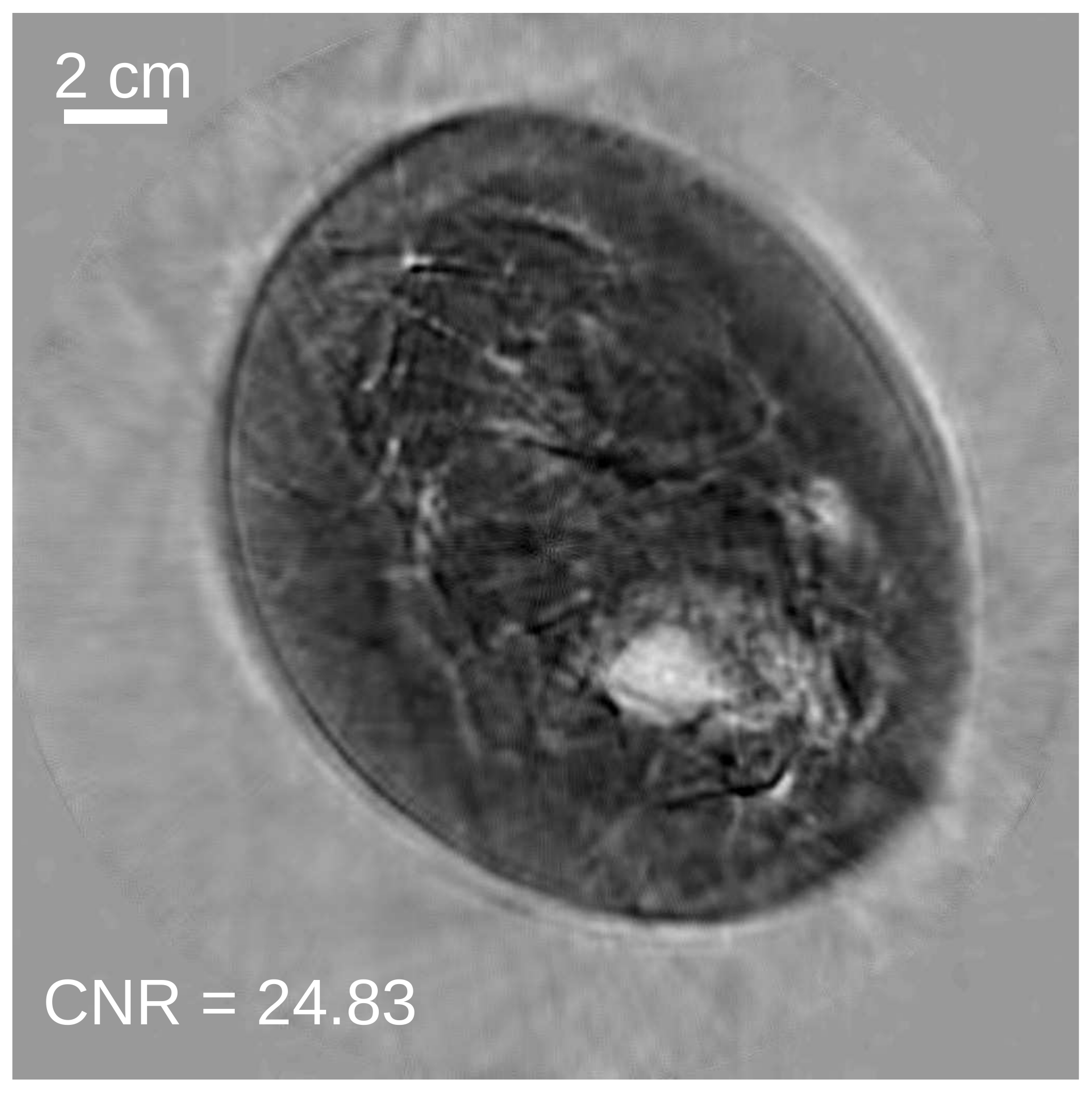}
					\caption{}
				\end{subfigure}%
				\begin{subfigure}[b]{0.25\textwidth}
					\includegraphics[width=\textwidth]{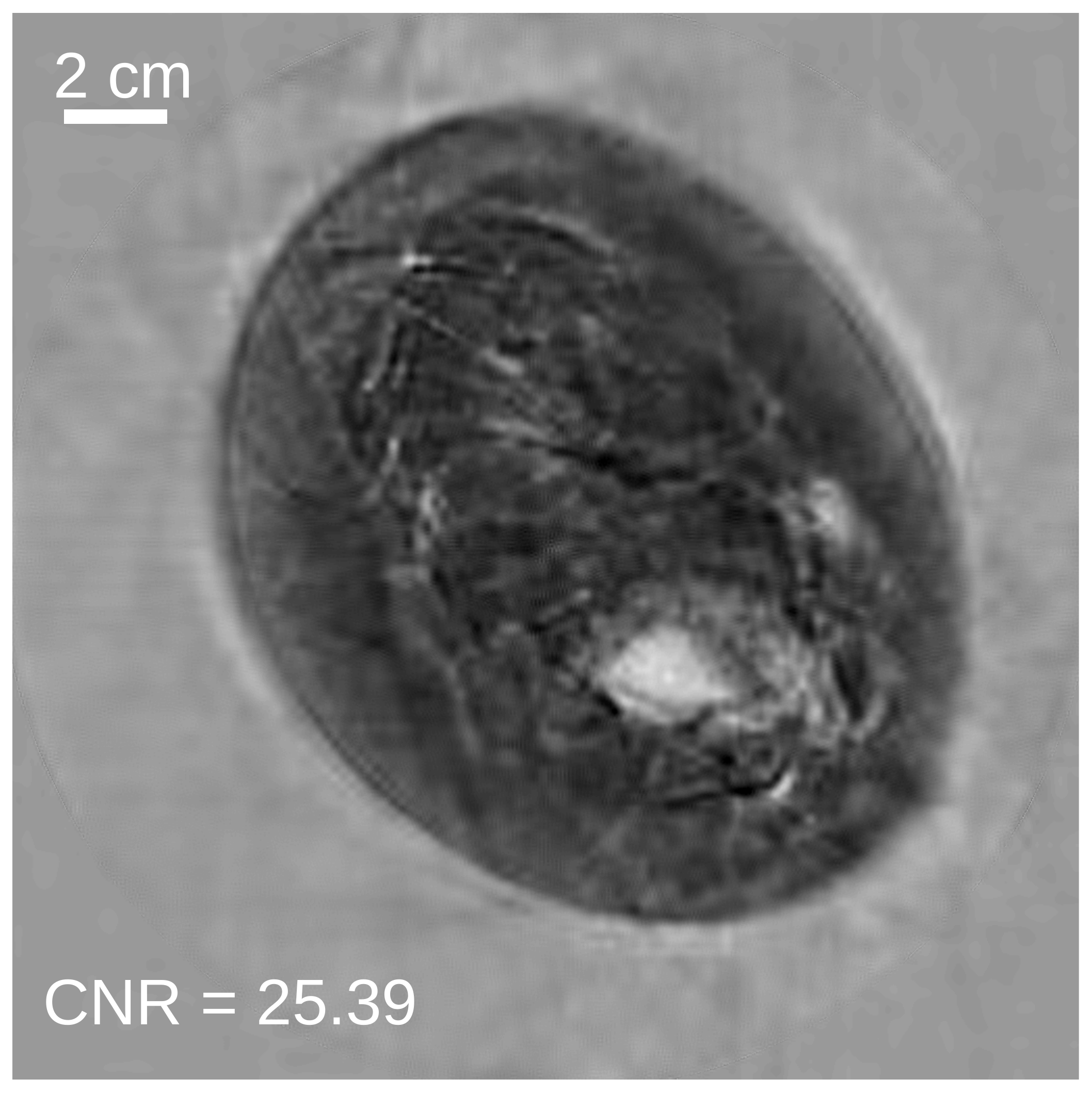}
					\caption{}
				\end{subfigure}%
				\begin{subfigure}[b]{0.25\textwidth}
					\includegraphics[width=\textwidth]{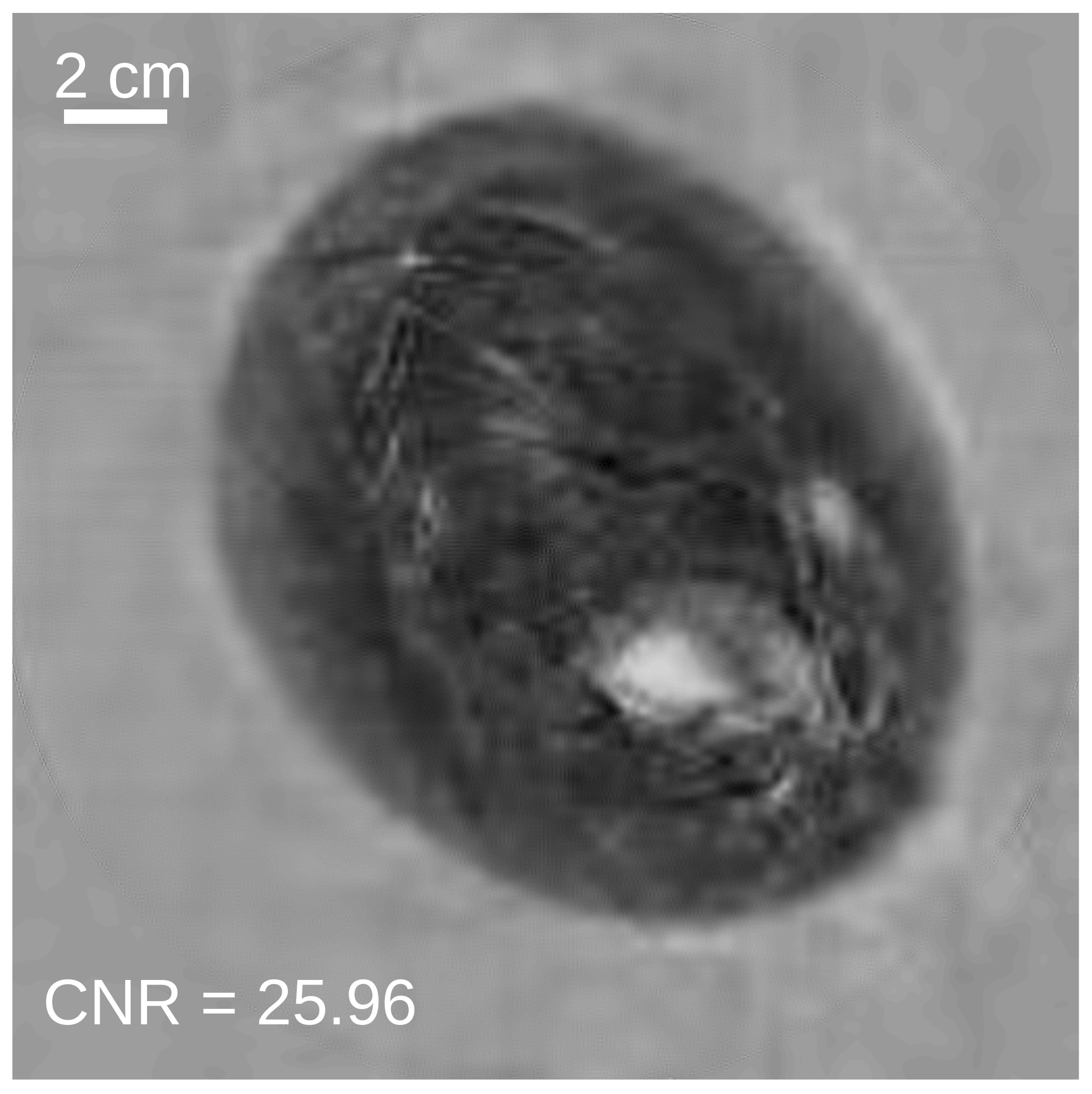}
					\caption{}
				\end{subfigure}
				\caption{Images reconstructed by use of the weighted RDA method with a wavelet-based penalty and regularization parameter values of (a)~$3 \times 10^{-10}$, (b)~$1 \times 10^{-9}$, (c)~$3 \times 10^{-9}$, and (d)~$1 \times 10^{-8}$. Images are shown after 100 iterations and in a grayscale window of [1.38, 1.60] mm/$\mu$s. }
				\label{fig:clin_wavelet}
			\end{figure*}

		As discussed previously, the RDA method allows natural incorporation of non-smooth penalties. This may allow the optimization problem be designed more optimally for a given image reconstruction task. While the determination of an optimal choice of regularization function (let alone the design of the entire optimization problem) is outside the scope of this work, in Fig.~\ref{fig:clin_wavelet}, we show results corresponding to an alternative non-smooth penalty in order to emphasize the flexibility of this approach. The regularization function was chosen to be 
		\begin{align}
			\mathcal{R}\left(\mathbf{c}\right) = \|\boldsymbol{\Phi} \mathbf{c}\|_1 ,
		\end{align}
		where $\boldsymbol{\Phi}$ is the 2-D wavelet transform of the object and the mother wavelet was the 12-tap Daubechies wavelet \cite{mallat_wavelet_2009}. The wavelet transform was computed by use of the GNU Scientific Library \cite{galassi_gnu_????}. Images reconstructed with several regularization parameter values are shown. 
		
%			\begin{figure}[ht]
%				\includegraphics[width=0.5\textwidth]{duric_phant_cnr.png}
%				\caption{Plot of contrast-to-noise ratio (CNR) as a function of the regularization parameter value for SGD with a constant step size (solid) and the weighted RDA method (dashed).}
%%				\caption{(Top row) Images reconstructed by use of SGD with a line search after (a) 20, (b) 50, (c) 100, and (d) 250 iterations with a regularization parameter value of 100. (Bottom row) Images reconstructed by use of weighted RDA after (e) 20, (f) 50, (g) 100, and (h) 250 iterations with a regularization parameter value of 50.}
%			\end{figure}
			
	\section{Summary}
	Waveform inversion with source encoding can produce high-resolution sound speed images without the computational burden of other time-domain waveform inversion approaches. Estimates of the sound speed distribution can be obtained using this method by minimizing an objective function consisting of a data fidelity term and a regularization term. While this optimization problem can be solved using stochastic gradient descent, use of a structured optimization method, such as the regularized dual averaging method, provides several advantages. First, it exploits knowledge of the structure of the cost function to separate the stochastic data fidelity term from the deterministic regularization term. This appears to result in more effective regularization. In the case of the TV semi-norm, noise is more effectively reduced while preserving the accuracy and contrast of the reconstructed images. Second, it does not assume that all terms in the regularization function are differentiable, allowing natural incorporation of non-smooth penalties, such as the total variation semi-norm. Third, it exploits information from past iterations when determining the search direction. This allows the method to employ a line search while avoiding overfitting a particular realization of the encoding vector. This allows a fast initial convergence rate without sacrificing image quality. This was demonstrated through computer-simulation studies involving a numerical breast phantom, generation of a bias-variance curve, and experimental studies involving clinical data.
	
	Some reconstruction parameters were not strictly optimized, particularly for the clinical results. Similar results to those presented could potentially be obtained with coarser temporal or spatial sampling rates. In addition, the number of measurements kept as part of the data filling strategy may not be optimal. The optimal number of measurements will depend on the object and the degree of model mismatch and measurement noise. Further tuning of these parameters could lead to improved performance.
	
	Opportunities for further improvement exist. The acoustic model employed in the calculation of the data fidelity term ignores a number of important factors that could lead to artifacts in the reconstructed images. In particular, the model ignores acoustic attenuation and dispersion and out-of-plane scattering. Since the assumed imaging model is 2-D, scattering out of the plane defined by the transducer ring array is not modeled. It also treats the transducers as ideal point detectors and emitters. Additional investigation of the numerical properties of this approach remains a topic for future study. As noted previously, the frequency content of the excitation pulse and the strength of the acoustic heterogeneities have a sizable impact on the reconstructed images \cite{wang_waveform_2015}. Comparison with other image reconstruction methods is also needed, e.g. \cite{sandhu_frequency_2015,wiskin_non-linear_2012,abdullah_approximate_1999,gemmeke_wave_2016}. 

	\begin{table}[!t]
		\caption{Summary of image reconstruction parameters}
		\label{table:recon_pars}
		\centering
		\begin{tabular}[h]{l | c | c}
			\hline
			Parameter & Simulation & Experimental \\ \hline
			Number of pixels & $1024 \times 1024$ & $2560 \times 2560$ \\
			Grid spacing [mm] & 0.5 & 0.2 \\
			Number of time points & 1800 & 3500 \\
			Sampling frequency [MHz] & 10 & 20 \\ 
			Number of transducers & 256 & 976 \\ \hline
		\end{tabular}
	\end{table}

\appendices
	\section{Line search for weighted RDA method} \label{sec:linesearch}
	The weights for the weighted RDA method were chosen via the line search method described by Alg.~\ref{alg:rda_linesearch}. Other line search methods may produce similar, or even superior, results. Each weight value considered for a given iteration requires $f\left(\vec{c}, \vec{w}\right)$ to be evaluated one additional time. Since $f\left(\vec{c}, \vec{w}\right)$ is evaluated for only one realization of the encoding vector, this requires only one additional wave solver run. This is the same computational cost as for the line search procedure employed for SGD. The goal of the line search procedure is to find weights that improve the convergence rate of the algorithm while minimizing the computational cost needed to select those weights. Thus, it is neither practical nor advisable to choose weights that most minimize the cost function at each iteration. Here, we decrease the weight by a factor of two if the stopping criterion for the line search is not satisfied. This factor can be adjusted to perform the line search more coarsely (larger factor) or more finely (smaller factor).
	\begin{algorithm}
		\caption{Line search for RDA method \label{alg:rda_linesearch}}
		\algsetup{indent=2em}
		\begin{algorithmic}[1]
			\REQUIRE $\vec{c}_0$, $A_{k-1}$, $\vec{w}_k$, $\vec{G}_k$, $\overbar{\vec{G}}_{k-1}$, $f\left(\vec{c}_k, \vec{w}_k\right)$, $\lambda$, $\alpha_{max}$
			\ENSURE $\alpha_k$ \COMMENT{Weight for $k$-th iteration.}
			\STATE{$\tilde{\alpha} \gets \alpha_{max}$} \COMMENT{$\alpha_{max}$ is the initial guess for the weight.}
			\STATE{$found \gets \FALSE$}
			\WHILE {\NOT $found$}
			\STATE{$\tilde{A} \gets A_{k-1} + \tilde{\alpha}$}
			\STATE{$\tilde{\vec{G}} \gets \left(1 - \frac{\tilde{\alpha}}{\tilde{A}}\right) \overbar{\vec{G}}_{k-1} + \frac{\tilde{\alpha}}{\tilde{A}} \vec{G}_k$} 
			\STATE{$\tilde{\mu} \gets \gamma \tilde{A}$} \COMMENT{Should be consistent with Alg.~\ref{alg:rda}.}
			\STATE{$ \tilde{\vec{c}} \gets \text{prox}_{\lambda \tilde{\mu} \mathcal{R}}\left(\vec{c}_{0} - \tilde{\mu} \tilde{\vec{G}}\right) $}
			\IF{$f\left(\tilde{\vec{c}}, \vec{w}_k\right) + \lambda \mathcal{R}\left(\tilde{\vec{c}}\right) < f\left(\vec{c}_k, \vec{w}_k\right) + \lambda \mathcal{R}\left(\vec{c}_k\right)$}
			\STATE{$found \gets \TRUE$}
			\ELSE
			\STATE{$\tilde{\alpha} \gets \tilde{\alpha}/2$}
			\ENDIF
			\ENDWHILE
			\STATE{$\alpha_k \gets \tilde{\alpha}$}
		\end{algorithmic}
	\end{algorithm}

\section*{Acknowledgments}

The authors would like to thank Fatima Anis for her assistance in reconstructing an image from the clinical breast data by use of the adjoint state method described in \cite{anis_investigation_2014}. Computations were performed using the facilities of the Washington University Center for High Performance Computing, which were partially funded by NIH grants 1S10RR022984-01A1 and 1S10OD018091-01. This work was supported in part by NIH awards CA1744601 and EB01696301
and NSF award DMS1614305.

\FloatBarrier


\end{document}